\documentclass[a4paper,12pt]{article}
\pdfoutput=1 
\usepackage{jheppub,amsmath,graphicx,hyperref,subcaption}
\usepackage[utf8]{inputenc}
\usepackage[T1]{fontenc}
\newcommand{\bea}{\begin{eqnarray}}
\newcommand{\eea}{\end{eqnarray}}

\usepackage[T1]{fontenc} % if needed

\title{\boldmath Explaining the hints for lepton flavour universality violation with three $S_2$ leptoquark generations}

\author[a,b]{Andreas Crivellin}
\author[c]{\!\!, Benjamin Fuks}
\author[d]{\! and \! Luc Schnell}

\affiliation[a]{Physik-Institut, Universit\"at Z\"urich,
	Winterthurerstrasse 190, CH-8057 Z\"urich, Switzerland}
\affiliation[b]{Paul Scherrer Institut, CH--5232 Villigen PSI, Switzerland}
\affiliation[c]{Laboratoire de Physique Th\'eorique et Hautes Energies (LPTHE), UMR 7589, Sorbonne Universit\'e et CNRS, 4 place Jussieu, 75252 Paris Cedex 05, France}
\affiliation[d]{Max Planck Institute for Physics, Föhringer Ring 6, 80805 München, Germany}

\emailAdd{andreas.crivellin@cern.ch}
\emailAdd{fuks@lpthe.jussieu.fr}
\emailAdd{schnell@mpp.mpg.de}

\abstract{Leptoquarks are prime candidates for explaining the intriguing hints for lepton flavour universality violation. In particular, the $SU(2)_L$ doublet of scalar leptoquarks $S_2$ is capable of providing an explanation for the tensions between the measurements and the Standard Model predictions in $(g-2)_\mu$, $b\to s\ell^+ \ell^-$ and $b\to c\tau\nu$ processes, as well as in non-resonant di-electron production. However, in the minimal setup with a single leptoquark generation, a common explanation for all these issues is not possible as this would lead to unacceptably large charged lepton flavour violation. We therefore propose a model with three generations of $S_2$, each coupling exclusively to a single lepton flavour, \textit{i.e.}~a model extending the Standard Model particle content by an electroquark, a muoquark and a tauquark. We show that after taking into account other constraints, such as those originating from electroweak precision observables and $\Delta F=2$ processes, it is possible to provide a combined explanation for all these hints of lepton flavour universality violation. Moreover, we find that the presence of the tauquark can generate a dimension-six ${\cal O}_9^U$ operator via off-shell photon penguin diagrams, which, together with the muoquark contribution, further improves the global fit to $b \to s \ell^+ \ell^-$ data. }

\begin{document} 
\preprint{ MPP-2022-30, PSI-PR-22-07, ZU-TH 08/22}
\maketitle
\flushbottom

\section{Introduction}

The Standard Model (SM) of particle physics has been tested extensively and confirmed within the last decades, culminating in the discovery of its last missing piece, the Higgs boson, in 2012 at the LHC~\cite{ATLAS:2012yve,CMS:2012qbp}. Nonetheless, it is evident that the SM cannot be the ultimate theory of nature, as it cannot for example account for dark matter or non-vanishing neutrino masses. However, these phenomena can be accounted for by new physics (NP), which could {\it a priori} lie within a wide energy range (\textit{i.e.} from the keV to the scale of Grand Unification), and is therefore not necessarily within the reach of current or even future colliders. 

Fortunately, in recent years exciting hints for lepton flavour universality violation (LFUV) beyond the SM have been accumulated (see  \textit{e.g.}~Refs.~\cite{Fischer:2021sqw,Crivellin:2021sff} for recent reviews), and an explanation of these anomalies requires TeV-scale new physics. This makes a discovery at the LHC possible, and an observation at future colliders even guaranteed. In particular, measurements of $b\to s\ell^+\ell^-$~\cite{CMS:2014xfa,Aaij:2015oid,Abdesselam:2016llu,Aaij:2017vbb,Aaij:2019wad,Aaij:2020nrf,LHCb:2021trn,BELLE:2019xld,Belle:2019oag,LHCb:2016ykl,LHCb:2021zwz,LHCb:2020zud} and $b\to c\tau\nu$~\cite{Lees:2012xj,Lees:2013uzd,Aaij:2015yra,Aaij:2017deq,Aaij:2017uff,Abdesselam:2019dgh} observables deviate from their corresponding SM predictions by more than $7\,\sigma$~\cite{Altmannshofer:2021qrr,Geng:2021nhg,Alguero:2021anc,Hurth:2021nsi,Kowalska:2019ley,Ciuchini:2021smi,DAmico:2017mtc,Arbey:2019duh,Kumar:2019nfv} (this tension being reduced to $4\!-\!5\sigma$ when including only LFUV observables~\cite{Altmannshofer:2021qrr,Geng:2021nhg,Alguero:2021anc,Hurth:2021nsi,Isidori:2021vtc}) and by more than $3\,\sigma$~\cite{HFLAV:2019otj} respectively. Moreover, measurements of the anomalous magnetic moment of the muon $(g-2)_\mu$~\cite{Muong-2:2006rrc,Muong-2:2021ojo} are known to deviate from the SM predictions by more than $4\,\sigma$ when following the community consensus~\cite{Aoyama:2020ynm}. The latter is based on the results of Refs.~\cite{Aoyama:2012wk,Aoyama:2019ryr,Czarnecki:2002nt,Gnendiger:2013pva,Davier:2017zfy,Keshavarzi:2018mgv,Colangelo:2018mtw,Hoferichter:2019gzf,Davier:2019can,Keshavarzi:2019abf,Kurz:2014wya,Melnikov:2003xd,Masjuan:2017tvw,Colangelo:2017fiz,Hoferichter:2018kwz,Gerardin:2019vio,Bijnens:2019ghy,Colangelo:2019uex,Blum:2019ugy,Colangelo:2014qya} that do not include the recent lattice findings of the Budapest-Marseilles-Wuppertal collaboration (BMWc) for the hadronic vacuum polarisation (HVP) contributions~\cite{Borsanyi:2020mff}. Whereas they render the SM predictions for $a_\mu$ compatible with data, the BMWc results are in tension with the HVP value determined from $e^+e^-\to$ hadrons data~\cite{Davier:2017zfy,Keshavarzi:2018mgv,Colangelo:2018mtw,Hoferichter:2019gzf,Davier:2019can,Keshavarzi:2019abf}. Furthermore, HVP also enters global electroweak (EW) fits~\cite{Passera:2008jk}, and its (indirect) determination leads to a value smaller than that obtained by the BMWc~\cite{Haller:2018nnx}. Therefore, making use of the BMWc predictions for the HVP contributions to $a_\mu$ would increase the tension originating from the EW fits~\cite{Crivellin:2020zul,Keshavarzi:2020bfy}, and we opted for omitting it until the situation gets clarified.

Prime examples that can address most, if not even all, LFUV anomalies are models featuring leptoquarks (LQs), hypothetical new particles with common couplings to quarks and leptons. LQs were originally proposed in the Pati-Salam model~\cite{Pati:1974yy} and in Grand Unified Theories (GUTs)~\cite{Georgi:1974sy,Dimopoulos:1980hn,Senjanovic:1982ex,Frampton:1989fu,Witten:1985xc}, and it has been shown that the anomalies in $b\to s\ell^+\ell^-$~\cite{Alonso:2015sja, Calibbi:2015kma, Hiller:2016kry, Bhattacharya:2016mcc, Buttazzo:2017ixm, Barbieri:2015yvd, Barbieri:2016las, Calibbi:2017qbu, Crivellin:2017dsk, Bordone:2018nbg, Kumar:2018kmr, Crivellin:2018yvo, Crivellin:2019szf, Cornella:2019hct, Bordone:2019uzc, Bernigaud:2019bfy,Aebischer:2018acj,Fuentes-Martin:2019ign,Popov:2019tyc,Fajfer:2015ycq,Blanke:2018sro,deMedeirosVarzielas:2019lgb,deMedeirosVarzielas:2015yxm,Crivellin:2019dwb,Saad:2020ihm,Saad:2020ucl,Gherardi:2020qhc,DaRold:2020bib,Greljo:2021xmg}, $R(D^{(*)})$~\cite{Alonso:2015sja, Calibbi:2015kma, Fajfer:2015ycq, Bhattacharya:2016mcc, Buttazzo:2017ixm, Barbieri:2015yvd, Barbieri:2016las, Calibbi:2017qbu, Bordone:2017bld, Bordone:2018nbg, Kumar:2018kmr, Biswas:2018snp, Crivellin:2018yvo, Blanke:2018sro, Heeck:2018ntp,deMedeirosVarzielas:2019lgb, Cornella:2019hct, Bordone:2019uzc,Sahoo:2015wya, Chen:2016dip, Dey:2017ede, Becirevic:2017jtw, Chauhan:2017ndd, Becirevic:2018afm, Popov:2019tyc,Fajfer:2012jt, Deshpande:2012rr, Freytsis:2015qca, Bauer:2015knc, Li:2016vvp, Zhu:2016xdg, Popov:2016fzr, Deshpand:2016cpw, Becirevic:2016oho, Cai:2017wry, Altmannshofer:2017poe, Kamali:2018fhr, Mandal:2018kau, Azatov:2018knx, Wei:2018vmk, Angelescu:2018tyl, Kim:2018oih, Aydemir:2019ynb, Crivellin:2019qnh, Yan:2019hpm,Crivellin:2017zlb, Marzocca:2018wcf,Fuentes-Martin:2020pww, Bigaran:2019bqv,Crivellin:2019dwb,Saad:2020ihm,Dev:2020qet,Saad:2020ucl,Altmannshofer:2020axr,Fuentes-Martin:2020bnh,Gherardi:2020qhc,DaRold:2020bib}, $b\to c\tau\nu$~\cite{Alonso:2015sja, Calibbi:2015kma, Fajfer:2015ycq, Bhattacharya:2016mcc, Buttazzo:2017ixm, Barbieri:2015yvd, Barbieri:2016las, Calibbi:2017qbu, Bordone:2017bld, Bordone:2018nbg, Kumar:2018kmr, Biswas:2018snp, Crivellin:2018yvo, Blanke:2018sro, Heeck:2018ntp,deMedeirosVarzielas:2019lgb, Cornella:2019hct, Bordone:2019uzc,Sahoo:2015wya, Chen:2016dip, Dey:2017ede, Becirevic:2017jtw, Chauhan:2017ndd, Becirevic:2018afm, Popov:2019tyc,Fajfer:2012jt, Deshpande:2012rr, Freytsis:2015qca, Bauer:2015knc, Li:2016vvp, Zhu:2016xdg, Popov:2016fzr, Deshpand:2016cpw, Becirevic:2016oho, Cai:2017wry, Altmannshofer:2017poe, Kamali:2018fhr, Mandal:2018kau, Azatov:2018knx, Wei:2018vmk, Angelescu:2018tyl, Kim:2018oih, Aydemir:2019ynb, Crivellin:2019qnh, Yan:2019hpm,Crivellin:2017zlb, Marzocca:2018wcf,Bigaran:2019bqv,Crivellin:2019dwb,Saad:2020ihm,Dev:2020qet,Saad:2020ucl,Altmannshofer:2020axr,Fuentes-Martin:2020bnh,Gherardi:2020qhc,DaRold:2020bib,Endo:2021lhi,Belanger:2021smw,Lee:2021jdr}, and $(g-2)_\mu$~\cite{Bauer:2015knc,Djouadi:1989md, Chakraverty:2001yg,Cheung:2001ip,Popov:2016fzr,Chen:2016dip,Biggio:2016wyy,Davidson:1993qk,Couture:1995he,Mahanta:2001yc,Queiroz:2014pra,ColuccioLeskow:2016dox,Chen:2017hir,Das:2016vkr,Crivellin:2017zlb,Cai:2017wry,Crivellin:2018qmi,Kowalska:2018ulj,Dorsner:2019itg,Crivellin:2019dwb,DelleRose:2020qak,Saad:2020ihm,Bigaran:2020jil,Dorsner:2020aaz,Fuentes-Martin:2020bnh,Gherardi:2020qhc,Babu:2020hun,Crivellin:2020tsz,Marzocca:2021azj,Wang:2021uqz,Perez:2021ddi} can all be explained by them. Moreover, LQs can provide an explanation for the CMS excess in non-resonant di-electron production~\cite{Crivellin:2021egp,Crivellin:2021bkd}. 

In this context, models including a $SU(2)_L$ leptoquark doublet $S_2$ that transforms under the SM gauge group $SU(3)_c \times SU(2)_L \times U(1)_Y$ as $\left(\mathbf{3}, \mathbf{2}, \frac{7}{6}\right)$, from now on called $\Phi_2$ (in the literature also called $R_2$), are particularly interesting. This LQ can restore the agreement between theory and data for $b\to s\ell^+\ell^-$ observables via a $W$-box contribution~\cite{Becirevic:2017jtw}, for $b\to c\tau\nu$ observables and the excess~\cite{Crivellin:2021rbf} in CMS di-lepton data~\cite{CMS:2021ctt} via tree-level contributions~\cite{Crivellin:2021egp,Crivellin:2021bkd}, and for $(g-2)_\mu$ via an $m_t/m_\mu$ chirally enhanced effect. Furthermore, the tension in the difference of the $B\to D^*\ell\nu$ forward-backward asymmetries ($\Delta A_{\rm FB}$)~\cite{Belle:2018ezy,Bobeth:2021lya} can be softened~\cite{Crivellin:2020mjs}, and the global EW fit can be improved through the generation of a shift in the $W$-boson mass predictions~\cite{Crivellin:2020ukd}, where a constructive effect is currently preferred~\cite{deBlas:2021wap}, via LQ interactions with the Higgs field~\cite{Crivellin:2021ejk}.

However, a combined explanation of the flavour anomalies in the minimal setup with a single leptoquark $S_2$ is not possible as this would lead to unacceptably large charged lepton flavour violation (LFV). In order to avoid this, we propose to extend the SM by three generations of leptoquarks, like the three generations of squarks in minimal $R$-parity violating supersymmetry~\cite{Barger:1989rk,Barbier:2004ez}. In this setup each LQ flavour couples to the corresponding lepton flavour, {\it i.e.}~the electroquark $\Phi_{2,e}$ coupling to electrons, the muoquark $\Phi_{2,\mu}$ to muons and the tauquark $\Phi_{2,\tau}$ to tau leptons. While this does not introduce additional degrees of freedom concerning the couplings to fermions, it decouples the LQ interactions with the different lepton generations from each others, rendering joint explanations of the hints for LFUV possible.  

We proceed by defining our three-generation model of $S_2$ in Section~\ref{sec:setup}, then discuss the most relevant observables in Section~\ref{sec:observables}. Section~\ref{sec:phenomenology} contains our phenomenological analysis, including a statistical analysis, and we conclude in Section~\ref{sec:conclusion}. 

\section{Setup}
\label{sec:setup}
Minimal extensions of the SM with a single $SU(2)_L$ doublet scalar LQ $\Phi_2$ have been studied in detail in the literature~\cite{Becirevic:2017jtw, Popov:2019tyc, Angelescu:2021lln, Angelescu:2018tyl, Crivellin:2021egp, Crivellin:2021bkd, He:2021yck, Bigaran:2020jil, Bigaran:2021kmn, deMedeirosVarzielas:2015yxm, Sahoo:2015wya, Kamali:2018fhr, Chakraverty:2001yg, Cheung:2001ip, Queiroz:2014pra, ColuccioLeskow:2016dox, Chen:2017hir, Kowalska:2018ulj, Dorsner:2020aaz, Crivellin:2020tsz, Iguro:2020keo}. In such models, the most general LQ Lagrangian reads~\cite{Crivellin:2021ejk}
\begin{equation}
	\begin{aligned}
		\mathcal{L}_\text{LQ} &= \bigg(Y_{ij}^{RL} \bar{u}_i \left[\Phi_2\cdot L_j \right] + Y_{ij}^{LR} \left[\bar{Q}_i e_j \Phi_2\right] + \text{ H.c.}\bigg)
		- \bigg(M^2 + Y^{H(1)} \Big[ H^\dagger H \Big]\bigg) \Phi_2^\dagger \Phi_2 \\
		&\phantom{===} - Y^{H(3)} \Big[H \cdot \Phi_2 \Big]^\dagger \Big[ H \cdot \Phi_2 \Big] + \mathcal{L}_{4\Phi} \,,
\end{aligned}
\label{eq:1_LQ_Lagrangian}
\end{equation}
where $Q_i \sim \left(\mathbf{3}, \mathbf{2}, \frac{1}{6}\right)$,  $L_i \sim \left(\mathbf{1}, \mathbf{2}, -\frac{1}{2}\right)$, $e_i \sim \left(\mathbf{1}, \mathbf{1}, -1 \right)$, $u_i \sim \left(\mathbf{3}, \mathbf{1}, \frac{2}{3}\right)$ and $H \sim \left(\mathbf{1}, \mathbf{2}, \frac{1}{2}\right)$ are the usual SM fields, $i,j = 1,2,3$ are flavour indices and the dot stands for the invariant product of two fields lying in the fundamental representation of $SU(2)_L$. Considering the version of the above Lagrangian after EW symmetry breaking, we  refer to the SM fermions by their usual names $q \in \{u,c,t,d,s,b\}$, $\ell \in \{e, \mu, \tau \}$ and $\nu_\ell \in \{ \nu_e, \nu_\mu, \nu_\tau \}$. Moreover, the Lagrangian term $\mathcal{L}_{4 \Phi}$ contains the LQ quartic interactions~\cite{Crivellin:2021ejk} that are not relevant for our analysis.

The couplings $Y^{LR}_{ij}$ and $Y^{RL}_{ij}$ are \textit{a priori} arbitrary complex matrices in the flavour space, leading in general to charged lepton flavour violation. As outlined in the introduction, explaining anomalies related to different lepton generations at the same time leads to unacceptably large effects in charged lepton flavour violating observables. However, one can avoid this by assigning a lepton flavour number to $\Phi_2$. In fact, in Refs.~\cite{Davighi:2020qqa,Greljo:2021npi} a $L_\mu-L_\tau$ symmetry~\cite{He:1990pn,Foot:1990mn,He:1991qd} was used to impose that $\Phi_2$ only interacts with second-generation leptons. However, this model can at most address muon-related anomalies, and it might be considered unnatural to give the muon such a special treatment because the tau lepton is even more massive. 

Therefore, we propose to introduce three generations of leptoquarks $\Phi_2$, so that the field $\Phi_{2,\ell}$ now carries a generation index $\ell = 1,2,3$, or equivalently and interchangeably $\ell = e,\mu,\tau$. In addition, we require that $\Phi_{2,\ell}$ only interacts with the lepton flavour $\ell$. While we remain agnostic about the specific underlying mechanism that enforces this, it could for instance again be achieved via a $L_\mu-L_\tau$ symmetry by assigning the charges $0$, $1$ and $-1$ to $\Phi_{2,e}$, $\Phi_{2,\mu}$ and $\Phi_{2,\tau}$, respectively. In addition,  the assignment of lepton flavours to LQs automatically avoids proton decay to all orders in perturbation theory, as it forbids di-quark couplings, in case this coupling would be allowed by the other quantum numbers.

In this setup, the LQ interaction Lagrangian reads
\begin{equation}
	\begin{aligned}
		\mathcal{L}_\text{LQ} &= \sum_\ell \bigg( Y_{i\ell}^{RL} \bar{u}_i \left[\Phi_{2,\ell} \cdot L_\ell \right] + Y_{i\ell}^{LR} \left[\bar{Q}_i e_\ell \Phi_{2,\ell}\right] + \text{ H.c.} \bigg)\\
		&\phantom{===}- \bigg(M_\ell^2 + Y^{H(1)}_\ell \Big[H^\dagger H \Big]\bigg) \Phi_{2,\ell}^\dagger \Phi_{2,\ell}
		 - Y_\ell^{H(3)} \Big[H \cdot \Phi_{2,\ell} \Big]^\dagger \Big[H \cdot \Phi_{2,\ell} \Big] + \mathcal{L}_{4\Phi}  \,.
\end{aligned}
\label{eq:3_LQ_Lagrangian}
\end{equation}
Comparing Eq.~(\ref{eq:3_LQ_Lagrangian}) to Eq.~(\ref{eq:1_LQ_Lagrangian}), it is apparent that we do not introduce any additional degrees of freedom in the LQ couplings to fermions compared to the minimal model with a single $\Phi_2$.

Once the Higgs doublet $H$ acquires its vacuum expectation value $v$, the Yukawa terms generate mass matrices for quarks and leptons. Here we assume that the lepton Yukawa matrices are diagonal in the basis of Eq.~(\ref{eq:3_LQ_Lagrangian}) such that no charged LFV is induced by EW symmetry breaking. This means that lepton flavour is only broken by the tiny neutrino masses, with negligible consequences in our phenomenological analysis. Concerning  quarks, we choose to work in the down-type quark basis, so that CKM matrix elements only appear in couplings involving left-handed up-type quarks. We therefore define
\begin{equation}
\hat{Y}^{LR}_{i\ell} \equiv V^{\rm CKM}_{ij} Y^{LR}_{j \ell} 
\end{equation}
for brevity. 

\section{Observables}
\label{sec:observables}

In the following, we discuss the most relevant observables allowing to test and constrain our model. For the low-energy precision and flavour observables we match the full LQ theory onto the weak effective theory (WET) whose Lagrangian is generically written as
\begin{equation}
	\mathcal{L}_{\text{eff}} = \sum_i \mathcal{C}_i \mathcal{O}_i \,.
\end{equation}
We refer to the manual of \texttt{flavio}~\cite{Straub:2018kue}, a package that we employ in our phenomenological analysis, for a precise definition of the operators.

\subsection{$R_{D^{(*)}}$ anomalies}

We consider the ratios 
\begin{equation}
\begin{aligned}
	R_{D^{(*)}} = \left. \frac{\text{Br}(B \to D^{(*)} \tau \bar{\nu})}{\text{Br}(B \to D^{(*)} \ell \bar{\nu})} \right |_{\ell \in \{e, \mu \}}\,,
\end{aligned}
\end{equation}
whose current experimental averages read
\begin{equation}
\begin{aligned}
  R^\text{exp}_{D} &= 0.346(31) ~\text{\cite{Lees:2013uzd, Huschle:2015rga, Abdesselam:2019dgh}} 
  \qquad\text{and}\qquad
  R^\text{exp}_{D^*}& = 0.296(16) ~\text{\cite{Lees:2013uzd, Huschle:2015rga, Abdesselam:2019dgh, Hirose:2016wfn}} \,.
\end{aligned}
\end{equation}
These values are compared to our theoretical predictions whose SM component is provided by~\texttt{flavio}~\cite{MILC:2015uhg,Na:2015kha,Fajfer:2012vx,FlavourLatticeAveragingGroup:2019iem,Iguro:2020keo},
\begin{equation}
\begin{aligned}
  R^\text{SM}_{D} &=0.297(8)
  \qquad\text{and}\qquad
  R^\text{SM}_{D^*} &= 0.245(8) \,,
\end{aligned}
\end{equation}
although more accurate predictions have been calculated in the meantime~\cite{HFLAV:2019otj,Bigi:2016mdz,Gambino:2019sif,Bordone:2019vic}. At low energy, the new physics contributions are described by the WET operators 
\begin{equation}
\begin{aligned}
  \left(\mathcal{O}_{S_L} \right)_{bc \tau \nu_\tau} =&\  -\frac{4G_F}{\sqrt{2}} V^{\rm CKM}_{23} \left( \bar{c} P_L b \right) \left(\bar{\tau} P_L \nu_\tau \right) \,,
  \\
  \left(\mathcal{O}_{T} \right)_{bc \tau \nu_\tau} =&\ -\frac{4G_F}{\sqrt{2}} V^{\rm CKM}_{23} \left( \bar{c} \sigma^{\mu\nu} P_L b \right) \left(\bar{\tau} \sigma_{\mu\nu} P_L \nu_\tau \right)\,, \\
\end{aligned}
\label{eq:wilson_operators_bctaunu}
\end{equation}
where $P_L$ and $P_R$ (for further reference) are the usual chirality projectors and $G_F$ is the Fermi constant. The corresponding Wilson coefficients are evaluated at the matching scale,
\begin{equation}
\left(\mathcal{C}_{S_L}\right)^\text{LQ}_{bc \tau \nu_\tau} =  4 \left(\mathcal{C}_{T}\right)^\text{LQ}_{bc \tau \nu_\tau} = - \frac{\sqrt{2}}{4 G_F V^{\rm CKM}_{23}} \left( \frac{Y^{RL}_{2\tau} Y^{LR*}_{3\tau}}{2M_\tau^2} \right)\,.
\label{eq:matching_bctaunu}
\end{equation}
While WET renormalisation group (RG) running is accounted for by \texttt{flavio}, we additionally multiply the predictions by appropriate correction factors to account for RG running from $M_\tau$ to the electroweak scale $M_W$ in the SM Effective Field Theory (SMEFT). These factors are determined using the {\sc Python} package \texttt{wilson}~\cite{Aebischer:2018bkb}\footnote{Whereas QCD matching corrections exist~\cite{Aebischer:2018acj}, we do not include them in our analysis as the package \texttt{wilson} is restricted to one-loop RG evolution.}, and are equal to 0.97 and 0.93 for $\mathcal{C}_{S_L}$ and $\mathcal{C}_T$ respectively. They modify the ratio between $\mathcal{C}_{S_L}$ and $\mathcal{C}_T$ at the low scale, and are thus relevant in our analysis.

\subsection{Rare $B$ decays to kaons ($b \to s \ell^+ \ell^-$)}
\label{sec:bsll_tauquark}
\begin{figure}
    \hspace{70px}
	\begin{subfigure}[b]{0.3\textwidth}
	    \begin{center}
		\includegraphics[width=0.8\textwidth]{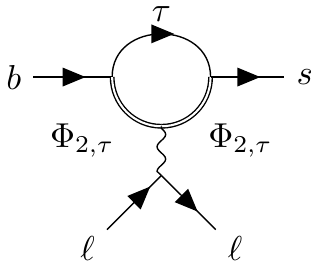}
		\end{center}
		\caption*{($\mathcal{C}_9$-1)}
	\end{subfigure} \hspace{20px}
	\begin{subfigure}[b]{0.3\textwidth}
	    \begin{center}
		\includegraphics[width=0.8\textwidth]{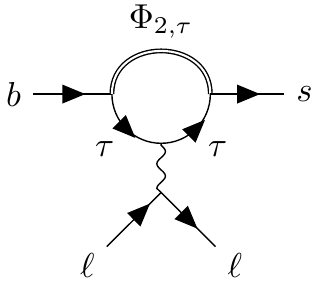}
		\end{center}
		\caption*{($\mathcal{C}_9$-2)}
	\end{subfigure} \newline
	\begin{subfigure}[b]{0.3\textwidth}
	    \begin{center}
		\includegraphics[width=0.8\textwidth]{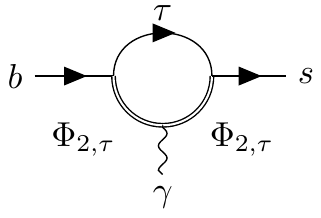}
		\end{center}
		\vspace{5px}
		\caption*{($\mathcal{C}_7$-1)}
	\end{subfigure} \hfill
	\begin{subfigure}[b]{0.3\textwidth}
	    \begin{center}
		\includegraphics[width=0.8\textwidth]{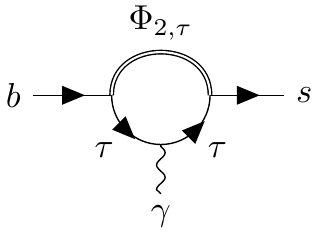}
		\end{center}
		\vspace{5px}
		\caption*{($\mathcal{C}_7$-2)}
	\end{subfigure}	\hfill
	\begin{subfigure}[b]{0.3\textwidth}
	    \begin{center}
		\includegraphics[width=0.8\textwidth]{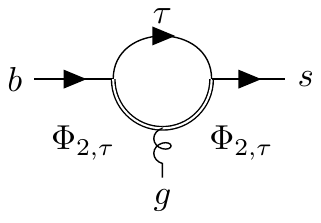}
		\end{center}
		\vspace{5px}
		\caption*{($\mathcal{C}_8$)}
	\end{subfigure}	
	\caption{Feynman diagrams depicting the one-loop contributions of $\Phi_{2, \tau}$ to the Wilson coefficients $\left(\mathcal{C}_9\right)_{bs\ell \ell}$, $\left(\mathcal{C}_7\right)_{bs}$ and  $\left(\mathcal{C}_8\right)_{bs}$, that play an important role in the global fit to $b \to s \ell^+ \ell^-$ data. }
	\label{fig:Bsll_feynman}
\end{figure}

The LHCb and Belle collaborations determined ratios $R^{[q_1^2, q_2^2]}_{K^{(*)}}$ of $B$-meson branching fractions into muons and electrons for different intervals $q_1^2 \leq q^2 \leq q_2^2$, the SM predicting values close to one in all energy bins~\cite{Bordone:2016gaq, Isidori:2020acz, Serra:2016ivr, Capdevila:2017ert, Bharucha:2015bzk, Jager:2014rwa}. In our analysis, we use the measurements selected in the package \texttt{smelli}~\cite{Aebischer:2018iyb} for the $B^+\to K^{(*)}\ell^+\ell^-$ and $B_0\to K^{(*)} \ell^+\ell^-$ decays, 
\begin{equation}
\begin{aligned}
  R_K^{\text{exp} [1.1, 6.0]} &= 0.846^{+0.042}_{-0.039}{}^{+0.013}_{-0.012}  ~ \text{\cite{LHCb:2021trn}} \,,\\
  R_{K_s^0}^{\text{exp} [1.1, 6.0]} &= 0.66^{+0.20}_{-0.14}{}^{+0.02}_{-0.04}  ~ \text{\cite{LHCb:2021lvy}} \,,\\
  R_{K^{*+}}^{\text{exp} [0.045, 6.0]} &= 0.70^{+0.18}_{-0.13}{}^{+0.03}_{-0.04}  ~ \text{\cite{LHCb:2021lvy}} \,,
\end{aligned}
\end{equation}
that we complement with the complete set of observables related to rare $B$-decay into kaons included both in the programs \texttt{flavio} and \texttt{smelli}.

After integrating out all leptoquarks in our model, the relevant effective operators for $b \to s \ell^+ \ell^-$ decays are given by
\begin{align}
\begin{aligned}
\left(\mathcal{O}_{7} \right)_{bs} &= \frac{4 G_F}{\sqrt{2}} V^{\rm CKM}_{33} \big(V^{\rm CKM}_{32}\big)^* \frac{e}{16 \pi^2} m_b \left( \bar{s} \sigma^{\mu\nu} P_R b \right) F_{\mu \nu}  \,, \\
\left(\mathcal{O}_{8} \right)_{bs} &= \frac{4 G_F}{\sqrt{2}} V^{\rm CKM}_{33} \big(V^{\rm CKM}_{32}\big)^* \frac{g_s}{16 \pi^2} m_b \left( \bar{s} \sigma^{\mu \nu} T^a P_R b \right) G^a_{\mu \nu} \,, \\
\left(\mathcal{O}_9 \right)_{bs \ell \ell} &= \frac{4 G_F}{\sqrt{2}} V^{\rm CKM}_{33} \big(V^{\rm CKM}_{32}\big)^* \frac{e^2}{16 \pi^2} \left( \bar{s} \gamma^\mu P_L b \right) \left(\bar{\ell} \gamma_\mu \ell \right) \,, \\
\left(\mathcal{O}_{10} \right)_{bs \ell \ell} &= \frac{4 G_F}{\sqrt{2}} V^{\rm CKM}_{33} \big(V^{\rm CKM}_{32}\big)^* \frac{e^2}{16 \pi^2} \left( \bar{s} \gamma^\mu P_L b \right) \left(\bar{\ell} \gamma_\mu \gamma_5 \ell \right)\,,
\end{aligned}
\end{align}
where $e$ stands for the electromagnetic coupling constant, and $F_{\mu\nu}$ and $G_{\mu\nu}$ are the photon and gluon field strength tensors. While the anomalies in $b \to s \ell^+ \ell^-$ observables mainly point towards new physics interacting with muons, the presence of a tauquark $\Phi_{2, \tau}$ in the model yields a flavour-universal contribution to $\mathcal{C}_9$ via the off-shell photon penguin diagrams shown in the first line of Figure~\ref{fig:Bsll_feynman}. Calling it $\mathcal{C}_9^U$, it is given in the leading-logarithmic approximation by
\begin{equation}
\begin{aligned}
\left( \mathcal{C}_9^U \right)^\text{LQ} = \frac{\sqrt{2}}{4 G_F} \frac{1}{V^{\rm CKM}_{33} \big(V^{\rm CKM}_{32}\big)^*} \left(\frac{Y^{LR}_{2\tau} Y^{LR*}_{3\tau} \log\left( \frac{M_\tau^2}{m_b^2}\right)}{3 M_\tau^2} \right) \,.
\end{aligned}
\label{eq:C_9Umatching}
\end{equation}
Additionally, the $\mathcal{C}_7$ and $\mathcal{C}_8$ Wilson coefficients are extracted from the diagrams shown in the second line of Figure~\ref{fig:Bsll_feynman}, and read
\begin{equation}
\left(\mathcal{C}_7\right)^\text{LQ}_{bs} = \frac83 \left(\mathcal{C}_8\right)^\text{LQ}_{bs} = \frac{\sqrt{2}}{4 G_F} \frac{1}{V^{\rm CKM}_{33} \big(V^{\rm CKM}_{32}\big)^*}\left( \frac{Y^{LR}_{2\tau} Y^{LR*}_{3\tau}}{9 M_\tau^2} \right)\,,
\label{eq:C_7C_8matching}\end{equation}
following \texttt{flavio}'s sign conventions for covariant derivatives (see Appendix A.3 of Ref.~\cite{Aebischer:2018iyb}). While $\mathcal{C}_8$ does not contribute to the $b \to s \ell^+ \ell^-$ observables directly, it mixes into $\mathcal{C}_7$ when RG running from the scale $M_\tau$ to $m_b$ is accounted for,
\begin{equation}
\begin{pmatrix}
\left(\mathcal{C}_7 \right )_{bs}
\\ 
\left(\mathcal{C}_8 \right )_{bs}
\end{pmatrix}_\text{4.2 GeV} = 
\begin{pmatrix}
0.49 & 0.12 \\ 
0.0055 & 0.54
\end{pmatrix}
\begin{pmatrix}
\left(\mathcal{C}_7 \right )_{bs} \\ 
\left(\mathcal{C}_8 \right )_{bs}
\end{pmatrix}_\text{1.7 TeV} \,.
\end{equation}

In addition to these tauquark effects in $b \to s \ell^+ \ell^-$, we get well-known contributions from the muoquark $\Phi_{2, \mu}$. Writing $\mathcal{C}_9^{\ell} \equiv \left(\mathcal{C}_9\right)_{bs\ell \ell}$ and $\mathcal{C}_{10}^{\ell} \equiv \left(\mathcal{C}_{10}\right)_{bs\ell \ell}$ to simplify the notation (for further reference), the tree-level matching to the WET gives
\begin{align}
  \left(\mathcal{C}_9^\mu\right)^\text{LQ} &= \left(\mathcal{C}_{10}^\mu\right)^\text{LQ} =   -\frac{\sqrt{2}}{4 G_F} \frac{1}{V^{\rm CKM}_{33} \big(V^{\rm CKM}_{32}\big)^*} \frac{16 \pi^2}{e^2} \left(\frac{Y^{LR}_{2\mu} Y^{LR*}_{3\mu}}{4M_{\mu}^2} \right) \,.
\end{align}
whereas the one-loop contributions~\cite{Becirevic:2017jtw} are
\begin{align}
  \left(\mathcal{C}^\mu_9\right)^\text{LQ} = - \left(\mathcal{C}^\mu_{10}\right)^\text{LQ} = \sum^3_{j,k = 1} \frac{V^{\rm CKM}_{j3} \big(V^{\rm CKM}_{k2}\big)^*}{V^{\rm CKM}_{33} \big(V^{\rm CKM}_{32}\big)^*} Y^{RL}_{k\mu} Y^{RL*}_{j\mu} \mathcal{F}_1(x_{j}, x_{k}, x_\mu) \,,
\end{align}
with $x_{\mu} = M_{\mu}^2/m_W^2$, $x_{i} = m_{u_i}^2/m_W^2$ and~\cite{Crivellin:2020mjs}
\begin{equation}
\mathcal{F}_1 (x,y,z) = \frac{\sqrt{xy}}{8 e^2} \Bigg [ 
   \frac{y(y\!-\!4) \log y}{(y\!-\!1)(x\!-\!y)(y\!-\!z)}
 + \frac{x(x\!-\!4) \log x}{(x\!-\!1)(y\!-\!x)(x\!-\!z)} 
 - \frac{z(z\!-\!4) \log z}{(z\!-\!1)(z\!-\!x)(z\!-\!y)} \Bigg ] \,.
\end{equation}

\subsection{$B$ and $D$ decays into tau leptons ($b \to s \tau^+\tau^-$ and $D_s \to \tau \nu$)}
The structure of the tauquark couplings that yield contributions to $b \to c \tau \nu$ and $b \to s \tau^+ \tau^-$ decay observables also leads to effects in the branching ratios $\text{Br}(B \to K^{(*)} \tau^+ \tau^-)$, $\text{Br}(B_s \to \tau^+ \tau^-)$ and $\text{Br}(D_s \to \tau \nu)$. In the SM, their values are given by $1.7(2)\times 10^{-7}$, $7.8(3)\times 10^{-7}$ and $0.0532(4)$ according to \texttt{flavio}. 

The LQ effects from our model in $B \to K^{(*)} \tau^+ \tau^-$ and $B_s \to \tau^+ \tau^-$ decay observables are embedded in the WET coefficients of Eqs.~(\ref{eq:C_9Umatching}) and~(\ref{eq:C_7C_8matching}). In contrast, the $D_s \to \tau \nu$ decay is sensitive to the $\left( {\cal O}_{S_L}\right)^\text{LQ}_{sc\tau \nu_\tau}$ and $\left( {\cal O}_T\right)^\text{LQ}_{sc\tau \nu_\tau}$ operators, whose associated Wilson coefficients can be derived from Eq.~(\ref{eq:matching_bctaunu}) by replacing the quark index 3 with 2. We finally have
\begin{equation}
    \text{Br}(B_s \to \tau^+ \tau^-)=\left|1 + \left(\mathcal{C}^\tau_{10} \right)^\text{LQ}/\left(\mathcal{C}^\tau_{10}\right)^{\rm SM} \right|^2 \,,
\end{equation}
while for $\text{Br}(B \to K^{(*)} \tau^+ \tau^-)$ we refer the reader to Ref.~\cite{Capdevila:2017iqn}.

\subsection{Difference in forward-backward asymmetries $\Delta A_{\text{FB}}$}
Ref.~\cite{Bobeth:2021lya} unveiled a tension of about $4\sigma$ between the SM predictions and data for the difference between the forward-backward asymmetry originating from $\bar{B} \to D^* \mu \bar{\nu}$ decays and that originating from $\bar{B} \to D^* e \bar{\nu}$  decays,
\begin{equation}
\begin{aligned}
\Delta A_\text{FB} = A_{\text{FB}}^{(\mu)} -  A_{\text{FB}}^{(e)} \,.
\end{aligned}
\end{equation}
In their fit based on Belle data, the authors of Ref.~\cite{Bobeth:2021lya} found an experimental value of
\begin{equation}
\begin{aligned}
\Delta A^\text{exp}_\text{FB} = 0.0349(89) ~ \text{\cite{Belle:2018ezy}} \,,
\end{aligned}
\end{equation}
which has to be compared to SM predictions exhibiting a negative value
\begin{equation}
\begin{aligned}
\Delta A^\text{SM}_\text{FB} = -0.0045(3) \,,
\end{aligned}
\end{equation}
as obtained using \texttt{flavio}.

Leptoquark solutions to this $\Delta A_\text{FB}$ anomaly have been discussed in Ref.~\cite{Carvunis:2021dss}. While models with an $S_2$ leptoquark cannot describe the data as well as models with an $S_1$ leptoquark, they can still significantly improve the fit. The relevant WET coefficients are the ones shown in Eq.~(\ref{eq:matching_bctaunu}), once we replace tau leptons $\tau$ with muons $\mu$.

\subsection{Anomalous magnetic moments and electric dipole moments of leptons}
\label{sec:AMMsEDMs}

Measurements of charged lepton anomalous magnetic moments ($a_\ell$) and electric dipole moments (EDMs) are both highly sensitive probes of new physics. The effects of $\Phi_2$ on the predictions for these observables are all related to the operator ${\cal O}_7$. Whereas we could in principle consider all three generations of leptons, we ignore tau leptons as the existing bounds are not constraining. We should however keep in mind that a polarised beam option at Belle~II~\cite{Liptak:2021opc} offers a possibility for a measurement of $a_\tau$ that could be sensitive to various new physics models~\cite{Bernabeu:2007rr,Bernabeu:2008ii,Crivellin:2021spu}, including ours. 

The current averaged values for the muon and electron anomalous magnetic moments read
\begin{equation}
 a_e^\text{exp} = 115\ 965\ 218.091(26)\times 10^{-11} ~\text{\cite{ParticleDataGroup:2020ssz}}
 \ \ \text{and}\ \ 
 a_\mu^\text{exp} = 116\ 592\ 061(41)\times 10^{-11} ~\text{\cite{Muong-2:2021ojo}}\,,
\end{equation}
where the latter measurement includes Run 1 data from the Fermilab Muon $g-2$ experiment. The corresponding SM predictions are given by
\begin{equation}
\begin{aligned}
 a_e^\text{SM, Cs} &= 115\ 965\ 218.161(23)\times 10^{-11} ~\text{\cite{Parker:2018vye}} \,, \\
 a_e^\text{SM, Rb} &= 115\ 965\ 218.0252(95)\times 10^{-11} ~\text{\cite{Morel:2020dww}} \,, \\
 a_\mu^\text{SM} &= 116\ 591\ 810(43)\times 10^{-11} ~\text{\cite{Aoyama:2020ynm}} \,.
 \end{aligned}
 \label{eq:AMM_theory}
\end{equation}
For $a_e$ we quote two sets of contradicting predictions, that are respectively based on the measurement of the fine-structure constant $\alpha^{-1}$ in Cs and in Rb atoms. While the disagreement between the experimental determination and the SM prediction for $a_e$ remains unclear, the deviation in $a_\mu$ is better established and currently amounts to $4.2 \sigma$. It is widely known as the $(g-2)_\mu$ anomaly. In our numerical analysis, we use the package \texttt{flavio} but consider both theoretical determinations of $a_e$ stated in Eq.~(\ref{eq:AMM_theory}) individually. In contrast, the measurements of the lepton EDMs have so far all yielded null results. Currently, the most stringent exclusion limits read
\begin{equation}
 d_e^\text{exp} < 0.11 \times 10^{-28} \text{ e cm} ~\text{\cite{ParticleDataGroup:2020ssz}}\,, \qquad
 d_\mu^\text{exp} < 1.8 \times 10^{-19}  \text{ e cm} ~\text{\cite{ParticleDataGroup:2020ssz}}\,.
\end{equation}
While the latter limit is currently not constraining, it is expected to be significantly improved in future experiments~\cite{Adelmann:2021udj,Aiba:2021bxe}.

As already mentioned, the sole relevant WET operator governing charged lepton's $g-2$ and EDMs is the ${\cal O}_7$ operator, defined by
\begin{equation}
\left( \mathcal{O}_7 \right)_{\ell \ell} = \frac{4G_F}{\sqrt{2}} \frac{e}{16 \pi^2} m_{\ell} \left( \bar{\ell} \sigma^{\mu \nu} P_R \ell \right) F_{\mu \nu}\,.
\end{equation}
Starting from our model featuring three generations of $\Phi_2$ LQs, the associated Wilson coefficient is obtained after integrating out the heavy LQs~\cite{Crivellin:2020mjs, Aebischer:2021uvt, Bigaran:2021kmn, Dorsner:2016wpm},
\begin{equation}
\left( \mathcal{C}_7 \right)^\text{LQ}_{\ell \ell}  \approx \frac{\sqrt{2}}{4 G_F m_\ell} \!\Bigg [\! -\!  \frac{N_c m_\ell}{8M_\ell^2} \sum_{i = 1}^3 \left( \left|Y_{i\ell}^{LR} \right|^2 \!+\! \left|Y_{i\ell}^{RL} \right|^2\right) \!+\! \frac{N_c}{12M_\ell^2} \sum_{i = 1}^3 m_{u_i} \hat{Y}_{i\ell}^{LR} Y_{i\ell}^{RL*}  \mathcal{E}_1\left( x_i \right)  \Bigg]\,,
\label{eq:C7_leptons}
\end{equation}
with $x_i = m_{u_i}^2/M_\ell^2$ and  $\mathcal{E}_1(x) \equiv 1 + 4 \log (x)$. The dominant contribution is the one included in the second term for $i=3$, {\it i.e.} the contribution enhanced by the large value of the top mass $m_t$. Predictions for $a_\ell$ and $d_\ell$ are then given by
\begin{equation}
 a_\ell^\text{LQ} = \frac{G_F m_\ell^2}{\sqrt{2} \pi^2}\  \text{Re} \big\{ \left( \mathcal{C}_7 \right)_{\ell \ell} \big\} \qquad\text{and}\qquad
 \left| d_\ell^\text{LQ} \right| = \frac{eG_F m_\ell}{2 \sqrt{2} \pi^2}\ \Big| \text{Im} \big\{ \left( \mathcal{C}_7 \right)_{\ell \ell} \big\} \Big| \,,
\end{equation}
with the dominant contribution being the one proportional to the real and imaginary part of $\hat{Y}_{3\ell}^{LR} Y_{3 \ell}^{RL*}$ respectively. In addition, the same combination of LQ Yukawa matrix elements contributes to a radiative shift in the charged lepton masses~\cite{Bigaran:2021kmn},
\begin{equation}
  m^\text{LQ}_\ell \approx - \frac{m_{t} N_c}{16 \pi^2} \mathcal{E}_3 \left( \frac{\mu^2}{M_\ell^2}, \frac{m_{t}^2}{M_\ell^2} \right) \hat{Y}^{LR}_{3\ell} Y^{RL*}_{3 \ell} 
  \ \ \text{with}\ \
\mathcal{E}_3\left( x, y \right) = \frac{1}{\epsilon} \!+\! 1 \!+\! \log \left(x \right) \!+\! y \log\left( y \right) \,,
\label{eq:radiative_mass}
\end{equation}
when we restrict ourselves to the contribution enhanced by the top mass.

\subsection{Parity violation observables}
By measuring  parity-violating interactions between electrons and nucleons, the nucleon weak charges can be determined. Currently, the best measurement of the weak charge of the proton, $Q_w^p$, comes from the $Q_{\rm weak}$ experiment~\cite{Qweak:2014xey,Carlini:2019ksi} at Jefferson Lab,
\begin{equation}
Q_w^\text{exp} (p) = 0.0704(47)~\text{\cite{Qweak:2018tjf,Crivellin:2021bkd}}.
\end{equation} 
In addition, the most precise results from atomic parity violation experiments were obtained for $^{133}\text{Cs}$ atoms~\cite{Wood:1997zq, Guena:2004sq},
\begin{equation}
Q_w^\text{exp}\left( ^{133} \text{Cs}\right) = -72.94(43)~\text{\cite{Cadeddu:2021dqx}}.
\end{equation}
These weak charge measurements can be used to constrain the necessarily chiral quark-electron interactions induced by LQs. The corresponding effects in parity violation (PV) observables have already been studied in detail in Ref.~\cite{Crivellin:2021bkd}. The relevant WET operators are
\begin{equation}
\begin{aligned}
\left(\mathcal{O}_V^{LR} \right)_{qqee} &= \frac{4 G_F}{\sqrt{2}} \left(\bar{q} \gamma^\mu P_L q \right) \left(\bar{e} \gamma_\mu P_R e \right) \,, \\
\left(\mathcal{O}_V^{LR} \right)_{eeqq} &= \frac{4 G_F}{\sqrt{2}} \left(\bar{e} \gamma^\mu P_L e \right) \left(\bar{q} \gamma_\mu P_R q \right) \,,
\end{aligned}
\end{equation}
for $q = u,d$. Integrating out the heavy LQ fields in our model, we can derive the corresponding Wilson coefficients,
\begin{equation}
\begin{aligned}
\left(\mathcal{C}_V^{LR} \right)^\text{LQ}_{uuee} &=  \frac{-\sqrt{2}}{4 G_F} \frac{\left | \hat{Y}_{1e}^{LR}\right |^2}{2 M_e^2} \,,\qquad & \left(\mathcal{C}_V^{LR} \right)^\text{LQ}_{eeuu} &= \frac{-\sqrt{2}}{4 G_F} \frac{\left | Y_{1e}^{RL}\right |^2}{2 M_e^2}  \,, \\
\left(\mathcal{C}_V^{LR} \right)^\text{LQ}_{ddee} &=  \frac{-\sqrt{2}}{4 G_F} \frac{\left | Y_{1e}^{LR}\right |^2}{2 M_e^2}  \,,\qquad & \left(\mathcal{C}_V^{LR} \right)^\text{LQ}_{eedd} &= 0 \,. 
\end{aligned}
\label{eq:PV_LQ_matching}
\end{equation}

There, for our numerical analysis we use~\cite{Crivellin:2021egp}
\begin{equation}
\begin{aligned}
Q_w &= -2 \Bigg[ Z \left( 2 \mathcal{C}_{1u}^{e} + \mathcal{C}_{1d}^{e} \right) + N \left( \mathcal{C}_{1u}^{e} + 2\mathcal{C}_{1d}^{e} \right) \Bigg] \,,
\end{aligned}
\end{equation}
where $Z$ and $N$ are the atomic number and the number of neutrons in a nucleus respectively, and  where we have defined
\begin{equation}
\mathcal{C}_{1q}^{e} =  \mathcal{C}_{1q}^{e, \text{SM}} +  \mathcal{C}_{1q}^{e, \text{LQ}}\,.
\end{equation}
In this last relation,
\begin{equation}
  \mathcal{C}_{1u}^{e, \text{SM}} = -0.1888\,, \ 
  \mathcal{C}_{1d}^{e, \text{SM}} = 0.3419~\text{\cite{Erler:2013xha, ParticleDataGroup:2020ssz}}
  \ \ \text{and}\ \ 
  \mathcal{C}_{1q}^{e, \text{LQ}} = \Big[ \left(\mathcal{C}^{LR}_{V}\right)^\text{LQ}_{qqee} \!-\! \left(\mathcal{C}^{LR}_{V}\right)^\text{LQ}_{eeqq} \Big]\,.
\label{eq:PV_C1q}\end{equation}
We recall that for protons $Z=1$ and $N=0$, whereas for $^{133}\text{Cs}$ atoms we have $Z=55$ and $N=78$. In order to extract constraints on our model, we finally build a likelihood functions
\begin{equation}
\begin{aligned}
-2 \log \mathcal{L} = \frac{\left(Q_w^\text{exp} - Q_w \right)^2}{\sigma^2} \,,
\end{aligned}
\end{equation}
where $\sigma$ stands for the experimental resolution.

\subsection{$Z$-boson decays into leptons and neutrinos}
The Lagrangian describing the interaction of the $Z$ boson with left-handed and right-handed leptons can be generically written as
\begin{equation}\label{eq:LZ}
\delta \mathcal{L}_\text{eff}^Z = \frac{g}{c_w} \sum_{\ell} \bar{\ell} \gamma^\mu \big[g_{\ell_{L}} P_L + g_{\ell_{R}} P_R \big] \ell Z_\mu + \frac{g}{c_w} \sum_{\ell} \bar{\nu}_{\ell} \gamma^\mu \big[g_{\nu_{L}} P_L \big] \nu_\ell Z_\mu \,,
\end{equation}
where $g$ is the weak coupling constant and $c_w = \cos \theta_w$ is the cosine of the electroweak mixing angle $\theta_w$. SM predictions for the charged lepton sector lead to 
\begin{equation}
g_{\ell_L}^{\text{SM}} = -0.26919(20) \qquad\text{and}\qquad g_{\ell_R}^{\text{SM}} = 0.23208(17) \,,
\end{equation}
which can be compared with measurements at LEP~\cite{ALEPH:2005ab},
\begin{equation}
\begin{aligned}
g_{e_L}^{\text{exp}} &= -0.26963(30) \,,\qquad    & g_{e_R}^{\text{exp}} &= 0.23148(29) \,, \\
g_{\mu_L}^{\text{exp}} &= -0.2689(11) \,,\qquad   & g_{\mu_R}^{\text{exp}} &= 0.2323(13) \,, \\
g_{\tau_L}^{\text{exp}} &= -0.26930(58) \,,\qquad & g_{\tau_R}^{\text{exp}} &= 0.23274(62) \,.
\end{aligned}
\end{equation}
In our numerical analysis, we include the impact of our model on those couplings through the likelihood function
\begin{equation}
-2 \log \mathcal{L} \!=\! \sum_{\ell_{A},  \ell^\prime_{B}} \Bigg[ \Big( g_{\ell_{A}}^{\text{SM}} + \text{Re}\big\{g^\text{LQ}_{\ell_{A}}\big\} -  g_{\ell_{A}}^{\text{exp}} \Big) \Big(V^{-1}\Big)_{\ell_A \ell^\prime_B} 
 \Big( g_{\ell^\prime_{B}}^{\text{SM}} + \text{Re}\big\{ g^\text{LQ}_{\ell^\prime_{B}}\big\} -  g_{\ell^\prime_{B}}^{\text{exp}} \Big) \Bigg]\,,
\end{equation}
where $\ell, \ell^\prime \in \{e, \mu, \tau\}$, $A, B \in \{ L, R\}$ and $V$ stands for the covariance matrix including the experimental uncertainties as well as the correlations among the measurements. The expressions for $g^\text{LQ}_{\ell_{A}}$ quantifying the LQ contributions to the $Z$ couplings are given in Appendix~\ref{sec:Zcouplings}. 

The $Z$-boson coupling to neutrinos can be extracted from the LEP measurement of the effective number of neutrino generations~\cite{ALEPH:2005ab}
\begin{equation}
N_\nu^{\text{exp}} = 2.9840(82)\,.
\end{equation}
This last measurement can be numerically confronted to predictions from our model through the likelihood function
\begin{equation}
  -2\log \mathcal{L} = \frac{\left(N_\nu - N_\nu^{\text{exp}}\right)^2}{\sigma^2} \qquad\text{with}\qquad
  N_\nu = \sum_i \left( 1 + \frac{\text{Re}\{ g^\text{LQ}_{\nu_{\ell, L}} \}}{g_{\nu_{L}}^\text{SM}}\right)^2~\text{\cite{Arnan:2019olv}}\,,
\end{equation}
where $\sigma$ refers to the experimental uncertainty. In this expression, we consider the SM $Z$-boson coupling value that has been measured at LEP~\cite{ALEPH:2005ab},
\begin{equation}
g^{\text{SM}}_{\nu_{L}} = 0.50199(19)\,,
\end{equation}
and the expression for $g^\text{LQ}_{\nu_{\ell, L}}$ given in Appendix~\ref{sec:Zcouplings}.

\subsection{$\Delta F = 2$ meson mixing observables}
Measurements of $K^0 - \bar{K}^0$, $B_d-\bar{B}_d$, $B_s-\bar{B}_s$ and $D^0 - \bar{D}^0$ mixing parameters allow for the extraction of constraints on our LQ model. We use in our analysis three meson mass differences and a $D$-meson mixing parameter for which the associated measurements,
\begin{align}
\Delta M_{K^0} &= -3.484(6) \times 10^{-15} \text{ GeV} ~ \text{\cite{Aoki:2021kgd}} \,, \\
\Delta M_{B_d} &= -3.33(1) \times 10^{-13} \text{ GeV} ~ \text{\cite{HFLAV:2019otj}} \,, \\
\Delta M_{B_s} &= -1.168(1) \times 10^{-11} \text{ GeV} ~ \text{\cite{HFLAV:2019otj}} \,, \\
x_{D^0} &=  0.0035(15) ~ \text{\cite{Bona:2017gut}} \,,
\end{align}
are in good agreement with SM theory predictions. For $K^0$, $B_d$ and $B_s$ mixing the sign of $\Delta M$ is known, while this is not the case for $D^0$ mixing. Moreover, we further include the UTfit~\cite{UTfit:2006onp, UTfit:2007eik} likelihood for the phase $\phi_{B_s}$, that restricts the new physics contributions to the complex phase inherent to $B_s-\bar{B}_s$ mixing, as well as the $D^0-\bar D^0$ mixing phase $\Phi_{12}$ whose UTfit value is
\begin{equation}
    \Phi_{12} = \left(0.045 \pm 1.335 \right)^\circ \,.
\end{equation}

The relevant four-fermion operators affecting the $\Delta F = 2$ observables related to $D^0-\bar D^0$ mixing are
\begin{align}
	\left(\mathcal{O}_V^{AB}\right)_{ucuc} &= \left( \bar{c} \gamma^\mu P_A u \right)\left( \bar{c} \gamma_\mu P_B u \right) \,, \qquad
\label{eq:opAB}\end{align}
with $A,B \in \{L, R\}$, whilst those relevant for other meson mixings are obtained via straight-forward exchanges of the quark flavours in the expression of the operators~\eqref{eq:opAB}. At leading order, the Wilson coefficients associated with these operators are~\cite{Crivellin:2021lix}
\begin{equation}
\begin{aligned}
\left(\mathcal{C}_V^{LL}\right)^\text{LQ}_{ucuc} &= \frac{ - 1}{128{\pi ^2}}\sum \limits_{\ell}^{} \frac{\left(\hat Y_{1\ell}^{LR}\right)^2 \left(\hat Y_{2\ell}^{LR*}\right)^2 }{M_{\ell}^2} \,, \\
 \left(\mathcal{C}_V^{RR}\right)^\text{LQ}_{ucuc} &= \frac{ - 1}{64{\pi ^2}}\sum \limits_{\ell}^{} \frac{\left( Y_{1\ell}^{RL} \right)^2 \left( Y_{2\ell}^{RL*} \right)^2}{M_{\ell}^2} \,, \\
 \left(\mathcal{C}_V^{LR}\right)^\text{LQ}_{ucuc} &= \frac{ 1}{32{\pi ^2}}\sum \limits_{\ell}^{} \Bigg( \frac{\hat Y_{1\ell}^{LR} \hat Y_{2\ell}^{LR*} Y_{1\ell}^{RL} Y_{2\ell}^{RL*}}{M_{\ell}^2} \Bigg) \,,
\end{aligned} \label{eq:DDbarMatching}
\end{equation}
whereas for $B$ and $K$ meson mixing, the relevant Wilson coefficients read
\begin{equation}
  \left(\mathcal{C}_V^{LL}\right)^\text{LQ}_{d_i d_j d_i d_j} = \frac{ - 1}{128{\pi ^2}}\sum \limits_{\ell}^{} \frac{\left(Y_{j\ell}^{LR}\right)^2 \left(Y_{i\ell}^{LR*}\right)^2}{M_{\ell}^2} \,, \ \ 
 \left(\mathcal{C}_V^{RR}\right)^\text{LQ}_{d_i d_j d_i d_j} =  \left(\mathcal{C}_V^{LR}\right)^\text{LQ}_{d_i d_j d_i d_j} = 0 \,,
\end{equation}
for generation indices $(i,j) = (2,1)$, $(3,1)$ and $(3,2)$.

\subsection{Drell-Yan di-lepton searches at the LHC}

\label{sec:DY_CMS}
\begin{figure}
    \begin{center}
	\begin{subfigure}[b]{0.45\textwidth}
	    \begin{center}
		\includegraphics[width=0.6\textwidth]{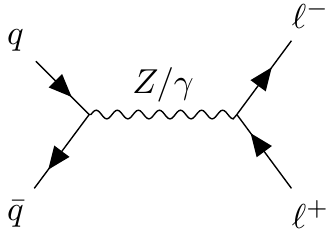}
		\caption*{(SM)}
		\end{center}
	\end{subfigure}
	\hspace{10px}
	\begin{subfigure}[b]{0.45\textwidth}
	    \begin{center}
		\includegraphics[width=0.6\textwidth]{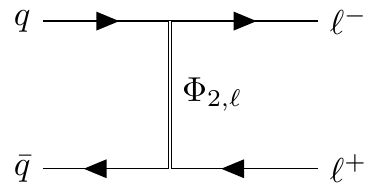} \vspace{5px}
		\caption*{(LQ)}
		\end{center}
	\end{subfigure}	
	\end{center}
	\caption{The leading-order SM and LQ diagrams (drawn with \texttt{TikZ-Feynman}~\cite{Ellis:2016jkw}) that contribute to di-lepton production $pp \to \ell^+ \ell^-$ at the LHC. While the SM contribution occurs via the $s$-channel exchange of a (virtual) photon or $Z$-boson, the LQ diagram on the right features a $t$-channel exchange and receives an enhancement at high di-lepton invariant masses.}
	\label{fig:DY_feynman}
\end{figure}

As illustrated by the neutral-current Feynman diagrams shown in Figure~\ref{fig:DY_feynman}, LQ exchanges can importantly contribute to Drell-Yan (DY) production at the LHC. Due to the relative energy enhancement of the LQ $t$-channel partonic amplitude relative to the SM one, measuring high-energy tails in $pp \to \ell^+ \ell^-$ and $pp \to \ell \nu$ processes at the LHC can potentially offer competitive bounds on the LQ couplings to fermions, even when parton density suppression is taken into account.

The CMS analysis of non-resonant di-lepton production~\cite{CMS:2021ctt} provided measurements for inclusive cross-section ratios in $n\!=\!9$ bins in the di-lepton invariant mass $m_{\ell \ell}$ included in the $[200, 3500]$~GeV range (with $\ell = e, \mu$),
\begin{equation}
	R_{\mu\mu / ee, n} \equiv \dfrac{\int_{\text{bin n}} \frac{d\sigma(pp \to \mu^+\mu^- )}{dm_{\mu \mu}} dm_{\mu \mu}}{\int_{\text{bin n}} \frac{d\sigma(pp \to e^+e^-)}{dm_{ee}} dm_{ee}}\,.
\end{equation}
In the CMS publication, the results are normalised to SM predictions obtained from Monte-Carlo (MC) simulations,
\begin{equation}
	R_{\mu \mu / ee, n}^{\text{data}}\  / \  R_{\mu \mu/ee, n}^{\text{MC}}\,.
\end{equation}
In this double ratio, many uncertainties cancel~\cite{Greljo:2017vvb}. Furthermore,  $R_{\mu\mu / ee}$ was normalised to unity in a combined bin ranging from 200 to 400 GeV to correct for the different detector sensitivity to electrons and muons. The CMS collaboration measured an excess in the di-electron channel, so that the ratios $R_{\mu \mu/ee, n}$ are smaller than one for large $m_{\ell \ell}$ values. This clearly points towards another source of LFUV. 

LQ contributions that could explain this excess have been studied in Refs.~\cite{Crivellin:2021bkd, Crivellin:2021egp, Crivellin:2021rbf}. Here we largely follow the analysis presented in the addendum to Ref.~\cite{Crivellin:2021egp}, but simulated the cross-section with the full dependence on the LQ propagator. We determined
\begin{equation}
\int_{\text{bin n}} \frac{d\sigma^\text{SM(+LQ)}(pp \to \ell^+ \ell^-)}{dm_{\ell \ell}} dm_{\ell \ell}\,,
\label{eq:DY_cross_section_simulation}
\end{equation}
at leading order using \texttt{MadGraph5\_aMC@NLO}~\cite{Alwall:2014hca} version 3.2.0 with the UFO~\cite{Degrande:2011ua} model \texttt{lqnlo\_v5}\footnote{Available from \url{https://www.uni-muenster.de/Physik.TP/research/kulesza/leptoquarks.html}.}~\cite{Borschensky:2020hot, Borschensky:2021jyk} and the PDF set \texttt{NNPDF40\_nlo\_as\_01180}~\cite{Ball:2021leu}.

On the other hand, the ATLAS collaboration also performed measurements of the high-$m_{\ell \ell}$ tail in di-lepton production, that we can thus use as an additional probe for our model. For electrons and muons we recast their non-resonant analysis presented in Ref.~\cite{ATLAS:2020yat}. There, a parametric background-model function is fitted to the di-lepton invariant mass distribution in the control regions $m_{ee} \in $ [280 GeV, 2200 GeV] and $m_{\mu \mu} \in $ [310 GeV, 2070~GeV] for electrons and muons, respectively, and then extrapolated to the signal regions (SRs) in which $m_{ee} \in $ [2200 GeV, 6000 GeV] and $m_{\mu \mu} \in $ [2070 GeV, 6000 GeV]. SM predictions for the expected number of events in the SRs are next compared with measurements to derive bounds on any new physics interfering constructively with SM DY production. Interestingly, the ATLAS collaboration has also found slightly more di-electron events and less di-muon events than expected.

For the tau-lepton channel, we recast the ATLAS search for heavy Higgs bosons decaying into two tau leptons that has been presented in Ref.~\cite{ATLAS:2020zms}. We consider events where both taus decay hadronically ($\tau_{\text{had}} \tau_{\text{had}}$), but we carry out a $b$-jet inclusive analysis. 

Finally, charged-current DY processes can also yield strong constraints on our model. In particular, limits from the process $pp \to \tau \nu$ are directly sensitive to the $\Phi_{2, \tau}$ coupling combination shown in Eq.~(\ref{eq:matching_bctaunu}). Bounds on the $S_2$ couplings to taus were presented in Ref.~\cite{Jaffredo:2021ymt} based on the latest 139~fb$^{-1}$ mono-tau search at the LHC~\cite{ATLAS:2021bjk}. The re-interpretation of the results of this analysis includes all contributing Feynman diagrams, and in particular those featuring a LQ propagator. The latter are found to be important for LQ masses in the TeV range, which corresponds to a configuration for which an effective field theory description is no longer valid. Limits derived from this $pp \to \tau \nu$ analysis have been found less constraining than the ones originating from the $pp \to \tau^+ \tau^-$ analysis. We therefore impose the former as a sharp cut on the LQ parameter space, without deriving in detail a full likelihood function. 

More information on all considered analyses and how we use them to constrain our model can be found in Appendix~\ref{app:lhc}. 

\subsection{Single-resonant leptoquark production (SRP)}
\label{sec:SRP} 
In Ref.~\cite{Buonocore:2020erb}, limits on LQ Yukawa couplings are derived based on LQ single-resonant production $\ell q \to \Phi_{2, \ell} \to \ell q$, where the initial-state lepton $\ell$ originates from the lepton density in the proton. The authors include $e + q_\text{light}$ and $\mu + q_\text{light}$ final states in their analysis, where $q_\text{light} \in \{ u,d,c,s \}$, and they further assume a minimal model containing a single $SU(2)_L$ LQ singlet that couples exclusively to one quark and one lepton generation. In our numerical analysis we generalise this setup and implement SRP exclusion limits as sharp cuts in the LQ parameter space. This makes sure that current SRP bounds are not violated through a choice of large enough LQ masses. 

Since the $\Phi_{2, \ell}$ states have multiple $SU(2)_L$ components and they potentially couple to multiple quark generations, we need to recast the limits of Ref.~\cite{Buonocore:2020erb} accordingly. In the minimal model and assuming $M_\ell \gg m_\ell, m_q$, the cross section for the process $\ell q_i \to {\Phi_{2, \ell}} \to \ell q_\text{light}$ is proportional to $\left| \lambda^q_{i\ell}\right|^2$, where $\lambda^q_{i\ell}$ for $q = u,d$ are the Yukawa couplings to the  $i^{\text{th}}$ quark generation and the lepton flavour $\ell$. Comparing this to our model, we define the effective couplings
\begin{equation}
\begin{aligned}
\left| \lambda_{i\ell}^{u, \text{eff}} \right|^2 &\equiv \left( \left| \hat{Y}^{LR}_{i\ell} \right|^2 + \left| Y^{RL}_{i\ell} \right|^2 \right) \dfrac{\sum_{j=1}^2 \left| \hat{Y}^{LR}_{j\ell} \right|^2 + \left| Y^{RL}_{j\ell} \right|^2 }{\sum_{j=1}^3 \left| \hat{Y}^{LR}_{j\ell} \right|^2 + \left| Y^{RL}_{j\ell} \right|^2} \,, \\
\left| \lambda_{i\ell}^{d, \text{eff}} \right|^2 &\equiv \left| Y^{LR}_{i\ell} \right|^2 \dfrac{\sum_{j=1}^2 \left| Y^{LR}_{j\ell} \right|^2 }{\sum_{j=1}^3 \left| Y^{LR}_{j\ell} \right|^2 + \left| Y^{RL}_{j\ell} \right|^2} \,, 
\end{aligned}
\end{equation}
such that $\sigma \sim \left| \lambda_{i\ell}^{q, \text{eff}} \right|^2$. Using the limits $\left| \lambda^{q, \text{lim}}_{i\ell}\right|$ derived in Ref.~\cite{Buonocore:2020erb}, the effective couplings of our model must satisfy
\begin{equation}
\sum_{i = 1}^2 \left| \dfrac{ \lambda_{i\ell}^{u, \text{eff}}}{\lambda_{i\ell}^{u, \text{lim}}} \right|^2 + \left| \dfrac{ \lambda_{i\ell}^{d, \text{eff}}}{\lambda_{i\ell}^{d, \text{lim}}} \right|^2 < 1\,,
\end{equation}
for $\ell = e, \mu$. Here we assume that the mass difference between the $SU(2)_L$ components of $\Phi_{2, \ell}$ is negligible, which holds for $v \ll M_\ell$.

\subsection{Leptoquark pair production (PP)}

\begin{center}
    \renewcommand{\arraystretch}{1.2}\setlength\tabcolsep{6pt}
	\begin{table}
		\begin{tabular}{c|c c c c c c c |}
			\cline{2-8}
			& \multicolumn{7}{|c|}{Coupling} \\
			\cline{2-8}
			& ~~$Y^{RL}_{1e}$ ~~ & ~~$Y^{RL}_{2e}$~~~~ & ~~$Y^{RL}_{3e}$~~ & ~~$Y^{RL}_{1\mu}$~~ & ~~$Y^{RL}_{2\mu}$~~~ & ~~$Y^{RL}_{3\mu}$~~ & ~~$Y^{RL}_{3\tau}$~~ \\
			\hline
			\multicolumn{1}{|c|}{Limit [GeV]} & $1790$ & $1760$ & $1480$ & $1730$ & $1690$ & $1470$ & 1440 \\
			\multicolumn{1}{|c|}{Analysis} & $e + q_\text{light}$ & $e + c$ & $e + t$ & $\mu + q_\text{light}$ & $\mu +c$ & $\mu + t$ & $\tau + t$ \\
			\multicolumn{1}{|c|}{Reference} & \cite{ATLAS:2020dsk} &  \cite{ATLAS:2020dsk} & \cite{ATLAS:2020xov} & \cite{ATLAS:2020dsk} &  \cite{ATLAS:2020dsk} & \cite{ATLAS:2020xov} & \cite{ATLAS:2021oiz} \\
			\multicolumn{1}{|c|}{$\beta_\text{eff}$} & $1.0$  &  $1.0$  & $1.0$ & $1.0$ & $1.0$ & 1.0 & 1.0 \\
			\hline
		\end{tabular}  \vspace{10px} \\
		
		\begin{tabular}{c|c c c c c c c |}
			\cline{2-8}
			%& \multicolumn{7}{|c|}{Coupling} \\
			%\cline{2-8}
			 & ~~$Y^{LR}_{1e}$~~ & ~~$Y^{LR}_{2e}$~~ & ~~$Y^{LR}_{3e}$~~ & ~~$Y^{LR}_{1\mu}$~~ & ~~$Y^{LR}_{2\mu}$~~ & ~~$Y^{LR}_{3\mu}$~~ & ~~$Y^{LR}_{3\tau}$~~ \\
			\hline
			\multicolumn{1}{|c|}{Limit [GeV]} & $1910$ & $1790$ & $1740$ & $1850$ & $1730$ & $1720$ & 1440 \\
			\multicolumn{1}{|c|}{Analysis} & $e + q_\text{light}$ & $e + q_\text{light}$ & $e + b$ & $\mu + q_\text{light}$ & $\mu + q_\text{light}$ & $\mu + b$ & $\tau + t$ \\
			\multicolumn{1}{|c|}{Reference} & \cite{ATLAS:2020dsk} &  \cite{ATLAS:2020dsk} & \cite{ATLAS:2020dsk} & \cite{ATLAS:2020dsk} &  \cite{ATLAS:2020dsk} & \cite{ATLAS:2020dsk} & \cite{ATLAS:2021oiz} \\
			\multicolumn{1}{|c|}{$\beta_\text{eff}$} & $1.9$  &  $1.0$  & $1.0$ & $1.9$ & $1.0$ & 1.0 & 1.0 \\
			\hline
		\end{tabular}
		\caption{95\% C.L.~limits on the LQ masses $M_\ell$ derived from the ATLAS pair production analyses targeting the processes $pp \to \Phi_{2,\ell} \Phi_{2,\ell} \to \ell q \ell q$. Only the couplings indicated in the table header are assumed to be non-zero. We state the ATLAS analysis that yields the most stringent limits for the respective scenario, and additionally indicate the corresponding $\beta_\text{eff}$ value. The light quarks $q_\text{light}$ consist of $u,d$ and $s$ quarks. No analysis exists for leptoquark states coupling to $\tau$ and one of the  $u$, $d$, $c$ or $s$ quarks.}
		\label{tab:PP_limits}
	\end{table}\renewcommand{\arraystretch}{1}
\end{center}

The ATLAS and CMS collaborations have published several analyses targeting the production of a pair of scalar LQs decaying into a specific $\ell + q$ or $\nu + q$ final state. Since the exclusion limits based on the former decay mode are more constraining and since $\beta \equiv \text{Br}\left(\Phi_{2,\ell} \to \ell + q \right) \geq 0.5$ in general, we solely focus on the $\ell q\ell q$ class of final states in our numerical analysis. We derive mass limits for the specific coupling structure that we consider, and make sure that the LQ masses $M_\ell$ lie well above these bounds. 

In analogy to Section~\ref{sec:SRP}, we make the choice of recasting exclusion limits obtained by the ATLAS collaboration, enforcing the fact that $\Phi_{2,\ell}$ multiplets have two $SU(2)_L$ components and that $\beta$ can be different from 1 for a specific choice of $\ell$ and $q$ flavours. Again assuming that the mass difference between the $SU(2)_L$ components of $\Phi_{2,\ell}$ is negligible, we introduce an ``effective'' $\beta_\text{eff}$ parameter,
\begin{equation}
\beta^2_\text{eff} \equiv  \left(\dfrac{\sum_{i \in U} \left| \hat{Y}^{LR}_{i\ell}\right|^2 + \left| Y^{RL}_{i\ell}\right|^2}{\sum_{i = 1}^{3}  \left| \hat{Y}^{LR}_{i\ell}\right|^2 + \left| Y^{RL}_{i\ell}\right|^2}\right)^2 \\
+ \left(\dfrac{\sum_{i \in D} \left| Y^{LR}_{i \ell}\right|^2}{\sum_{i = 1}^{3}  \left| Y^{LR}_{i\ell}\right|^2 + \left| Y^{RL}_{i\ell}\right|^2}\right)^2 \,,
\label{eq:effective_Br}
\end{equation}
for an analysis that would effectively include up-quark generations $U \subset \left \{1,2,3 \right \}$, down-quark generations $D \subset \left \{1,2,3 \right \}$ and charged-lepton generations $\ell = e, \mu, \tau$. Considering the $\beta$-dependence of the limits given in Refs.~\cite{ATLAS:2019qpq, ATLAS:2021oiz, ATLAS:2020dsk, ATLAS:2020xov}, we obtain the LQ mass limits presented in Table~\ref{tab:PP_limits}. 

\subsection{Oblique corrections}
The effect of vacuum-polarization amplitudes of EW gauge bosons can be parametrized by the oblique Peskin-Takeuchi parameters $S$, $T$ and $U$~\cite{Peskin:1990zt}. The authors of Ref.~\cite{Ellis:2018gqa} performed a global fit to electroweak precision observables including LEP~\cite{ALEPH:2005ab}, Tevatron~\cite{CDF:2013dpa} and LHC~\cite{ATLAS:2019fyu} data. While the fit is compatible with SM predictions, it can be improved via a positive contribution to $\Delta T$.

The oblique corrections from LQs were studied extensively in Ref.~\cite{Crivellin:2020ukd}. Their contributions to the Peskin-Takeuchi parameters read
\begin{equation}
S^\text{LQ} \approx - \frac{7 N_c v^2}{36 \pi} \sum_\ell  \frac{Y_\ell^{H(3)}}{M_\ell^2} \qquad\text{and}\qquad
T^\text{LQ} \approx + \frac{N_c v^2}{96 \pi^2 \alpha} \sum_\ell \left( \frac{Y_\ell^{H(3)}}{M_\ell} \right)^2\,.
\end{equation}

\section{Phenomenological Analysis}
\label{sec:phenomenology}

In this section we perform a global analysis of our model with three generations of the $\Phi_{2}$ leptoquark. For this we construct full likelihood functions for the LQ parameters whenever possible~\cite{AbdusSalam:2020rdj}. Implementing the formulas of Section~\ref{sec:observables}, we use for the numerical analysis the software packages \texttt{flavio}~\cite{Straub:2018kue} and \texttt{smelli}~\cite{Aebischer:2018iyb}. In order to maximize the log likelihood $-2 \log \mathcal{L}$, we employ the \texttt{optimize.fmin} algorithm of \texttt{scipy}~\cite{Virtanen:2019joe}.

\begin{figure}
	\begin{centering}
		\includegraphics[width=0.98\textwidth]{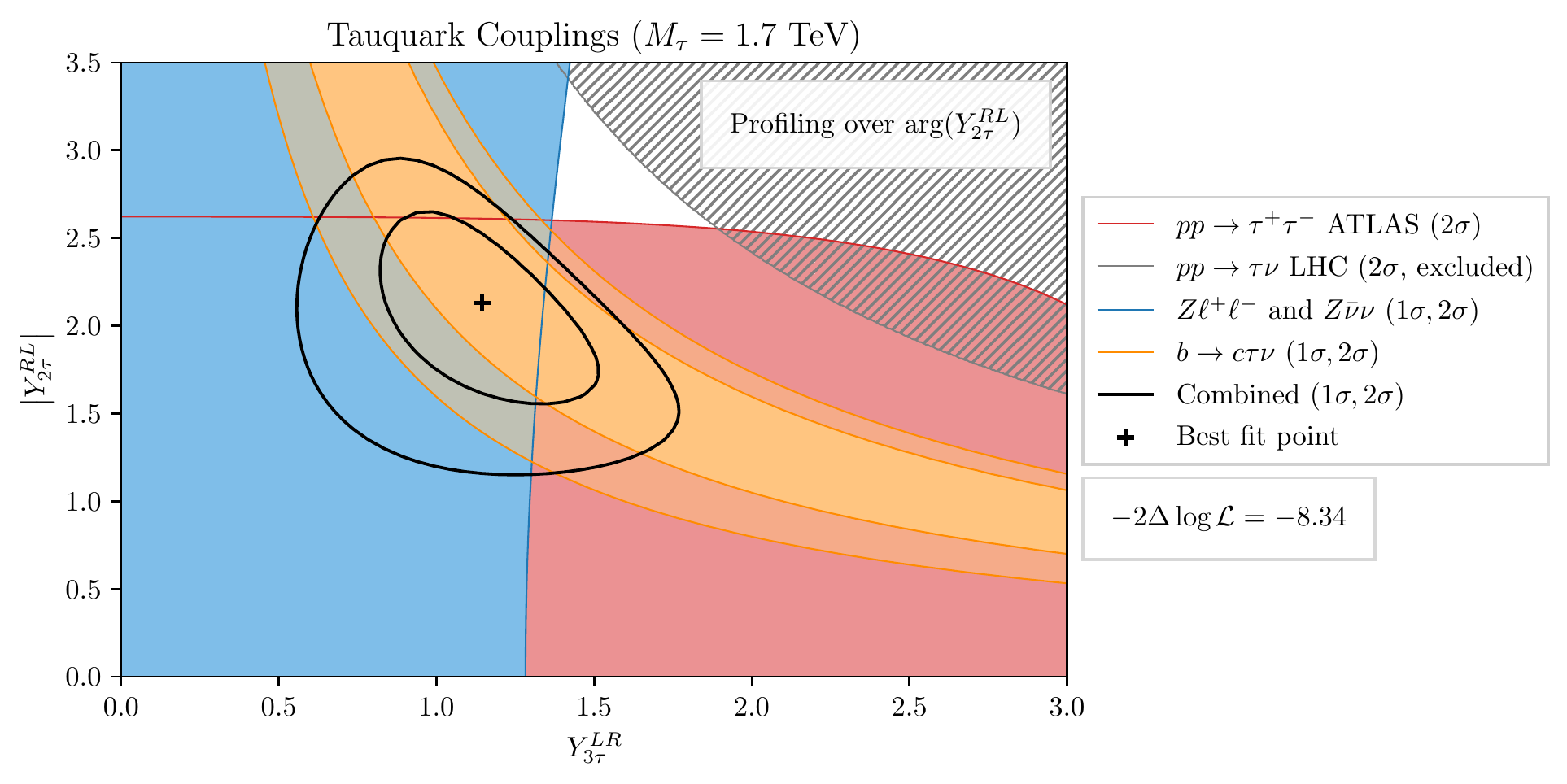}
			\end{centering}
	\caption{Preferred parameter space region providing an explanation for the $R_{D^{(*)}}$ anomalies (orange), shown in the $Y^{LR}_{3\tau}$--$Y^{RL}_{2\tau}$ plane. Our results assume that $Y^{LR}_{3\tau}$ is real and we profile over the phase of $Y^{RL}_{2\tau}$. We overlay to this contour the region favoured by $Z \tau^+ \tau^-$ and $Z \bar\nu\nu$ data (blue), as well as that favoured by the di-tau ATLAS DY analysis (red) and that excluded by the mono-tau search (hatched). Our results exhibit a mild tension.}
	\label{fig:RDRDstar_tau}
\end{figure}

\subsection{Tauquark}
\label{sec:taus_only}

As pointed out in Ref.~\cite{Angelescu:2021lln}, we can explain the $R_D$ and $R_{D^{*}}$ anomalies with the product of couplings $Y^{LR*}_{3\tau}Y^{RL}_{2\tau}$, given the presence of a large complex phase. This phase avoids interference with the SM, at the price that the couplings need to be large and the LQ mass needs to be low. We thus set $M_\tau=1.7$~TeV, which is compatible with the limits originating from LQ pair production given in Table~\ref{tab:PP_limits}. In fact, such a choice leads to a LQ pair-production cross section that is a factor 4 smaller than the corresponding bound. At the same time, a non-zero $Y^{LR}_{3\tau}$ value leads to an $m_t^2/m_Z^2$ enhancement in the $Z\tau^+\tau^-$ and $Z \bar\nu \nu$ couplings, and our non-zero $Y^{LR}_{2\tau}$ value contributes significantly, due to the second generation quarks involved, to non-resonant $pp \to \tau^+ \tau^-$ and $pp \to \tau \nu$ production. 

\begin{figure}
	\begin{center}
		\includegraphics[width=0.7\textwidth]{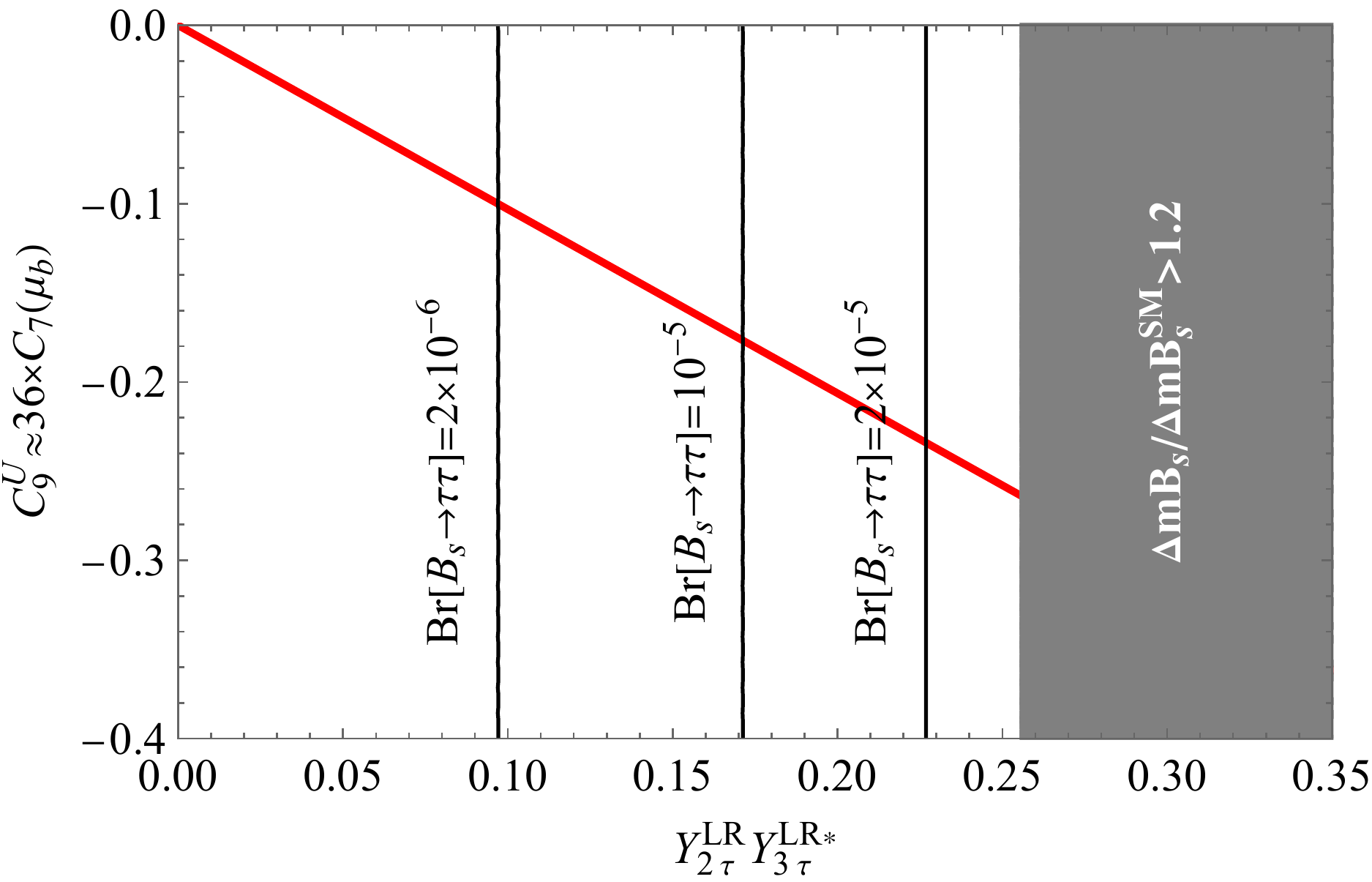}
	\caption{$\mathcal{C}_9^U\approx 36\times \mathcal{C}_7(\mu_b)$ as a function of $Y^{LR}_{2\tau} Y^{LR*}_{3\tau}$ for $M_\tau=1.7$ TeV. The contour lines indicate predicted values for the $B_s\to\tau^+\tau^-$ branching ratio, and in the grey region the effect in $B_s-\bar B_s$ mixing is higher than 20\% of the SM predictions. }
	\label{fig:Bsll-tauquark}
	\end{center}
\end{figure}

We display in Figure~\ref{fig:RDRDstar_tau} the preferred parameter space region that provides an explanation for the $R_{D^{(*)}}$ anomalies (orange), the results being projected in the  $Y_{3\tau}^{LR}$--$Y_{2\tau}^{RL}$ plane. As can be seen, this leads to strong bounds on the size of the tauquark couplings to fermions. In order to derive those constraints, we have profiled over the relative complex phase of the $Y^{LR*}_{3\tau}$ and $Y^{RL}_{2\tau}$ couplings, which we attribute to $Y^{RL}_{2\tau}$ (\textit{i.e.}~we assume that $Y_{3\tau}^{LR}$ is real). Despite of this new source of CP violation, EDM bounds are not (yet) constraining (see Appendix~\ref{sec:hadron_edms}). The figure also shows the parameter space region favoured by $Z\to \tau^+\tau^-$ and $Z\to \bar\nu\nu$ coupling data (blue) as well as by neutral-current and charged-current DY measurements at the LHC in the di-tau (red) and mono-tau (hatched) channel. While the $R_{D^{(*)}}$ anomalies can be partially explained, there is a mild tension with the EW fit (\textit{i.e.}~with $Z\to \tau^+\tau^-$ and $Z\to \bar\nu\nu$ data), and the DY di-tau bounds derived from recent measurements achieved by  the ATLAS collaboration. This leads to a combined likelihood difference
\begin{equation}
	-2 \Delta \log \mathcal{L} \equiv 2 \log \mathcal{L}_\text{SM} - 2 \log \mathcal{L} = -8.34\,,
\end{equation}
which corresponds to a pull of $2.1 \sigma$ for three degrees of freedom (d.o.f.).

As calculated in Section~\ref{sec:bsll_tauquark}, the loop-diagrams in Figure~\ref{fig:Bsll_feynman} induce lepton flavour violation universal effects in $b\to s\ell^+ \ell^-$ decays (via ${\cal O}_9^U$ and ${\cal O}_7$ operators) that are proportional to the product of couplings $Y^{LR}_{2\tau} Y^{LR*}_{3\tau}$. While these contributions cannot account for the $R_{K^{(*)}}$ anomalies, they are capable of explaining (partially) $P_5^\prime$ and $B_s\to\phi\mu^+\mu^-$ data. Moreover, as shown in the next subsection, this effect further improves the fit to $b\to s\ell^+ \ell^-$ data once the LFUV effects originating from the presence of the muoquark are included. On the other hand, the size of $Y^{LR}_{2\tau} Y^{LR*}_{3\tau}$ is limited by $B_s-\bar B_s$ mixing. While $Y^{LR*}_{3\tau}$ was already constrained by $Z\tau^+ \tau-$ and $Z \bar \nu \nu$ data (see above), a non-zero value of $Y^{LR}_{2\tau}$ additionally contributes to $D^0-\bar D^0 $ mixing. However, for $|Y^{LR}_{2\tau}|\approx 1/2 Y^{LR*}_{3\tau}$, $B_s-\bar B_s$ mixing provides the leading constraint and the corresponding bounds, together with their impact on the Wilson coefficients $\mathcal{C}_9^U$ and $\mathcal{C}_7$ are shown in Figure~\ref{fig:Bsll-tauquark}.

\subsection{Muoquark}
\label{sec:muons_only}

\begin{figure}
	\begin{raggedright}
		\includegraphics[width=0.99\textwidth]{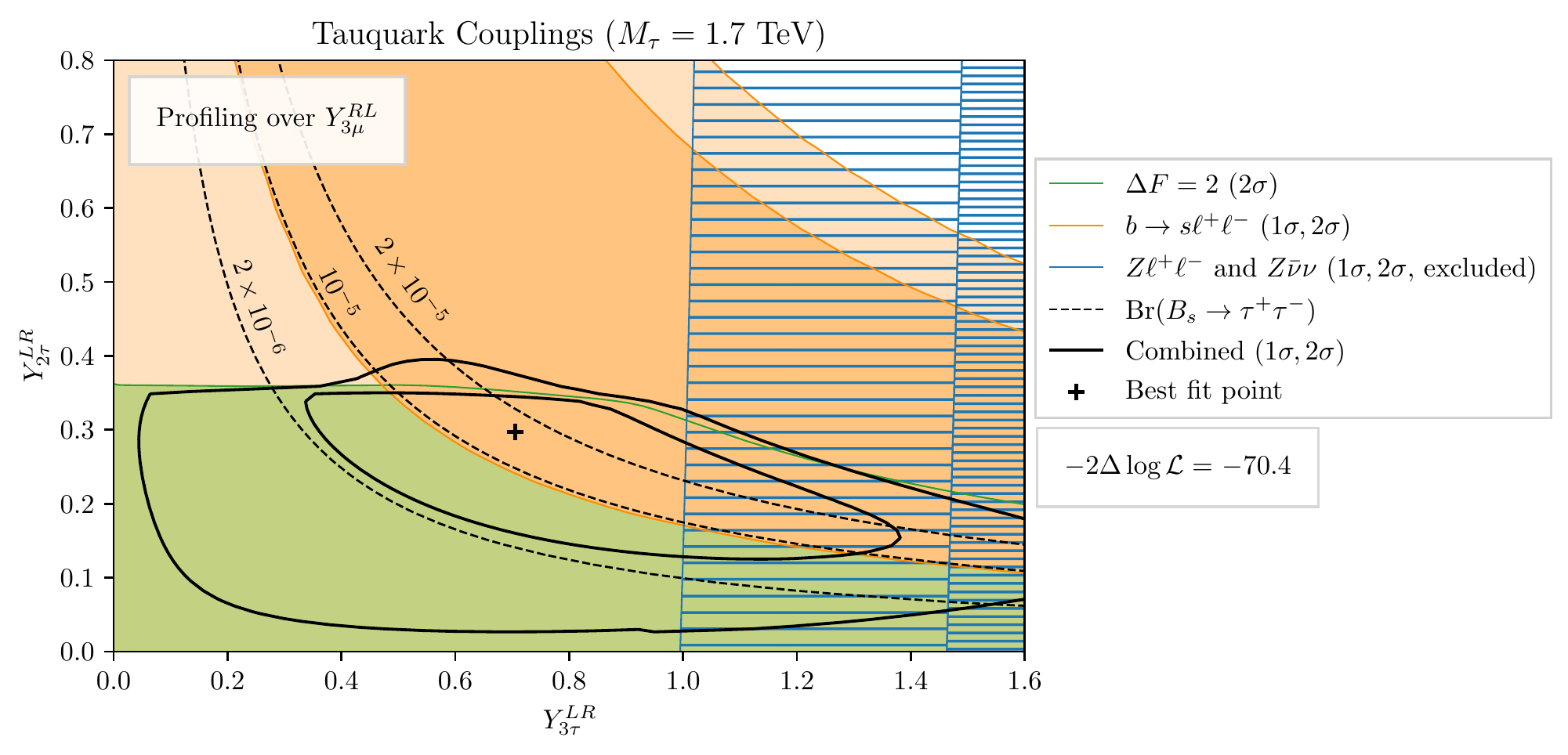}
		\includegraphics[width=0.99 \textwidth]{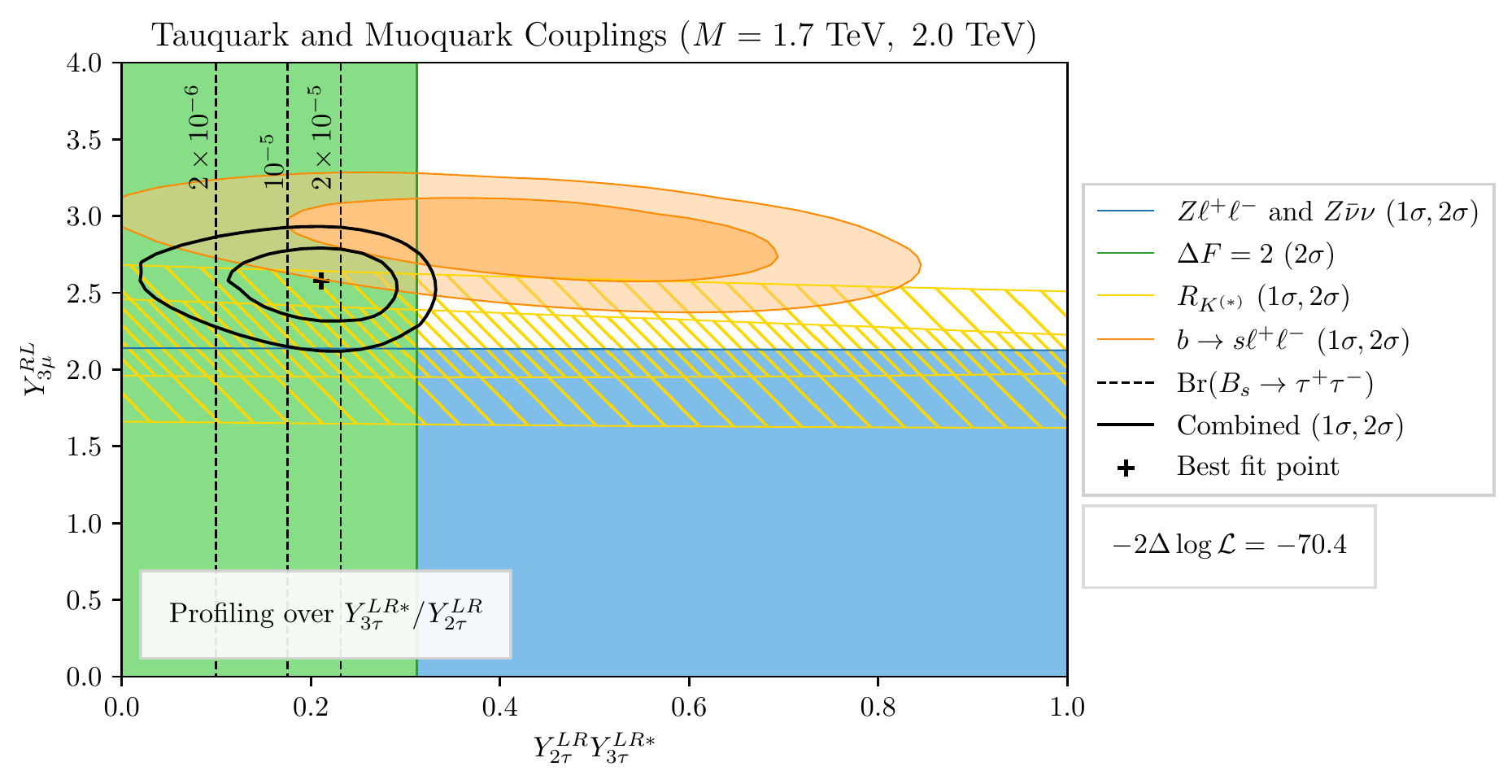}
	\caption{{\it Top}: Bounds on the $\Phi_{2, \tau}$ couplings $Y^{LR}_{3\tau}$ and $Y^{LR}_{2\tau}$ derived from the combined fit of $\Phi_{2, \tau}$ and $\Phi_{2, \mu}$ effects to $b \to s \ell^+ \ell^-$ observables (orange). Our results are obtained by profiling over the $\Phi_{2, \mu}$ contributions depending on the $Y^{RL}_{3\mu}$ coupling. The coupling $Y_{2\tau}^{LR}$ is tightly constrained by $D^0-\bar D^0$ mixing (green), and large $Y^{LR}_{3\tau}$ values worsen the fit to $B_s - \bar B_s$ mixing (green as well) and leads to tension with $Z \ell^+ \ell^-$ and $Z \bar \nu \nu$ data (blue). The figure also includes predictions for the $B_s \to \tau^+ \tau^-$ branching ratio. 
	{\it Bottom}: The same solution presented in terms of the product $Y^{LR}_{2\tau}Y^{LR*}_{3\tau}$ and the muoquark coupling $Y^{RL}_{3 \mu}$. We profile over the ratio $Y^{LR*}_{3\tau}/Y^{LR}_{2\tau}$ that is relevant for the bounds originating from $D^0 - \bar D^0$ mixing, as well as from the fit to $Z\tau^+ \tau^-$ and $Z \bar \nu \nu$ data.}
	\label{fig:Tau-mu-plots}
	\end{raggedright}
\end{figure}

Regarding $b \to s \ell^+ \ell^-$, an excellent fit to data can be obtained in new physics scenarios featuring an LFU $\mathcal{C}_9^U$ effect in addition to a LFUV violating effect $\mathcal{C}_9^\mu=-\mathcal{C}_{10}^\mu$~\cite{Alguero:2018nvb,Alguero:2019ptt}. In fact, this scenario even constitutes the two-dimensional one which gives the best fit to data~\cite{Alguero:2021anc}. This is precisely the setup realised in the considered model with three generations of LQs, in which the photon penguin contribution induced by the presence of the tauquark is combined with the $W$-box contribution involving the muoquark, so that $\mathcal{C}^\mu_9 = -\mathcal{C}^\mu_{10}$. We checked that the tree-level effect giving rise to $\mathcal{C}_9^\mu= +\mathcal{C}_{10}^\mu$ does not improve the fit, such that $Y^{LR*}_{3 \mu}Y^{LR}_{2 \mu}$ must be small. Furthermore, also the preferred value of $\mathcal{C}_9^e = +\mathcal{C}_{10}^e$ that is induced by the electroquark is consistent with zero once added to the other LQ contributions. As before, we choose the muoquark mass to be as low as possible while still satisfying the limits from LQ pair-production searches at the LHC by a clear margin. We hence fix $M_\mu = 2$~TeV.

In Figure~\ref{fig:Tau-mu-plots} we focus on the three couplings $Y^{LR}_{3 \tau}$, $Y^{LR}_{2 \tau}$ and $Y^{RL}_{3\mu}$ of the tauquark $\Phi_{2, \tau}$ and the muoquark $\Phi_{2, \mu}$, assuming them to be real. While these couplings are constrained by $B_s-\bar B_s$ mixing, $D^0-\bar D^0$ mixing, as well as by $Z \tau^+ \tau^-$ and $Z \bar \nu \nu$ coupling measurements, we can still significantly improve the fit to $b \to s \ell^+ \ell^-$ data. In fact, a likelihood value $-2\Delta \log{\cal L}=-70.4$, corresponding to a combined pull of $7.9 \sigma$ for three d.o.f., can be reached. Whereas the $W$-box contribution leading to $\mathcal{C}_9^\mu=-\mathcal{C}_{10}^\mu$ via a $Y^{RL}_{3\mu}$ coupling is in tension with $Z\mu^+\mu^-$ and $Z \bar \nu \nu$~\cite{Angelescu:2021lln} data, this tension is reduced once the contribution from the tauquark discussed in Section~\ref{sec:taus_only} is accounted for. Therefore, the combined effect of $\Phi_{2,\tau}$ and $\Phi_{2,\mu}$ leptoquarks does not only result in a better fit to $b \to s \ell^+ \ell^-$ data, but also weakens the bounds from EW precision observables as a smaller coupling $Y^{RL}_{3\mu}$ suffices.

\begin{figure}
	\begin{raggedright}
		\includegraphics[width=0.99\textwidth]{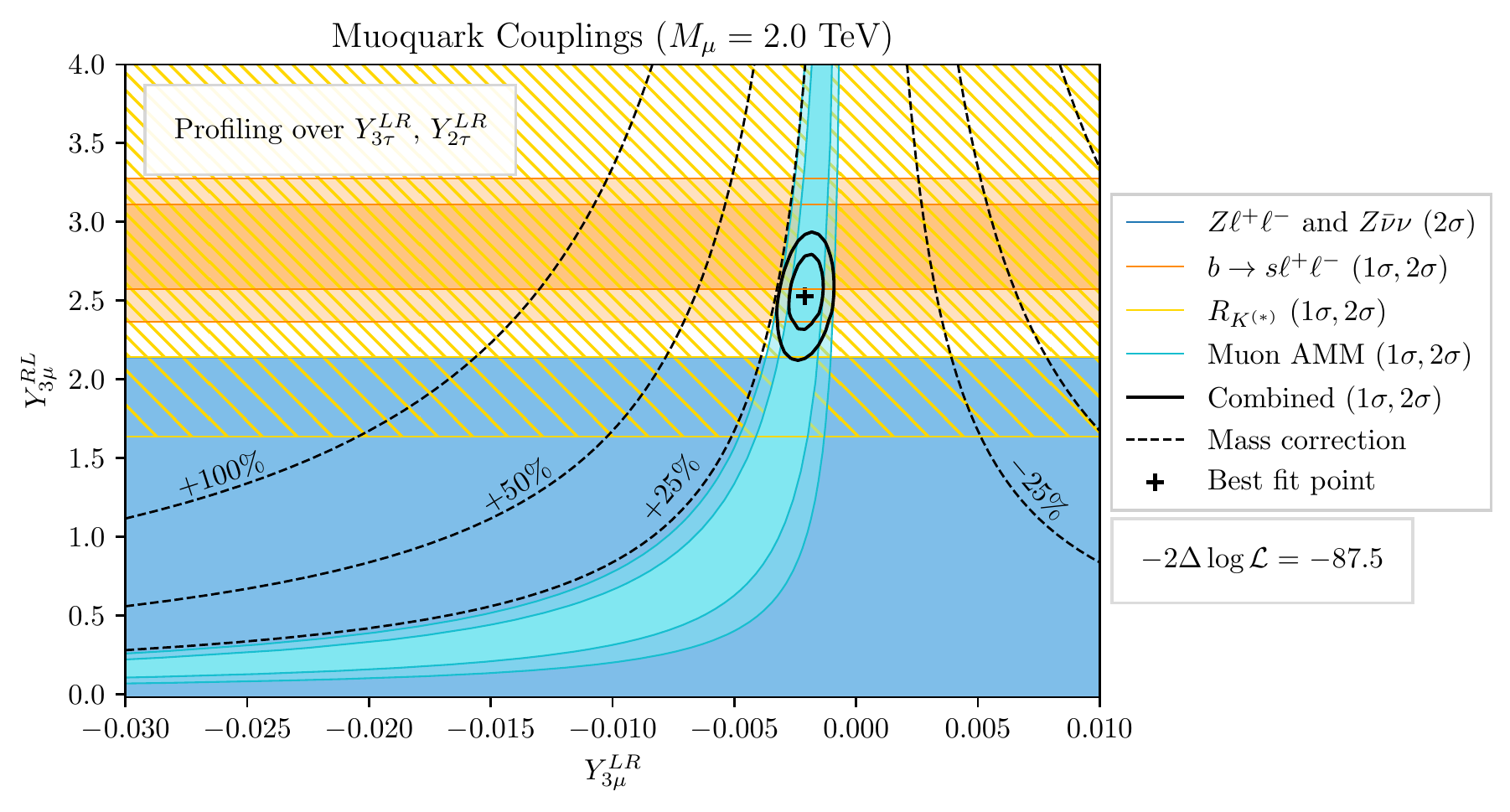}
	\caption{Simultaneous solution to the $b\to s \ell^+ \ell^-$ (orange) and $\left( g-2\right)_\mu$ (light blue) anomalies, as obtained by combining the effects of the couplings $Y^{LR}_{3\tau}$, $Y^{LR}_{2\tau}$, $Y^{RL}_{3\mu}$ and $Y^{LR}_{3\mu}$. We additionally show the relative radiative correction to the muon mass (black lines), that are derived by only taking into account only the finite part of the involved loop function. }
	\label{fig:RKRKstar_AMM_mu}
	\end{raggedright}
\end{figure}

We now include $(g-2)_\mu$ in our analysis. To provide an explanation to this anomaly, we need the combined effect of non-zero $Y^{RL}_{3\mu}$ and $Y^{LR}_{3\mu}$ couplings so that we can get the desired $m_t/m_\mu$ enhancement. Adding therefore $Y^{LR}_{3\mu}$ as a free parameter, but profiling over $Y_{3\tau}^{LR}$ and $Y_{2\tau}^{LR}$, we obtain the results of Figure~\ref{fig:RKRKstar_AMM_mu}. While we have assumed that all couplings are real, they can generally be complex. As consequence, a large muon EDM can be generated~\cite{Crivellin:2018qmi}. Our results additionally show that a chirally enhanced explanation for the $(g-2)_\mu$ anomaly leads to large radiative contributions to the muon mass as well as to the $h\to\mu\mu$ branching ratio~\cite{Crivellin:2021rbq}. While in order to measure the latter percent-level effect a future precision determination, like at FCC-hh~\cite{FCC:2018vvp}, would be necessary~\cite{Crivellin:2020tsz,Crivellin:2020mjs}, the contribution to the muon mass can be of order one (see the dashed isolines in Figure~\ref{fig:RKRKstar_AMM_mu}). However, this effect is not physical. It can be absorbed in a re-definition of the muon mass and thus only be bounded by fine-tuning arguments requiring the absence of large accidental cancellations. The combined log-likelihood difference is further decreased to $-87.5$, corresponding to a SM pull of $8.7\sigma$ for four d.o.f.

Finally, as shown in Ref.~\cite{Carvunis:2021dss}, the presence of a muoquark in the model can only weaken, but not fully explain, the anomaly in $\Delta A_\text{FB}$. In this case, a product of couplings $Y^{RL}_{2\mu} Y^{LR*}_{3\mu} \approx -1.4$ would lead to a pull of $1.8 \sigma$ for two d.o.f.\,. 

\subsection{Electroquark}
\label{sec:electrons_only}

\begin{figure}
	\begin{centering}
		\includegraphics[width=0.85\textwidth]{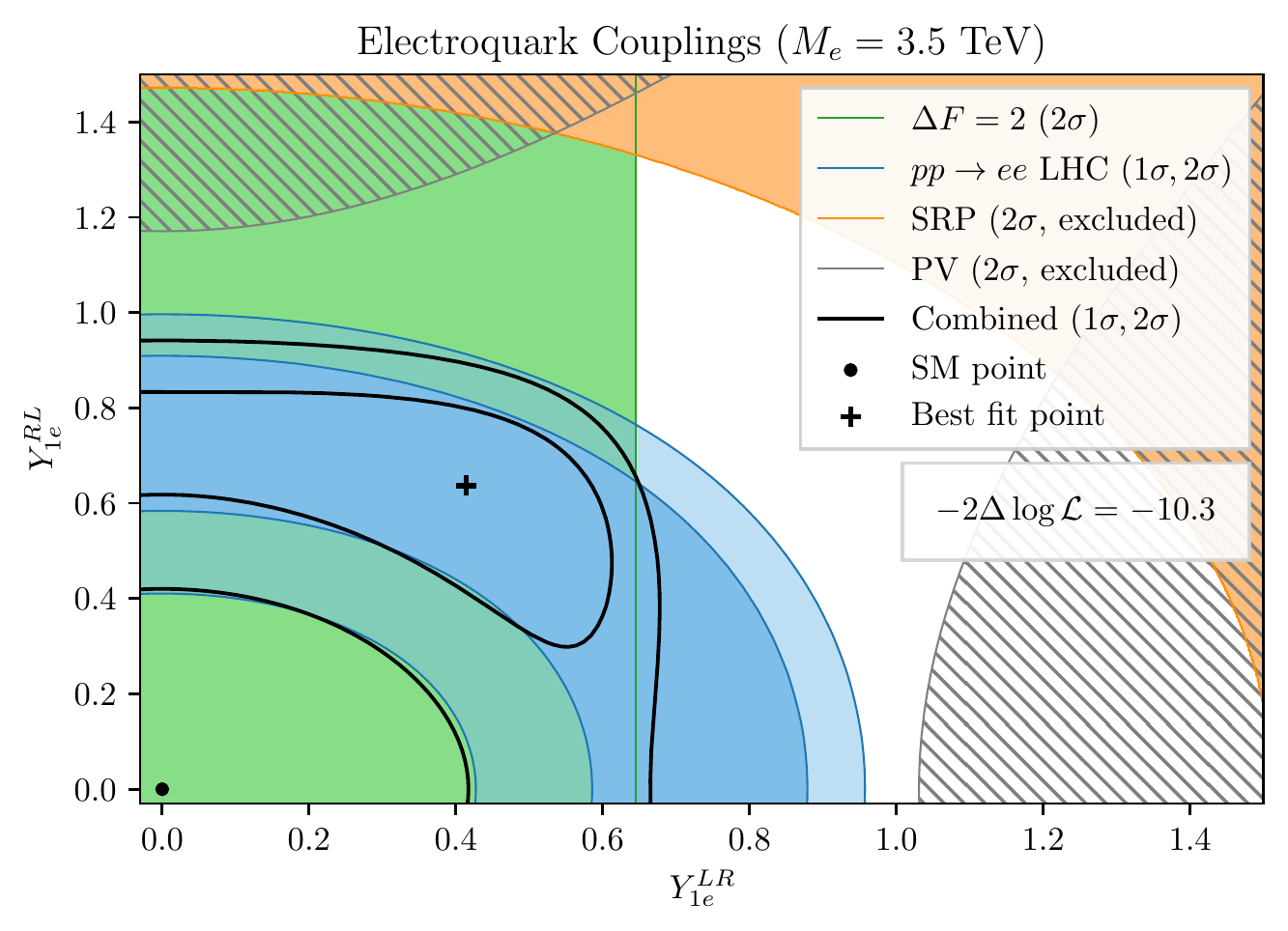}
	\caption{Parameter space region leading to a possible explanation for the excess in di-electrons found by the CMS collaboration (light blue). The solution via real couplings $Y^{LR}_{1e}$ and $Y_{1e}^{RL}$ prefers values of $\mathcal{O}(0.6)$, which also yields an improved fit to PV observables~\cite{Crivellin:2021bkd}. The limit from the electron EDM measurement additionally places a strong constraint on the complex phases of the couplings. } \label{fig:CMS_e}
		\end{centering}
\end{figure}
\begin{figure}
	\begin{centering}
		\includegraphics[width=0.99\textwidth]{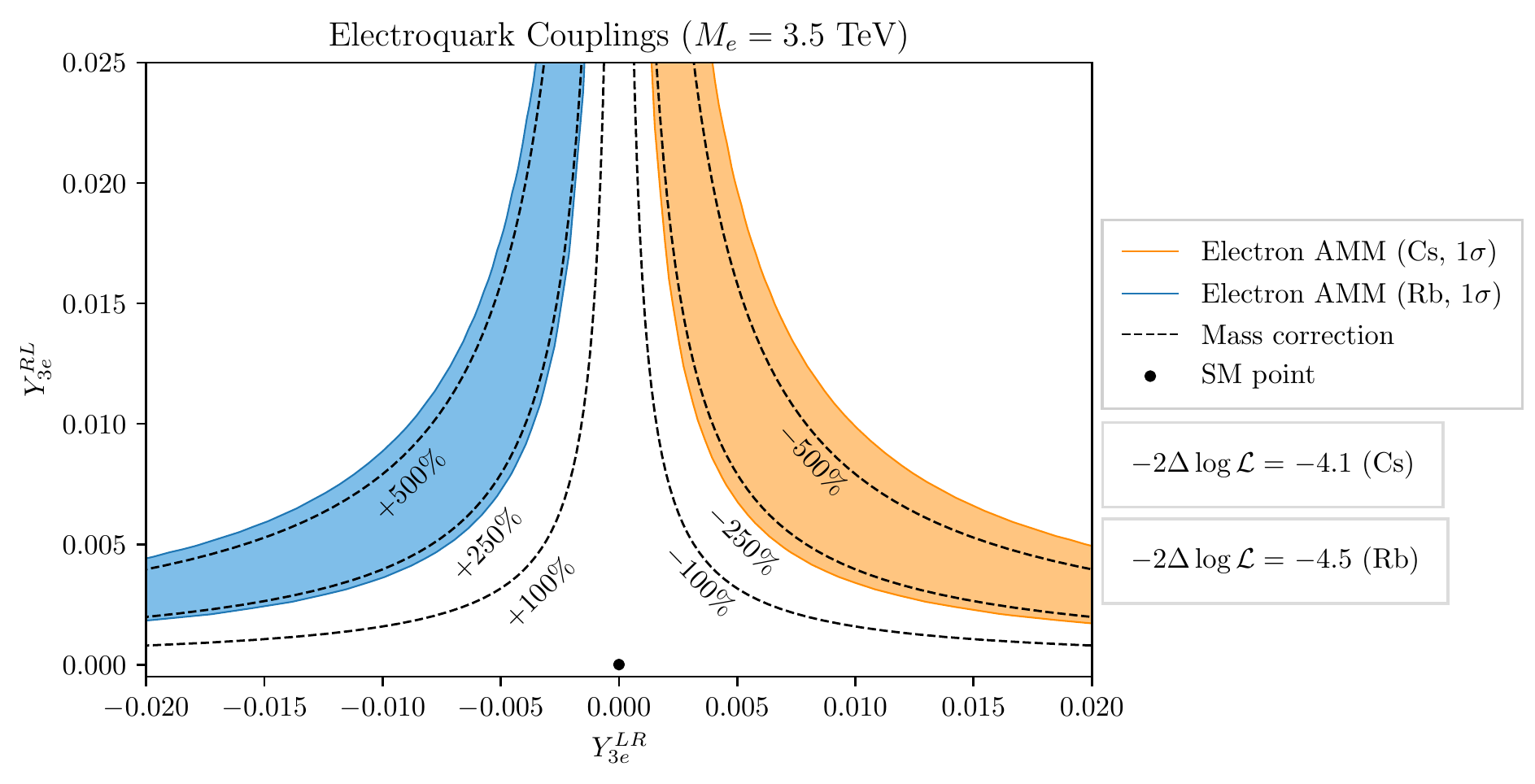}
	\caption{Potential effect of the $\Phi_{2,e}$ leptoquark on $(g-2)_e$ via the couplings  $Y^{LR}_{3e}$ and $Y^{RL}_{3e}$. We show the preferred parameter space regions for the two contradicting SM determinations of $\left(g-2\right)_e$ based on Rb (blue) and Cs (orange) atoms. In the preferred parameter space regions, sizeable loop corrections to the electron mass are generated, and the EDM constraint places strong limits on the complex phases of these couplings. }
	\label{fig:AMM_e}
	\end{centering}
\end{figure}

In the electron sector we aim to explain the CMS excess found in non-resonant di-electron production. In order to satisfy the bounds on the electroquark mass coming from its pair production at the LHC, $M_e$ should be of at least 2.1~TeV. However, we also have to consider SRP limits. The latter severely restrict the electroquark couplings to light quarks, which need to be large enough to yield a sizeable effect in non-resonant di-electron production to explain the deviations considered. Importantly, SRP involves on-shell LQs, while the corresponding effect on the non-resonant di-electron spectrum occurs from $t$-channel exchange (see the diagram on the right in Figure~\ref{fig:DY_feynman}). Therefore, by increasing the electroquark mass, the former bounds can be avoided independently of the LQ couplings while sizeable contributions to the latter process are still possible. We find that $M_e >3.5$ TeV is sufficient to avoid any SRP limits without requiring a too large Yukawa couplings when building an explanation for the non-resonant di-electron signal.

The corresponding preferred region of the parameter space is presented in the $Y^{LR}_{1e}$--$Y^{RL}_{1e}$ plane in Figure~\ref{fig:CMS_e}. Our results combine the CMS and ATLAS analyses of non-resonant di-electron production, and are shown in blue. Interestingly, these couplings can also achieve a slightly improved fit to the low-energy parity violation data~\cite{Crivellin:2021bkd}. Moreover, for sizeable $Y^{LR}_{1e}$ values the electron EDM limit places a strong constraint on $\text{arg}(Y^{LR}_{1e}Y^{RL*}_{1e})$. For the best-fit point in Figure~\ref{fig:CMS_e}, it has to be smaller than $\mathcal{O}(10^{-5})$.

Finally, the electroquark couplings $Y^{LR}_{3e}$ and $Y^{RL}_{3e}$ could in principle also lead to a deviation from SM predictions in $(g-2)_e$. The corresponding preferred regions in the parameter space are shown in Figure~\ref{fig:AMM_e} for the two contradicting determinations of $\alpha$ considered. In this case, significant radiative corrections to the electron mass arise, which can become larger than the measured mass itself. Additionally, the same couplings can also give rise to an electron EDM, placing strong constraints on the complex phases of the involved couplings. For the best fit values for the product $Y^{LR}_{3e}Y^{RL*}_{3e}$ (\textit{i.e.}~at the center of the blue and orange regions in Figure~\ref{fig:CMS_e}), the complex phase needs to be $\lesssim \mathcal{O}(10^{-5})$ in order to satisfy EDM constraints. 

\subsection{$W$ mass}
\begin{figure}
	\begin{raggedright}
		\includegraphics[width=0.99\textwidth]{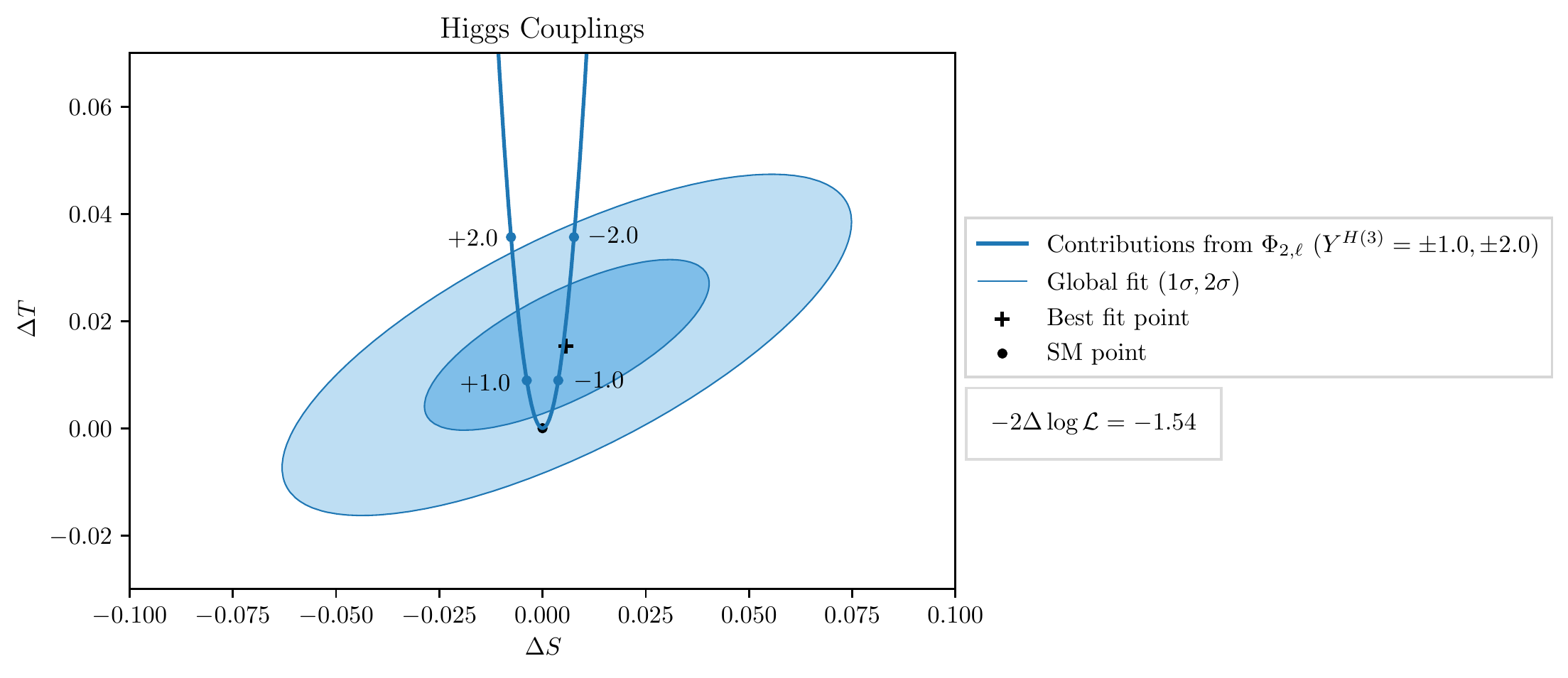}
	\caption{Global fit to the (difference of the) Peskin-Takeuchi parameters  $\Delta S$ and $\Delta T$. This fit can be improved via a positive contribution to both $\Delta S$ and $\Delta T$, which corresponds to negative couplings $Y^{H(3)}_\ell$. We indicate the position of the points $Y^{H(3)} = \pm 1, \pm 2$ when we assume that $Y^{H(3)}_e = Y^{H(3)}_\mu = Y^{H(3)}_\tau \equiv Y^{H(3)}$.}
	\label{fig:OC}
	\end{raggedright}
\end{figure}

Finally, we discuss the shift in the predictions for the $W$ mass generated by the presence of LQs in our model. We show the global fit~\cite{Ellis:2018gqa} to new physics contributions in the Peskin-Takeuchi parameters $S$ and $T$ in Figure~\ref{fig:OC} (for one d.o.f.), together with the effects that our model can yield. The agreement with data is improved for positive values for $\Delta S$ and $\Delta T$. As an illustration, the best fit point in our model is given by
\begin{equation}
Y_\ell^{H(3)} = -1.16 ~~~(\ell = e, \mu, \tau) \,,
\end{equation}
where we assume that $Y^{H(3)}_{e} = Y^{H(3)}_{\mu} = Y^{H(3)}_{\tau}$ for simplicity. The corresponding likelihood difference value is $-1.54$.

\section{Conclusions}
\label{sec:conclusion}

While it is known that a single $SU(2)_L$ doublet of scalar leptoquarks can explain $b\to s\ell^+\ell^-$ data, $R_{D^{(*)}}$ and $(g-2)_\mu$ separately, a common explanation is not possible. This would indeed require couplings to taus, muons and electrons simultaneously, which would lead to unacceptably large effects in charged lepton flavour violating processes. In order to overcome this obstacle, we proposed in this article to extend the SM by three generations of scalar $SU(2)_L$ doublets $\Phi_{2,e}$, $\Phi_{2,\mu}$ and $\Phi_{2,\tau}$, each of them carrying the corresponding lepton flavour number. In that way, we can have multiple sources of LFUV while the individual lepton flavours are still exactly conserved, as in the SM (with massless neutrinos). As a result, our setup can be distinguished from LQ scenarios with a single LQ already by considering low energy observables: It predicts that while effects in observables involving different lepton generations are possible, no sign of charged lepton flavour violating should be observed. Furthermore, assigning lepton flavour numbers to LQs automatically avoids bounds from proton decay experiments, as it forbids di-quark couplings.

In this setup we showed that:
\begin{itemize}
	\item The tauquark $\Phi_{2,\tau}$, despite effects in $Z\to\tau^+\tau^-$, $Z\to\bar\nu\nu$, $D^0-\bar D^0$ mixing and the bounds stemming from $\tau^+\tau^-$ searches at the LHC, can (mostly) explain $b\to c\tau\nu$ data in the presence of a complex phase. 
	\item  The presence of the tauquark $\Phi_{2,\tau}$ generates a universal operator ${\cal O}_9^U$ via off-shell photon penguin diagram contributions with tau leptons running in the loop. While the effect is bounded by $B_s-\bar B_s$ mixing and crucially depends on $M_\tau$ (\textit{i.e.}~the relative effect in $B_s-\bar B_s$ mixing is enhanced for larger LQ masses), a relevant contribution is possible.
	\item An excellent fit to  $b\to s\ell^+\ell^-$ data can be obtained via a combination of tauquark contributions (generating $\mathcal{C}_9^U$) and muoquark $W$-box contribution (generating $\mathcal{C}^\mu_9=-\mathcal{C}^\mu_{10}$). This leads to a likelihood difference $-2\Delta \log \mathcal{L} = -70.4$, which corresponds to $7.9 \sigma$ for three d.o.f.~.
	\item $\left(g-2\right)_\mu$ can be explained via an $m_t/m_\mu$ enhanced muoquark contribution.
	\item The electron excess appearing in the invariant-mass tails of DY di-electron production at the LHC $pp \to e^+e^-$ can be accounted for via a contribution from the electroquark alone, without violating any bounds from $D^0-\bar{D}^0$ mixing.
	\item The EW fit can be improved by a constructive contribution to the $W$-boson mass.
	\item A (possible) new physics effect in $(g-2)_e$ could be incorporated.
\end{itemize}

For the latter, the phase of the couplings in general needs to be precisely tuned in order to avoid the bounds from electron EDM. However, this obstacle could be resolved in our model by requiring that the lepton masses are generated at the loop level~\cite{Greljo:2021npi}. Such a radiative mass generation leads to an automatic phase alignment between the mass term and the dipole operators such that the electron EDM vanishes~\cite{Borzumati:1999sp,Crivellin:2010gw}.
\appendix

\section{Leptoquark effects in $Z$-boson couplings}
\label{sec:Zcouplings}

The $\Phi_{2, \ell}$ contributions $ g^\text{LQ}_{\ell_{A}} $ to the $Z$ boson couplings to leptons are given by~\cite{Crivellin:2020mjs, Arnan:2019olv}\footnote{Similar results for the diquark contribution to $Z \to \ell^+ \ell^-$ have been obtained in Ref.~\cite{Djouadi:1989me}.}
\begin{equation}
\begin{split}
g^\text{LQ}_{\ell_{L}} = &\ N_c \frac{Y^{RL}_{3\ell} Y^{RL*}_{3\ell}}{16 \pi^2} \Bigg[ \frac{x_{t, \ell} \left( x_{t, \ell} -1 - \log x_{t, \ell}\right)}{2\left(x_{t, \ell} - 1\right)^2}
+ \frac{x_{Z, \ell}}{12} \mathcal{G}_L(x_{t, \ell}, g^\text{tree}_{\ell_{L}}) \Bigg] \\
&\qquad + x_{Z, \ell} N_c \sum_{k = 1,2} \frac{Y^{RL}_{k\ell} Y^{RL*}_{k\ell}}{48 \pi^2} \Bigg[\frac{2s_w^2}{3} \ \left( \log x_{Z, \ell} - i\pi -\frac{1}{6} \right) + \frac{g^\text{tree}_{\ell_{L}}}{6}  \Bigg]\,, \\
g^\text{LQ}_{\ell_{R}} = &\ N_c \frac{\hat{Y}^{LR}_{3\ell} \hat{Y}^{LR*}_{3\ell}}{16 \pi^2} \Bigg[- \frac{x_{t, \ell} \left( x_{t, \ell} -1 - \log x_{t, \ell}\right)}{2\left(x_{t, \ell} - 1\right)^2}
+ \frac{x_{Z, \ell}}{12} \mathcal{G}_R(x_{t, \ell}, g^\text{tree}_{\ell_{R}}) \Bigg] \\
&\quad + x_{Z, \ell} N_c \sum_{k = 1,2} \frac{\hat{Y}^{LR}_{k\ell} \hat{Y}^{LR*}_{k\ell}}{48 \pi^2} \Bigg[\left(-\frac{1}{2} + \frac{2s_w^2}{3} \right) \left( \log x_{Z, \ell} - i\pi -\frac{1}{6} \right) + \frac{g^\text{tree}_{\ell_{R}}}{6}  \Bigg] \\ 
&\quad + x_{Z, \ell} N_c \sum_{k = 1,2,3} \frac{Y^{RL}_{k\ell} Y^{RL*}_{k\ell}}{48 \pi^2} \Bigg[\left(\frac{1}{2} - \frac{s_w^2}{3} \right) \left( \log x_{Z, \ell} - i\pi -\frac{1}{6} \right)+ \frac{g^\text{tree}_{\ell_{R}}}{6}   \Bigg] \,,
\end{split}
\end{equation}
where $x_{t, \ell} \equiv m_t^2/M_\ell^2$, $x_{Z, \ell} \equiv m_Z^2/M_\ell^2$ and the tree-level SM couplings are $g^\text{tree}_{\ell_{L}} = -\frac{1}{2} + s_w^2$ and $g^\text{tree}_{\ell_{R}} = s_w^2$. The functions $\mathcal{G}_L$ and $\mathcal{G}_R$ contain the $\mathcal{O}\left( x_Z \log x_t \right)$ terms that induce non-negligible corrections when the LQ mass is small. They read
\begin{equation}
\begin{split}
  \mathcal{G}_{L(R)} (x, g_\ell) =&\ g^\text{tree}_{u_{R(L)}} \frac{\left( x - 1\right) \left(5x^2 - 7x + 8 \right)
    - 2\left(x^3 + 2 \right)\log x}{\left(x -1 \right)^4} \\
  &\ + g^\text{tree}_{u_{L(R)}} \frac{\left( x - 1\right) \left(x^2 - 5x -2 \right) + 6x\log x}{\left(x -1 \right)^4} \\
  &\ + g_{\ell} \frac{\left( x - 1\right) \left(-11x^2 + 7x -2 \right) + 6x^3\log x}{3\left(x -1 \right)^4} \,,
\end{split}
\end{equation}
where $g^\text{tree}_{u_{L}} = \frac{1}{2} - \frac{2}{3}s_w^2$ and $g^\text{tree}_{u_{R}} = - \frac{2}{3}s_w^2$. The neutrino coupling $g^\text{LQ}_{\nu_{\ell, L}}$ can be derived from $g^\text{LQ}_{\ell_L}$ by replacing the tree-level coupling $g^\text{tree}_{\ell_L}$ by $g^\text{tree}_{\nu_{\ell, L}} = \frac{1}{2}$.

\section{Details of the LHC analyses}
\label{app:lhc}
\subsection{CMS non-resonant di-lepton production at the LHC}

We determined the cross sections in Eq.~(\ref{eq:DY_cross_section_simulation}) for each value of the couplings $Y^{LR}_{i\ell}, Y^{RL}_{i\ell} \in \{0.25, 0.5, 1.0, 2.0 \}$ individually, setting the other couplings to zero. We implemented a selection on the transverse momentum and pseudo-rapidity of the electrons (muons), namely $p_T> 35$~GeV (53 GeV) and $|\eta|< 2.5$ (2.4), which allows us to mimic the lepton candidate definitions of Ref.~\cite{Sirunyan:2021khd}. By fitting a polynomial in the Yukawa couplings to the resulting cross sections, we determined the contributions from the SM alone, the LQs alone, and the LQ-SM interference term for LQ-fermion couplings set to one. This provides enough information to calculate the different components of the cross section for general Yukawa coupling values through a rescaling of the results. 

\begin{figure}
	\begin{center}
		\includegraphics[width=0.99\textwidth]{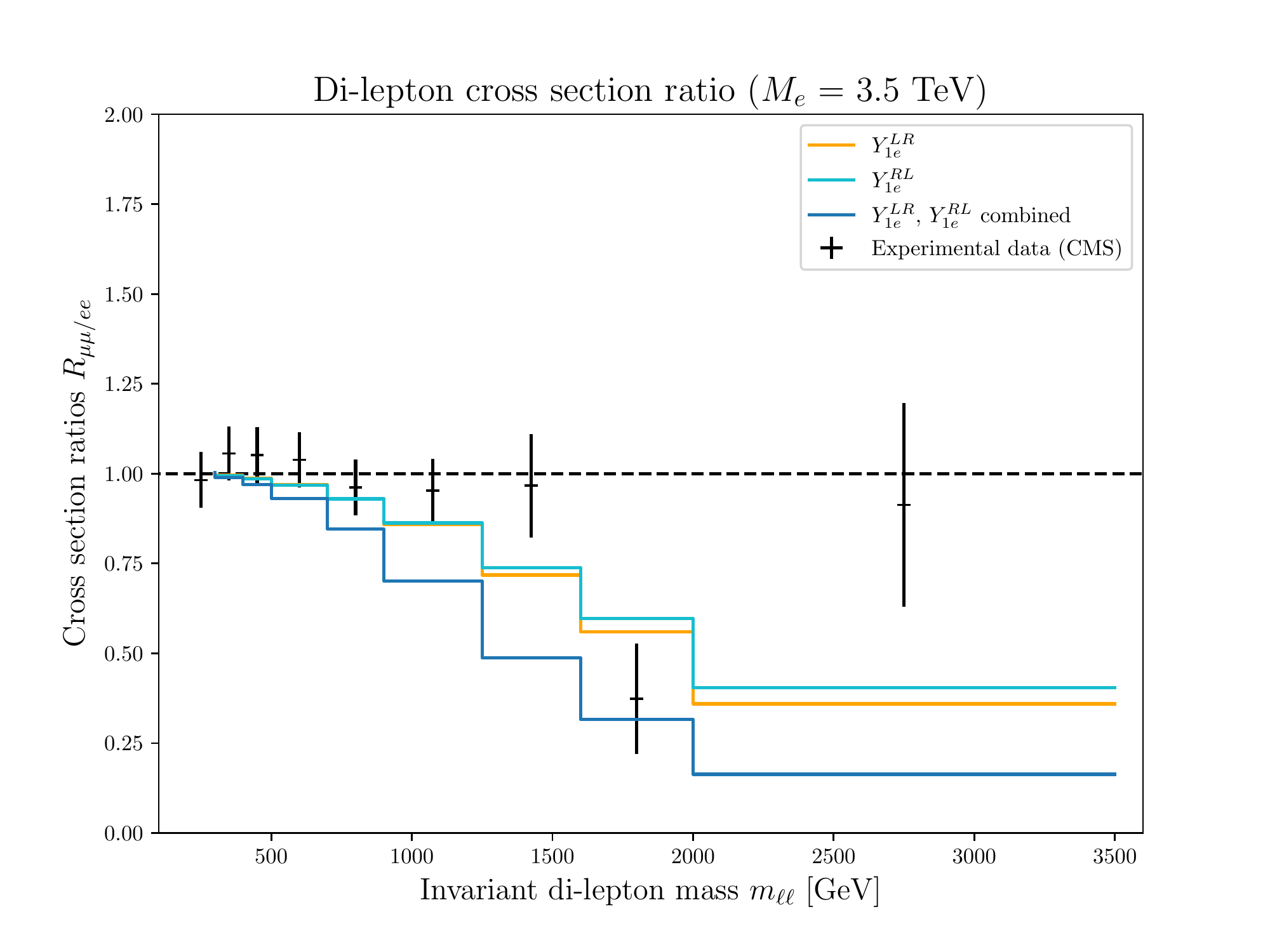}
	\end{center}
	\caption{The resulting values $R_{\mu\mu/ee}$ in the nine di-lepton mass bins. We show the distributions for an electroquark extension of the SM with $Y^{LR}_{1e} = 1.0$ and/or $Y^{RL}_{1e} = 1.0$. }
	\label{fig:CMS_cross_sections}
\end{figure}

Based on these cross sections we derive the ratios $R^{\text{LQ+SM}}_{\mu \mu/ee,n} \big/ R^{\text{SM}}_{\mu \mu/ee,n}$. The results for a specific scenario are shown in Figure~\ref{fig:CMS_cross_sections}, overlayed with the data from Ref.~\cite{Sirunyan:2021khd}. This allows us to build the likelihood function
\begin{equation}
-2\log \mathcal{L} = \sum_{n = 1}^9 \frac{\left( \dfrac{R_{\mu \mu / ee, n}^{\text{data}}}{R_{\mu \mu/ee, n}^{\text{MC}}} -  \dfrac{R_{\mu \mu / ee, n}^{\text{SM+LQ}}}{R_{\mu \mu / ee, n}^{\text{SM}}}\right)^2}{\sigma_n^2} \,,
\end{equation}
where $n$ runs over the nine $m_{\ell \ell}$ bins and $\sigma_n$ are the experimental uncertainties reported in Ref.~\cite{Sirunyan:2021khd}. 

\subsection{ATLAS non-resonant di-lepton production at the LHC}

In our analysis, we estimated the individual cross sections in the SRs chosen by the ATLAS collaboration using \texttt{MadGraph\_aMC@NLO}, analogously to the setup described in Section~\ref{sec:DY_CMS}. We implemented lepton $p_T$ cuts of 30 GeV (30 GeV) and $|\eta|$ cuts of 2.47 (2.5) on electron (muon) candidates. Based on the resulting cross sections for Yukawa coupling values of one, we extracted the ratios
\begin{equation}
\mu_\ell \equiv \dfrac{\int_{\text{SR}} \frac{d\sigma^\text{LQ+SM}}{d m_{\ell \ell}}\left(pp \to \ell^+ \ell^- \right) dm_{\ell \ell}}{\int_{\text{SR}} \frac{d\sigma^\text{SM}}{d m_{\ell \ell}}\left(pp \to \ell^+ \ell^- \right) dm_{\ell \ell}}
\end{equation}
with $\ell = e, \mu$ for general Yukawa coupling matrices $Y^{LR}$ and $Y^{RL}$. We then followed the statistical analysis achieved by the ATLAS collaboration, and built a likelihood function using a single-bin Poissonian counting-experiment approach. In the latter, the uncertainties are accounted for as Gaussian constraints that we profile over~\cite{ATLAS:2020yat, Junk:1999kv}.

\subsection{Non-resonant di-tau production at the LHC}
\label{sec:di-tau}

\begin{figure}
	\begin{center}
		\includegraphics[width=0.99\textwidth]{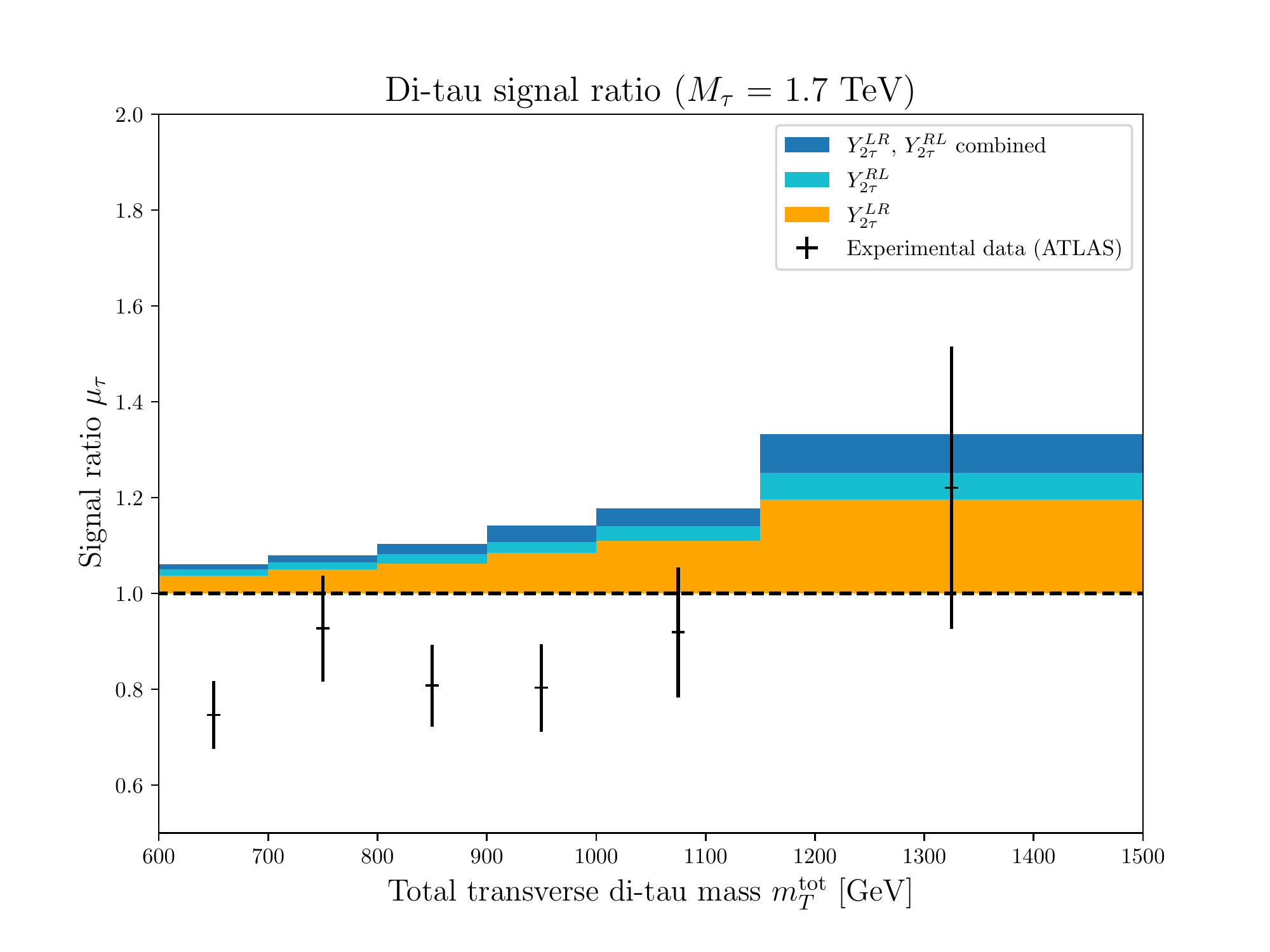}
	\caption{Signal ratios $\mu_\tau$ for the six $m_T^{\text{tot}}$ bins studied in the considered ATLAS analysis (for $m_T$ ranging from 600 to 1500 GeV). We show the contributions of a tauquark $\Phi_{2,\tau}$ with a mass $M_\tau$ = 1.7 TeV and Yukawa couplings $Y^{LR}_{2\tau} = 1.5$ and $Y^{RL}_{2\tau} = 1.5$. We distinguish between the contributions from $Y^{LR}_{2\tau}$ and those from $Y^{RL}_{2\tau}$, taking the two couplings individually, and include as well  results for a setup with both of them included.}
	\label{fig:CMS_cross_section_tau}
	\end{center}
\end{figure}

We focus on the production of a pair of tau leptons at the LHC when the di-tau system has a large invariant mass. We consider di-tau events populating the signal region of the ATLAS analysis of Ref.~\cite{ATLAS:2020yat}, and focus on the six highest bins in the total transverse mass  $m_T^{\text{tot}}$ bins defined by
\begin{equation}
\begin{aligned}
m_T^\text{tot} \equiv \sqrt{\left(p_T^{\tau_1} + p_T^{\tau_2} + E_T^\text{miss} \right)^2- \left( \mathbf{p}_T^{\tau_1} + \mathbf{p}_T^{\tau_1} + \mathbf{E}_T^{\text{miss}} \right)^2}\,.
\end{aligned}
\end{equation}
In this expression, $\mathbf{p}_T^{\tau_1}$ and  $\mathbf{p}_T^{\tau_2}$ are the transverse momenta of the visible daughter particles originating from the two hadronically-decaying taus, and $\mathbf{E}_T^{\text{miss}}$ is the total missing transverse momentum vector stemming from the unresolved particles including the daughter neutrinos. In our notation, $p_T^{\tau_1}$, $p_T^{\tau_2}$ and $E_T^{\text{miss}}$ are the respective moduli of the different two-vectors. The corresponding measurements are shown in Figure~\ref{fig:CMS_cross_section_tau} for $m_T^{\rm tot}$ bins defined by endpoints in  $\{600, 700, 800, 900, 1000, 1150, 1500\}$~GeV. % Note that in a $b$-tag exclusive analysis potentially more stringent limits on $Y^{LR}_{3\tau}$ could be derived, which we leave for future work. 

In our analysis, we generated partonic $pp \to \tau^+ \tau^-$ events using \texttt{MadGraph\_aMC@NLO}, analogously to what has been done in Section~\ref{sec:DY_CMS} but with loose cuts on the tau leptons ($p_T > 65$ GeV, $\left| \eta \right| < 2.5$) and on the invariant mass of the di-tau system $m_{\tau \tau} > 200$ GeV. We then showered these events with \texttt{Pythia} version 8.306~\cite{Sjostrand:2014zea} and analysed the resulting hadronic events using \texttt{MadAnalysis 5}~\cite{Conte:2012fm, Conte:2014zja, Conte:2018vmg} and a \texttt{FastJet}-based detector simulation~\cite{Cacciari:2011ma, Araz:2020lnp}. At the reconstructed level, we imposed $p_T > 160$ GeV and $|\eta| < 2.5$ for each hadronic tau and finally carried out the binning in $m_T^\text{tot}$. 

This yielded
\begin{equation}
\begin{aligned}
\mu_{\tau, n} \equiv \dfrac{\int_{\text{bin n}} \frac{d\sigma^\text{LQ+SM}}{d m^\text{tot}_{T}}\left(pp \to \tau_\text{had}^+ \tau_\text{had}^- \right) dm^\text{tot}_{T}}{\int_{\text{bin n}} \frac{d\sigma^\text{SM}}{d m^\text{tot}_{T}}\left(pp \to \tau_\text{had}^+ \tau_\text{had}^- \right) dm^\text{tot}_{T}} \times R_n\,,
\end{aligned}
\end{equation}
for the $n = 1, \dots, 6$ bins indicated above, and for general Yukawa coupling matrices $Y^{LR}$ and  $Y^{RL}$. The factor $R_n \in [0,1]$ accounts for the additional SM backgrounds originating from multijet production and the charged-current process $p p\to W \to \tau \nu$ that are small but not negligible compared to the background from the Drell-Yan process $p p\to (Z/\gamma)^\ast \to \tau^+ \tau^-$~\cite{ATLAS:2020zms}. The resulting signal ratios for a specific scenario are given in Figure~\ref{fig:CMS_cross_section_tau}. Finally, we again built a likelihood using a multi-bin Poissonian counting-experiment approach with Gaussian constraints for the uncertainties.

% ––––––––––––––––––––––––––––––––
% –-––--–  Hadron EDMS 	–––––––––
% ––––––––––––––––––––––––––––––––
\section{Electric dipole moments of hadrons}
\label{sec:hadron_edms}

\begin{table}[t!]
	\begin{center}
		\begin{tabular}{c|cccc|}
			\cline{2-5}
			& \multicolumn{4}{|c|}{Coupling} \\
			\cline{2-5}
			& $X_{12}$ & $X_{22}$ & $X_{13}$ & $X_{23}$ \\
			\hline
			\multicolumn{1}{|c|}{Fundamental limits} & $<9\times 10^{-4}$ & -- & $<7\times 10^{-5}$ & -- \\
			\multicolumn{1}{|c|}{Measurement} & $d_{n}$ & -- &$d_{n}$ & -- \\
			\hline
			\multicolumn{1}{|c|}{Strict limits} & $4\times 10^{-4}$ & 0.5 (3) & $3\times 10^{-5}$ & $3\times 10^{-2} (0.2)$  \\
			\multicolumn{1}{|c|}{Measurement} & $d_{\text{Hg}}$ & $d_{\text{Hg}}$ &$d_{\text{Hg}}$ & $d_{\text{Hg}}$ \\
			\hline
		\end{tabular}
		\end{center}
		\caption{Limits on the LQ couplings to fermions coming from EDM measurements in hadrons. We only show the constraints that are more stringent than the ones originating from lepton EDM measurements. }
		\label{tab:EDM_limits}
	\end{table}

The experiments targeting the observation of EDMs in neutrons~\cite{Pendlebury:2015lrz, Baker:2006ts, nEDM:2020crw} and Hg atoms~\cite{Griffith:2009zz, Graner:2016ses} have so far yielded null results. For our model, the most relevant resulting upper limits (at 90\% C.L.) are given by
\begin{equation}
\begin{aligned}
d_n&<1.8 \times 10^{-26}  \text{e cm} \qquad\text{and}\qquad  d_{\text{Hg}} & < 6.3\times 10^{-30}  \text{e cm}\,.
\end{aligned}
\end{equation}
Ref.~\cite{Dekens:2018bci} presented a detailed study of the LQ effects in the EDMs of hadrons. In Table~\ref{tab:EDM_limits} we list the subset of their limits that are more stringent than the ones coming from the lepton EDMs, extracting the bounds on the quantity
\begin{equation}
X_{i\ell} \equiv \text{Im} \left[ \frac{\hat{Y}^{LR}_{i\ell} Y^{RL*}_{i\ell}}{M_\ell^2} \right] \,,
\end{equation}
for $i, \ell = 1,2,3$. The authors of Ref.~\cite{Dekens:2018bci} presented two sets of limits. The first one is based on a ``pessimistic'' approach (fundamental limits), where all matrix elements are varied within their admissible theoretical ranges without assuming any probability distribution. This method is called the range-fit method that is suitable to rule out models. The second set of limits is based on a more ``optimistic'' approach (strict limits), where the theoretical uncertainties are neglected and the central values of the hadronic, nuclear, and atomic matrix elements are assumed to be true. The real constraints are likely to lie between these two extremes. The table additionally includes, in parentheses,  the limits that can be obtained after neglecting the contributions from charm tensor charge.

Whereas the strictest limit for $X_{23}$ in Table~\ref{tab:EDM_limits} would rule out the $\Phi_{2, \tau}$ solution to $R_{D^{(*)}}$, the authors of Ref.~\cite{Dekens:2018bci} noted that due to the large theoretical uncertainties inherent to the derivation of this limit it is too early to draw strong conclusions regarding the viability of the $\Phi_{2, \tau}$ solution. It will however be interesting to see whether next-generation $d_n$ or $d_\text{Hg}$ experiments will be able to detect EDM signals. 

\acknowledgments
The work of A.C. is supported by a Professorship Grant (PP00P2\_176884) of the National Science Foundation. We thank Yuta Takahashi and  Arne Christoph Reimers for useful discussions regarding the PP limits, Ulrich Haisch for pointing out the EDM constraints and his help regarding the \texttt{MadGraph\_aMC@NLO} simulations as well as Diego Guadagnoli, Alexandre Carvunis and Claudio Andrea Manzari for sharing the \texttt{flavio} implementations of $\Delta A_{\rm FB}$ and $R_{K_s^0}$ with us.

\bibliographystyle{JHEP}
\bibliography{main.bib}

\providecommand{\href}[2]{#2}\begingroup\raggedright\begin{thebibliography}{100}

\bibitem{ATLAS:2012yve}
{\scshape ATLAS} collaboration, \emph{{Observation of a new particle in the
  search for the Standard Model Higgs boson with the ATLAS detector at the
  LHC}}, \href{https://doi.org/10.1016/j.physletb.2012.08.020}{\emph{Phys.
  Lett. B} {\bfseries 716} (2012) 1}
  [\href{https://arxiv.org/abs/1207.7214}{{\ttfamily 1207.7214}}].

\bibitem{CMS:2012qbp}
{\scshape CMS} collaboration, \emph{{Observation of a New Boson at a Mass of
  125 GeV with the CMS Experiment at the LHC}},
  \href{https://doi.org/10.1016/j.physletb.2012.08.021}{\emph{Phys. Lett. B}
  {\bfseries 716} (2012) 30} [\href{https://arxiv.org/abs/1207.7235}{{\ttfamily
  1207.7235}}].

\bibitem{Fischer:2021sqw}
O.~Fischer et~al., \emph{{Unveiling Hidden Physics at the LHC}},
  \href{https://arxiv.org/abs/2109.06065}{{\ttfamily 2109.06065}}.

\bibitem{Crivellin:2021sff}
A.~Crivellin and M.~Hoferichter, \emph{{Hints of lepton flavor universality
  violations}}, \href{https://doi.org/10.1126/science.abk2450}{\emph{Science}
  {\bfseries 374} (2021) 1051}
  [\href{https://arxiv.org/abs/2111.12739}{{\ttfamily 2111.12739}}].

\bibitem{CMS:2014xfa}
{\scshape CMS, LHCb} collaboration, \emph{{Observation of the rare
  $B^0_s\to\mu^+\mu^-$ decay from the combined analysis of CMS and LHCb data}},
  \href{https://doi.org/10.1038/nature14474}{\emph{Nature} {\bfseries 522}
  (2015) 68} [\href{https://arxiv.org/abs/1411.4413}{{\ttfamily 1411.4413}}].

\bibitem{Aaij:2015oid}
{\scshape LHCb} collaboration, \emph{{Angular analysis of the $B^{0} \to K^{*0}
  \mu^{+} \mu^{-}$ decay using 3 fb$^{-1}$ of integrated luminosity}},
  \href{https://doi.org/10.1007/JHEP02(2016)104}{\emph{JHEP} {\bfseries 02}
  (2016) 104} [\href{https://arxiv.org/abs/1512.04442}{{\ttfamily
  1512.04442}}].

\bibitem{Abdesselam:2016llu}
{\scshape Belle} collaboration, \emph{{Angular analysis of $B^0 \to
  K^\ast(892)^0 \ell^+ \ell^-$}},  in \emph{{LHC Ski 2016}: {A First Discussion
  of 13 TeV Results}}, 4, 2016
  [\href{https://arxiv.org/abs/1604.04042}{{\ttfamily 1604.04042}}].

\bibitem{Aaij:2017vbb}
{\scshape LHCb} collaboration, \emph{{Test of lepton universality with $B^{0}
  \rightarrow K^{*0}\ell^{+}\ell^{-}$ decays}},
  \href{https://doi.org/10.1007/JHEP08(2017)055}{\emph{JHEP} {\bfseries 08}
  (2017) 055} [\href{https://arxiv.org/abs/1705.05802}{{\ttfamily
  1705.05802}}].

\bibitem{Aaij:2019wad}
{\scshape LHCb} collaboration, \emph{{Search for lepton-universality violation
  in $B^+\to K^+\ell^+\ell^-$ decays}},
  \href{https://doi.org/10.1103/PhysRevLett.122.191801}{\emph{Phys. Rev. Lett.}
  {\bfseries 122} (2019) 191801}
  [\href{https://arxiv.org/abs/1903.09252}{{\ttfamily 1903.09252}}].

\bibitem{Aaij:2020nrf}
{\scshape LHCb} collaboration, \emph{{Measurement of $CP$-Averaged Observables
  in the $B^{0}\rightarrow K^{*0}\mu^{+}\mu^{-}$ Decay}},
  \href{https://doi.org/10.1103/PhysRevLett.125.011802}{\emph{Phys. Rev. Lett.}
  {\bfseries 125} (2020) 011802}
  [\href{https://arxiv.org/abs/2003.04831}{{\ttfamily 2003.04831}}].

\bibitem{LHCb:2021trn}
{\scshape LHCb} collaboration, \emph{{Test of lepton universality in
  beauty-quark decays}},  \href{https://arxiv.org/abs/2103.11769}{{\ttfamily
  2103.11769}}.

\bibitem{BELLE:2019xld}
{\scshape BELLE} collaboration, \emph{{Test of lepton flavor universality and
  search for lepton flavor violation in $B \rightarrow K\ell \ell$ decays}},
  \href{https://doi.org/10.1007/JHEP03(2021)105}{\emph{JHEP} {\bfseries 03}
  (2021) 105} [\href{https://arxiv.org/abs/1908.01848}{{\ttfamily
  1908.01848}}].

\bibitem{Belle:2019oag}
{\scshape Belle} collaboration, \emph{{Test of Lepton-Flavor Universality in
  ${B\to K^\ast\ell^+\ell^-}$ Decays at Belle}},
  \href{https://doi.org/10.1103/PhysRevLett.126.161801}{\emph{Phys. Rev. Lett.}
  {\bfseries 126} (2021) 161801}
  [\href{https://arxiv.org/abs/1904.02440}{{\ttfamily 1904.02440}}].

\bibitem{LHCb:2016ykl}
{\scshape LHCb} collaboration, \emph{{Measurements of the S-wave fraction in
  $B^{0}\rightarrow K^{+}\pi^{-}\mu^{+}\mu^{-}$ decays and the
  $B^{0}\rightarrow K^{\ast}(892)^{0}\mu^{+}\mu^{-}$ differential branching
  fraction}}, \href{https://doi.org/10.1007/JHEP11(2016)047}{\emph{JHEP}
  {\bfseries 11} (2016) 047}
  [\href{https://arxiv.org/abs/1606.04731}{{\ttfamily 1606.04731}}].

\bibitem{LHCb:2021zwz}
{\scshape LHCb} collaboration, \emph{{Branching Fraction Measurements of the
  Rare $B^0_s\rightarrow\phi\mu^+\mu^-$ and $B^0_s\rightarrow
  f_2^\prime(1525)\mu^+\mu^-$- Decays}},
  \href{https://doi.org/10.1103/PhysRevLett.127.151801}{\emph{Phys. Rev. Lett.}
  {\bfseries 127} (2021) 151801}
  [\href{https://arxiv.org/abs/2105.14007}{{\ttfamily 2105.14007}}].

\bibitem{LHCb:2020zud}
{\scshape LHCb, ATLAS, CMS} collaboration, \emph{{Combination of the ATLAS, CMS
  and LHCb results on the $B^0_{(s)} \to \mu^+ \mu^-$ decays}}, .

\bibitem{Lees:2012xj}
{\scshape BaBar} collaboration, \emph{{Evidence for an excess of $\bar{B} \to
  D^{(*)} \tau^-\bar{\nu}_\tau$ decays}},
  \href{https://doi.org/10.1103/PhysRevLett.109.101802}{\emph{Phys. Rev. Lett.}
  {\bfseries 109} (2012) 101802}
  [\href{https://arxiv.org/abs/1205.5442}{{\ttfamily 1205.5442}}].

\bibitem{Lees:2013uzd}
{\scshape BaBar} collaboration, \emph{{Measurement of an Excess of $\bar{B} \to
  D^{(*)}\tau^- \bar{\nu}_\tau$ Decays and Implications for Charged Higgs
  Bosons}}, \href{https://doi.org/10.1103/PhysRevD.88.072012}{\emph{Phys. Rev.
  D} {\bfseries 88} (2013) 072012}
  [\href{https://arxiv.org/abs/1303.0571}{{\ttfamily 1303.0571}}].

\bibitem{Aaij:2015yra}
{\scshape LHCb} collaboration, \emph{{Measurement of the ratio of branching
  fractions $\mathcal{B}(\bar{B}^0 \to
  D^{*+}\tau^{-}\bar{\nu}_{\tau})/\mathcal{B}(\bar{B}^0 \to
  D^{*+}\mu^{-}\bar{\nu}_{\mu})$}},
  \href{https://doi.org/10.1103/PhysRevLett.115.111803}{\emph{Phys. Rev. Lett.}
  {\bfseries 115} (2015) 111803}
  [\href{https://arxiv.org/abs/1506.08614}{{\ttfamily 1506.08614}}].

\bibitem{Aaij:2017deq}
{\scshape LHCb} collaboration, \emph{{Test of Lepton Flavor Universality by the
  measurement of the $B^0 \to D^{*-} \tau^+ \nu_{\tau}$ branching fraction
  using three-prong $\tau$ decays}},
  \href{https://doi.org/10.1103/PhysRevD.97.072013}{\emph{Phys. Rev. D}
  {\bfseries 97} (2018) 072013}
  [\href{https://arxiv.org/abs/1711.02505}{{\ttfamily 1711.02505}}].

\bibitem{Aaij:2017uff}
{\scshape LHCb} collaboration, \emph{{Measurement of the ratio of the $B^0 \to
  D^{*-} \tau^+ \nu_{\tau}$ and $B^0 \to D^{*-} \mu^+ \nu_{\mu}$ branching
  fractions using three-prong $\tau$-lepton decays}},
  \href{https://doi.org/10.1103/PhysRevLett.120.171802}{\emph{Phys. Rev. Lett.}
  {\bfseries 120} (2018) 171802}
  [\href{https://arxiv.org/abs/1708.08856}{{\ttfamily 1708.08856}}].

\bibitem{Abdesselam:2019dgh}
{\scshape Belle} collaboration, \emph{{Measurement of $\mathcal{R}(D)$ and
  $\mathcal{R}(D^{\ast})$ with a semileptonic tagging method}},
  \href{https://arxiv.org/abs/1904.08794}{{\ttfamily 1904.08794}}.

\bibitem{Altmannshofer:2021qrr}
W.~Altmannshofer and P.~Stangl, \emph{{New physics in rare B decays after
  Moriond 2021}},
  \href{https://doi.org/10.1140/epjc/s10052-021-09725-1}{\emph{Eur. Phys. J. C}
  {\bfseries 81} (2021) 952}
  [\href{https://arxiv.org/abs/2103.13370}{{\ttfamily 2103.13370}}].

\bibitem{Geng:2021nhg}
L.-S.~Geng, B.~Grinstein, S.~J\"ager, S.-Y.~Li, J.~Martin~Camalich and
  R.-X.~Shi, \emph{{Implications of new evidence for lepton-universality
  violation in b\textrightarrow{}s\ensuremath{\ell}+\ensuremath{\ell}-
  decays}}, \href{https://doi.org/10.1103/PhysRevD.104.035029}{\emph{Phys. Rev.
  D} {\bfseries 104} (2021) 035029}
  [\href{https://arxiv.org/abs/2103.12738}{{\ttfamily 2103.12738}}].

\bibitem{Alguero:2021anc}
M.~Alguer\'o, B.~Capdevila, S.~Descotes-Genon, J.~Matias and M.~Novoa-Brunet,
  \emph{{$b\to s\ell\ell$ Global Fits after $R_{K_S}$ and $R_{K^{*+}}$}},  4,
  2021 [\href{https://arxiv.org/abs/2104.08921}{{\ttfamily 2104.08921}}].

\bibitem{Hurth:2021nsi}
T.~Hurth, F.~Mahmoudi, D.M.~Santos and S.~Neshatpour, \emph{{More indications
  for lepton nonuniversality in $b\to s \ell^+ \ell^-$}},
  \href{https://doi.org/10.1016/j.physletb.2021.136838}{\emph{Phys. Lett. B}
  {\bfseries 824} (2022) 136838}
  [\href{https://arxiv.org/abs/2104.10058}{{\ttfamily 2104.10058}}].

\bibitem{Kowalska:2019ley}
K.~Kowalska, D.~Kumar and E.M.~Sessolo, \emph{{Implications for new physics in
  $b\rightarrow s \mu \mu $ transitions after recent measurements by Belle and
  LHCb}}, \href{https://doi.org/10.1140/epjc/s10052-019-7330-2}{\emph{Eur.
  Phys. J. C} {\bfseries 79} (2019) 840}
  [\href{https://arxiv.org/abs/1903.10932}{{\ttfamily 1903.10932}}].

\bibitem{Ciuchini:2021smi}
M.~Ciuchini, M.~Fedele, E.~Franco, A.~Paul, L.~Silvestrini and M.~Valli,
  \emph{{New Physics without bias: Charming Penguins and Lepton Universality
  Violation in $b \to s \ell^+ \ell^-$ decays}},
  \href{https://arxiv.org/abs/2110.10126}{{\ttfamily 2110.10126}}.

\bibitem{DAmico:2017mtc}
G.~D'Amico, M.~Nardecchia, P.~Panci, F.~Sannino, A.~Strumia, R.~Torre et~al.,
  \emph{{Flavour anomalies after the $R_{K^*}$ measurement}},
  \href{https://doi.org/10.1007/JHEP09(2017)010}{\emph{JHEP} {\bfseries 09}
  (2017) 010} [\href{https://arxiv.org/abs/1704.05438}{{\ttfamily
  1704.05438}}].

\bibitem{Arbey:2019duh}
A.~Arbey, T.~Hurth, F.~Mahmoudi, D.M.~Santos and S.~Neshatpour, \emph{{Update
  on the b\textrightarrow{}s anomalies}},
  \href{https://doi.org/10.1103/PhysRevD.100.015045}{\emph{Phys. Rev. D}
  {\bfseries 100} (2019) 015045}
  [\href{https://arxiv.org/abs/1904.08399}{{\ttfamily 1904.08399}}].

\bibitem{Kumar:2019nfv}
D.~Kumar, K.~Kowalska and E.M.~Sessolo, \emph{{Global Bayesian Analysis of new
  physics in $b \to s \mu\mu$ transitions after Moriond-2019}},  in \emph{{17th
  Conference on Flavor Physics and CP Violation}}, 6, 2019
  [\href{https://arxiv.org/abs/1906.08596}{{\ttfamily 1906.08596}}].

\bibitem{Isidori:2021vtc}
G.~Isidori, D.~Lancierini, P.~Owen and N.~Serra, \emph{{On the significance of
  new physics in $b\to s \ell^+ \ell^-$ decays}},
  \href{https://doi.org/10.1016/j.physletb.2021.136644}{\emph{Phys. Lett. B}
  {\bfseries 822} (2021) 136644}
  [\href{https://arxiv.org/abs/2104.05631}{{\ttfamily 2104.05631}}].

\bibitem{HFLAV:2019otj}
{\scshape HFLAV} collaboration, \emph{{Averages of b-hadron, c-hadron, and
  $\tau $-lepton properties as of 2018}},
  \href{https://doi.org/10.1140/epjc/s10052-020-8156-7}{\emph{Eur. Phys. J. C}
  {\bfseries 81} (2021) 226}
  [\href{https://arxiv.org/abs/1909.12524}{{\ttfamily 1909.12524}}].

\bibitem{Muong-2:2006rrc}
{\scshape Muon g-2} collaboration, \emph{{Final Report of the Muon E821
  Anomalous Magnetic Moment Measurement at BNL}},
  \href{https://doi.org/10.1103/PhysRevD.73.072003}{\emph{Phys. Rev. D}
  {\bfseries 73} (2006) 072003}
  [\href{https://arxiv.org/abs/hep-ex/0602035}{{\ttfamily hep-ex/0602035}}].

\bibitem{Muong-2:2021ojo}
{\scshape Muon g-2} collaboration, \emph{{Measurement of the Positive Muon
  Anomalous Magnetic Moment to 0.46 ppm}},
  \href{https://doi.org/10.1103/PhysRevLett.126.141801}{\emph{Phys. Rev. Lett.}
  {\bfseries 126} (2021) 141801}
  [\href{https://arxiv.org/abs/2104.03281}{{\ttfamily 2104.03281}}].

\bibitem{Aoyama:2020ynm}
T.~Aoyama et~al., \emph{{The anomalous magnetic moment of the muon in the
  Standard Model}},
  \href{https://doi.org/10.1016/j.physrep.2020.07.006}{\emph{Phys. Rept.}
  {\bfseries 887} (2020) 1} [\href{https://arxiv.org/abs/2006.04822}{{\ttfamily
  2006.04822}}].

\bibitem{Aoyama:2012wk}
T.~Aoyama, M.~Hayakawa, T.~Kinoshita and M.~Nio, \emph{{Complete Tenth-Order
  QED Contribution to the Muon g-2}},
  \href{https://doi.org/10.1103/PhysRevLett.109.111808}{\emph{Phys. Rev. Lett.}
  {\bfseries 109} (2012) 111808}
  [\href{https://arxiv.org/abs/1205.5370}{{\ttfamily 1205.5370}}].

\bibitem{Aoyama:2019ryr}
T.~Aoyama, T.~Kinoshita and M.~Nio, \emph{{Theory of the Anomalous Magnetic
  Moment of the Electron}},
  \href{https://doi.org/10.3390/atoms7010028}{\emph{Atoms} {\bfseries 7} (2019)
  28}.

\bibitem{Czarnecki:2002nt}
A.~Czarnecki, W.J.~Marciano and A.~Vainshtein, \emph{{Refinements in
  electroweak contributions to the muon anomalous magnetic moment}},
  \href{https://doi.org/10.1103/PhysRevD.67.073006}{\emph{Phys. Rev. D}
  {\bfseries 67} (2003) 073006}
  [\href{https://arxiv.org/abs/hep-ph/0212229}{{\ttfamily hep-ph/0212229}}].

\bibitem{Gnendiger:2013pva}
C.~Gnendiger, D.~St\"ockinger and H.~St\"ockinger-Kim, \emph{{The electroweak
  contributions to $(g-2)_\mu$ after the Higgs boson mass measurement}},
  \href{https://doi.org/10.1103/PhysRevD.88.053005}{\emph{Phys. Rev. D}
  {\bfseries 88} (2013) 053005}
  [\href{https://arxiv.org/abs/1306.5546}{{\ttfamily 1306.5546}}].

\bibitem{Davier:2017zfy}
M.~Davier, A.~Hoecker, B.~Malaescu and Z.~Zhang, \emph{{Reevaluation of the
  hadronic vacuum polarisation contributions to the Standard Model predictions
  of the muon $g-2$ and ${\alpha (m_Z^2)}$ using newest hadronic cross-section
  data}}, \href{https://doi.org/10.1140/epjc/s10052-017-5161-6}{\emph{Eur.
  Phys. J. C} {\bfseries 77} (2017) 827}
  [\href{https://arxiv.org/abs/1706.09436}{{\ttfamily 1706.09436}}].

\bibitem{Keshavarzi:2018mgv}
A.~Keshavarzi, D.~Nomura and T.~Teubner, \emph{{Muon $g-2$ and $\alpha(M_Z^2)$:
  a new data-based analysis}},
  \href{https://doi.org/10.1103/PhysRevD.97.114025}{\emph{Phys. Rev. D}
  {\bfseries 97} (2018) 114025}
  [\href{https://arxiv.org/abs/1802.02995}{{\ttfamily 1802.02995}}].

\bibitem{Colangelo:2018mtw}
G.~Colangelo, M.~Hoferichter and P.~Stoffer, \emph{{Two-pion contribution to
  hadronic vacuum polarization}},
  \href{https://doi.org/10.1007/JHEP02(2019)006}{\emph{JHEP} {\bfseries 02}
  (2019) 006} [\href{https://arxiv.org/abs/1810.00007}{{\ttfamily
  1810.00007}}].

\bibitem{Hoferichter:2019gzf}
M.~Hoferichter, B.-L.~Hoid and B.~Kubis, \emph{{Three-pion contribution to
  hadronic vacuum polarization}},
  \href{https://doi.org/10.1007/JHEP08(2019)137}{\emph{JHEP} {\bfseries 08}
  (2019) 137} [\href{https://arxiv.org/abs/1907.01556}{{\ttfamily
  1907.01556}}].

\bibitem{Davier:2019can}
M.~Davier, A.~Hoecker, B.~Malaescu and Z.~Zhang, \emph{{A new evaluation of the
  hadronic vacuum polarisation contributions to the muon anomalous magnetic
  moment and to $\mathbf{\boldsymbol\alpha(m_Z^2)}$}},
  \href{https://doi.org/10.1140/epjc/s10052-020-7792-2}{\emph{Eur. Phys. J. C}
  {\bfseries 80} (2020) 241}
  [\href{https://arxiv.org/abs/1908.00921}{{\ttfamily 1908.00921}}].

\bibitem{Keshavarzi:2019abf}
A.~Keshavarzi, D.~Nomura and T.~Teubner, \emph{{$g-2$ of charged leptons,
  $\alpha (M^2_Z)$ , and the hyperfine splitting of muonium}},
  \href{https://doi.org/10.1103/PhysRevD.101.014029}{\emph{Phys. Rev. D}
  {\bfseries 101} (2020) 014029}
  [\href{https://arxiv.org/abs/1911.00367}{{\ttfamily 1911.00367}}].

\bibitem{Kurz:2014wya}
A.~Kurz, T.~Liu, P.~Marquard and M.~Steinhauser, \emph{{Hadronic contribution
  to the muon anomalous magnetic moment to next-to-next-to-leading order}},
  \href{https://doi.org/10.1016/j.physletb.2014.05.043}{\emph{Phys. Lett. B}
  {\bfseries 734} (2014) 144}
  [\href{https://arxiv.org/abs/1403.6400}{{\ttfamily 1403.6400}}].

\bibitem{Melnikov:2003xd}
K.~Melnikov and A.~Vainshtein, \emph{{Hadronic light-by-light scattering
  contribution to the muon anomalous magnetic moment revisited}},
  \href{https://doi.org/10.1103/PhysRevD.70.113006}{\emph{Phys. Rev. D}
  {\bfseries 70} (2004) 113006}
  [\href{https://arxiv.org/abs/hep-ph/0312226}{{\ttfamily hep-ph/0312226}}].

\bibitem{Masjuan:2017tvw}
P.~Masjuan and P.~Sanchez-Puertas, \emph{{Pseudoscalar-pole contribution to the
  $(g_{\mu}-2)$: a rational approach}},
  \href{https://doi.org/10.1103/PhysRevD.95.054026}{\emph{Phys. Rev. D}
  {\bfseries 95} (2017) 054026}
  [\href{https://arxiv.org/abs/1701.05829}{{\ttfamily 1701.05829}}].

\bibitem{Colangelo:2017fiz}
G.~Colangelo, M.~Hoferichter, M.~Procura and P.~Stoffer, \emph{{Dispersion
  relation for hadronic light-by-light scattering: two-pion contributions}},
  \href{https://doi.org/10.1007/JHEP04(2017)161}{\emph{JHEP} {\bfseries 04}
  (2017) 161} [\href{https://arxiv.org/abs/1702.07347}{{\ttfamily
  1702.07347}}].

\bibitem{Hoferichter:2018kwz}
M.~Hoferichter, B.-L.~Hoid, B.~Kubis, S.~Leupold and S.P.~Schneider,
  \emph{{Dispersion relation for hadronic light-by-light scattering: pion
  pole}}, \href{https://doi.org/10.1007/JHEP10(2018)141}{\emph{JHEP} {\bfseries
  10} (2018) 141} [\href{https://arxiv.org/abs/1808.04823}{{\ttfamily
  1808.04823}}].

\bibitem{Gerardin:2019vio}
A.~G\'erardin, H.B.~Meyer and A.~Nyffeler, \emph{{Lattice calculation of the
  pion transition form factor with $N_f=2+1$ Wilson quarks}},
  \href{https://doi.org/10.1103/PhysRevD.100.034520}{\emph{Phys. Rev. D}
  {\bfseries 100} (2019) 034520}
  [\href{https://arxiv.org/abs/1903.09471}{{\ttfamily 1903.09471}}].

\bibitem{Bijnens:2019ghy}
J.~Bijnens, N.~Hermansson-Truedsson and A.~Rodr\'\i{}guez-S\'anchez,
  \emph{{Short-distance constraints for the HLbL contribution to the muon
  anomalous magnetic moment}},
  \href{https://doi.org/10.1016/j.physletb.2019.134994}{\emph{Phys. Lett. B}
  {\bfseries 798} (2019) 134994}
  [\href{https://arxiv.org/abs/1908.03331}{{\ttfamily 1908.03331}}].

\bibitem{Colangelo:2019uex}
G.~Colangelo, F.~Hagelstein, M.~Hoferichter, L.~Laub and P.~Stoffer,
  \emph{{Longitudinal short-distance constraints for the hadronic
  light-by-light contribution to $(g-2)_\mu$ with large-$N_c$ Regge models}},
  \href{https://doi.org/10.1007/JHEP03(2020)101}{\emph{JHEP} {\bfseries 03}
  (2020) 101} [\href{https://arxiv.org/abs/1910.13432}{{\ttfamily
  1910.13432}}].

\bibitem{Blum:2019ugy}
T.~Blum, N.~Christ, M.~Hayakawa, T.~Izubuchi, L.~Jin, C.~Jung et~al.,
  \emph{{Hadronic Light-by-Light Scattering Contribution to the Muon Anomalous
  Magnetic Moment from Lattice QCD}},
  \href{https://doi.org/10.1103/PhysRevLett.124.132002}{\emph{Phys. Rev. Lett.}
  {\bfseries 124} (2020) 132002}
  [\href{https://arxiv.org/abs/1911.08123}{{\ttfamily 1911.08123}}].

\bibitem{Colangelo:2014qya}
G.~Colangelo, M.~Hoferichter, A.~Nyffeler, M.~Passera and P.~Stoffer,
  \emph{{Remarks on higher-order hadronic corrections to the muon
  g\ensuremath{-}2}},
  \href{https://doi.org/10.1016/j.physletb.2014.06.012}{\emph{Phys. Lett. B}
  {\bfseries 735} (2014) 90} [\href{https://arxiv.org/abs/1403.7512}{{\ttfamily
  1403.7512}}].

\bibitem{Borsanyi:2020mff}
S.~Borsanyi et~al., \emph{{Leading hadronic contribution to the muon magnetic
  moment from lattice QCD}},
  \href{https://doi.org/10.1038/s41586-021-03418-1}{\emph{Nature} {\bfseries
  593} (2021) 51} [\href{https://arxiv.org/abs/2002.12347}{{\ttfamily
  2002.12347}}].

\bibitem{Passera:2008jk}
M.~Passera, W.J.~Marciano and A.~Sirlin, \emph{{The Muon g-2 and the bounds on
  the Higgs boson mass}},
  \href{https://doi.org/10.1103/PhysRevD.78.013009}{\emph{Phys. Rev. D}
  {\bfseries 78} (2008) 013009}
  [\href{https://arxiv.org/abs/0804.1142}{{\ttfamily 0804.1142}}].

\bibitem{Haller:2018nnx}
J.~Haller, A.~Hoecker, R.~Kogler, K.~M\"onig, T.~Peiffer and J.~Stelzer,
  \emph{{Update of the global electroweak fit and constraints on
  two-Higgs-doublet models}},
  \href{https://doi.org/10.1140/epjc/s10052-018-6131-3}{\emph{Eur. Phys. J. C}
  {\bfseries 78} (2018) 675}
  [\href{https://arxiv.org/abs/1803.01853}{{\ttfamily 1803.01853}}].

\bibitem{Crivellin:2020zul}
A.~Crivellin, M.~Hoferichter, C.A.~Manzari and M.~Montull, \emph{{Hadronic
  Vacuum Polarization: $(g-2)_\mu$ versus Global Electroweak Fits}},
  \href{https://doi.org/10.1103/PhysRevLett.125.091801}{\emph{Phys. Rev. Lett.}
  {\bfseries 125} (2020) 091801}
  [\href{https://arxiv.org/abs/2003.04886}{{\ttfamily 2003.04886}}].

\bibitem{Keshavarzi:2020bfy}
A.~Keshavarzi, W.J.~Marciano, M.~Passera and A.~Sirlin, \emph{{Muon $g-2$ and
  $\Delta \alpha$ connection}},
  \href{https://doi.org/10.1103/PhysRevD.102.033002}{\emph{Phys. Rev. D}
  {\bfseries 102} (2020) 033002}
  [\href{https://arxiv.org/abs/2006.12666}{{\ttfamily 2006.12666}}].

\bibitem{Pati:1974yy}
J.C.~Pati and A.~Salam, \emph{{Lepton Number as the Fourth Color}},
  \href{https://doi.org/10.1103/PhysRevD.10.275}{\emph{Phys. Rev. D} {\bfseries
  10} (1974) 275}.

\bibitem{Georgi:1974sy}
H.~Georgi and S.L.~Glashow, \emph{{Unity of All Elementary Particle Forces}},
  \href{https://doi.org/10.1103/PhysRevLett.32.438}{\emph{Phys. Rev. Lett.}
  {\bfseries 32} (1974) 438}.

\bibitem{Dimopoulos:1980hn}
S.~Dimopoulos, S.~Raby and L.~Susskind, \emph{{Light Composite Fermions}},
  \href{https://doi.org/10.1016/0550-3213(80)90215-1}{\emph{Nucl. Phys. B}
  {\bfseries 173} (1980) 208}.

\bibitem{Senjanovic:1982ex}
G.~Senjanovic and A.~Sokorac, \emph{{Light Leptoquarks in SO(10)}},
  \href{https://doi.org/10.1007/BF01574858}{\emph{Z. Phys. C} {\bfseries 20}
  (1983) 255}.

\bibitem{Frampton:1989fu}
P.H.~Frampton and B.-H.~Lee, \emph{{SU(15) GRAND UNIFICATION}},
  \href{https://doi.org/10.1103/PhysRevLett.64.619}{\emph{Phys. Rev. Lett.}
  {\bfseries 64} (1990) 619}.

\bibitem{Witten:1985xc}
E.~Witten, \emph{{Symmetry Breaking Patterns in Superstring Models}},
  \href{https://doi.org/10.1016/0550-3213(85)90603-0}{\emph{Nucl. Phys. B}
  {\bfseries 258} (1985) 75}.

\bibitem{Alonso:2015sja}
R.~Alonso, B.~Grinstein and J.~Martin~Camalich, \emph{{Lepton universality
  violation and lepton flavor conservation in $B$-meson decays}},
  \href{https://doi.org/10.1007/JHEP10(2015)184}{\emph{JHEP} {\bfseries 10}
  (2015) 184} [\href{https://arxiv.org/abs/1505.05164}{{\ttfamily
  1505.05164}}].

\bibitem{Calibbi:2015kma}
L.~Calibbi, A.~Crivellin and T.~Ota, \emph{{Effective Field Theory Approach to
  $b\to s\ell\ell^{(')}$, $B\to K^{(*)}\nu\overline{\nu}$ and $B\to
  D^{(*)}\tau\nu$ with Third Generation Couplings}},
  \href{https://doi.org/10.1103/PhysRevLett.115.181801}{\emph{Phys. Rev. Lett.}
  {\bfseries 115} (2015) 181801}
  [\href{https://arxiv.org/abs/1506.02661}{{\ttfamily 1506.02661}}].

\bibitem{Hiller:2016kry}
G.~Hiller, D.~Loose and K.~Sch\"onwald, \emph{{Leptoquark Flavor Patterns \& B
  Decay Anomalies}}, \href{https://doi.org/10.1007/JHEP12(2016)027}{\emph{JHEP}
  {\bfseries 12} (2016) 027}
  [\href{https://arxiv.org/abs/1609.08895}{{\ttfamily 1609.08895}}].

\bibitem{Bhattacharya:2016mcc}
B.~Bhattacharya, A.~Datta, J.-P.~Gu\'evin, D.~London and R.~Watanabe,
  \emph{{Simultaneous Explanation of the $R_K$ and $R_{D^{(*)}}$ Puzzles: a
  Model Analysis}}, \href{https://doi.org/10.1007/JHEP01(2017)015}{\emph{JHEP}
  {\bfseries 01} (2017) 015}
  [\href{https://arxiv.org/abs/1609.09078}{{\ttfamily 1609.09078}}].

\bibitem{Buttazzo:2017ixm}
D.~Buttazzo, A.~Greljo, G.~Isidori and D.~Marzocca, \emph{{B-physics anomalies:
  a guide to combined explanations}},
  \href{https://doi.org/10.1007/JHEP11(2017)044}{\emph{JHEP} {\bfseries 11}
  (2017) 044} [\href{https://arxiv.org/abs/1706.07808}{{\ttfamily
  1706.07808}}].

\bibitem{Barbieri:2015yvd}
R.~Barbieri, G.~Isidori, A.~Pattori and F.~Senia, \emph{{Anomalies in
  $B$-decays and $U(2)$ flavour symmetry}},
  \href{https://doi.org/10.1140/epjc/s10052-016-3905-3}{\emph{Eur. Phys. J. C}
  {\bfseries 76} (2016) 67} [\href{https://arxiv.org/abs/1512.01560}{{\ttfamily
  1512.01560}}].

\bibitem{Barbieri:2016las}
R.~Barbieri, C.W.~Murphy and F.~Senia, \emph{{B-decay Anomalies in a Composite
  Leptoquark Model}},
  \href{https://doi.org/10.1140/epjc/s10052-016-4578-7}{\emph{Eur. Phys. J. C}
  {\bfseries 77} (2017) 8} [\href{https://arxiv.org/abs/1611.04930}{{\ttfamily
  1611.04930}}].

\bibitem{Calibbi:2017qbu}
L.~Calibbi, A.~Crivellin and T.~Li, \emph{{Model of vector leptoquarks in view
  of the $B$-physics anomalies}},
  \href{https://doi.org/10.1103/PhysRevD.98.115002}{\emph{Phys. Rev. D}
  {\bfseries 98} (2018) 115002}
  [\href{https://arxiv.org/abs/1709.00692}{{\ttfamily 1709.00692}}].

\bibitem{Crivellin:2017dsk}
A.~Crivellin, D.~M\"uller, A.~Signer and Y.~Ulrich, \emph{{Correlating lepton
  flavor universality violation in $B$ decays with $\mu\to e\gamma$ using
  leptoquarks}}, \href{https://doi.org/10.1103/PhysRevD.97.015019}{\emph{Phys.
  Rev. D} {\bfseries 97} (2018) 015019}
  [\href{https://arxiv.org/abs/1706.08511}{{\ttfamily 1706.08511}}].

\bibitem{Bordone:2018nbg}
M.~Bordone, C.~Cornella, J.~Fuentes-Mart\'\i{}n and G.~Isidori,
  \emph{{Low-energy signatures of the $\mathrm{PS}^3$ model: from $B$-physics
  anomalies to LFV}},
  \href{https://doi.org/10.1007/JHEP10(2018)148}{\emph{JHEP} {\bfseries 10}
  (2018) 148} [\href{https://arxiv.org/abs/1805.09328}{{\ttfamily
  1805.09328}}].

\bibitem{Kumar:2018kmr}
J.~Kumar, D.~London and R.~Watanabe, \emph{{Combined Explanations of the $b \to
  s \mu^+ \mu^-$ and $b \to c \tau^- {\bar\nu}$ Anomalies: a General Model
  Analysis}}, \href{https://doi.org/10.1103/PhysRevD.99.015007}{\emph{Phys.
  Rev. D} {\bfseries 99} (2019) 015007}
  [\href{https://arxiv.org/abs/1806.07403}{{\ttfamily 1806.07403}}].

\bibitem{Crivellin:2018yvo}
A.~Crivellin, C.~Greub, D.~M\"uller and F.~Saturnino, \emph{{Importance of Loop
  Effects in Explaining the Accumulated Evidence for New Physics in B Decays
  with a Vector Leptoquark}},
  \href{https://doi.org/10.1103/PhysRevLett.122.011805}{\emph{Phys. Rev. Lett.}
  {\bfseries 122} (2019) 011805}
  [\href{https://arxiv.org/abs/1807.02068}{{\ttfamily 1807.02068}}].

\bibitem{Crivellin:2019szf}
A.~Crivellin and F.~Saturnino, \emph{{Explaining the Flavor Anomalies with a
  Vector Leptoquark (Moriond 2019 update)}},
  \href{https://doi.org/10.22323/1.352.0163}{\emph{PoS} {\bfseries DIS2019}
  (2019) 163} [\href{https://arxiv.org/abs/1906.01222}{{\ttfamily
  1906.01222}}].

\bibitem{Cornella:2019hct}
C.~Cornella, J.~Fuentes-Martin and G.~Isidori, \emph{{Revisiting the vector
  leptoquark explanation of the B-physics anomalies}},
  \href{https://doi.org/10.1007/JHEP07(2019)168}{\emph{JHEP} {\bfseries 07}
  (2019) 168} [\href{https://arxiv.org/abs/1903.11517}{{\ttfamily
  1903.11517}}].

\bibitem{Bordone:2019uzc}
M.~Bordone, O.~Cat\`a and T.~Feldmann, \emph{{Effective Theory Approach to New
  Physics with Flavour: General Framework and a Leptoquark Example}},
  \href{https://doi.org/10.1007/JHEP01(2020)067}{\emph{JHEP} {\bfseries 01}
  (2020) 067} [\href{https://arxiv.org/abs/1910.02641}{{\ttfamily
  1910.02641}}].

\bibitem{Bernigaud:2019bfy}
J.~Bernigaud, I.~de~Medeiros~Varzielas and J.~Talbert, \emph{{Finite Family
  Groups for Fermionic and Leptoquark Mixing Patterns}},
  \href{https://doi.org/10.1007/JHEP01(2020)194}{\emph{JHEP} {\bfseries 01}
  (2020) 194} [\href{https://arxiv.org/abs/1906.11270}{{\ttfamily
  1906.11270}}].

\bibitem{Aebischer:2018acj}
J.~Aebischer, A.~Crivellin and C.~Greub, \emph{{QCD improved matching for
  semileptonic B decays with leptoquarks}},
  \href{https://doi.org/10.1103/PhysRevD.99.055002}{\emph{Phys. Rev. D}
  {\bfseries 99} (2019) 055002}
  [\href{https://arxiv.org/abs/1811.08907}{{\ttfamily 1811.08907}}].

\bibitem{Fuentes-Martin:2019ign}
J.~Fuentes-Mart\'\i{}n, G.~Isidori, M.~K\"onig and N.~Selimovi\'c,
  \emph{{Vector Leptoquarks Beyond Tree Level}},
  \href{https://doi.org/10.1103/PhysRevD.101.035024}{\emph{Phys. Rev. D}
  {\bfseries 101} (2020) 035024}
  [\href{https://arxiv.org/abs/1910.13474}{{\ttfamily 1910.13474}}].

\bibitem{Popov:2019tyc}
O.~Popov, M.A.~Schmidt and G.~White, \emph{{$R_2$ as a single leptoquark
  solution to $R_{D^{(*)}}$ and $R_{K^{(*)}}$}},
  \href{https://doi.org/10.1103/PhysRevD.100.035028}{\emph{Phys. Rev. D}
  {\bfseries 100} (2019) 035028}
  [\href{https://arxiv.org/abs/1905.06339}{{\ttfamily 1905.06339}}].

\bibitem{Fajfer:2015ycq}
S.~Fajfer and N.~Ko\v{s}nik, \emph{{Vector leptoquark resolution of $R_K$ and
  $R_{D^{(*)}}$ puzzles}},
  \href{https://doi.org/10.1016/j.physletb.2016.02.018}{\emph{Phys. Lett. B}
  {\bfseries 755} (2016) 270}
  [\href{https://arxiv.org/abs/1511.06024}{{\ttfamily 1511.06024}}].

\bibitem{Blanke:2018sro}
M.~Blanke and A.~Crivellin, \emph{{$B$ Meson Anomalies in a Pati-Salam Model
  within the Randall-Sundrum Background}},
  \href{https://doi.org/10.1103/PhysRevLett.121.011801}{\emph{Phys. Rev. Lett.}
  {\bfseries 121} (2018) 011801}
  [\href{https://arxiv.org/abs/1801.07256}{{\ttfamily 1801.07256}}].

\bibitem{deMedeirosVarzielas:2019lgb}
I.~de~Medeiros~Varzielas and J.~Talbert, \emph{{Simplified Models of Flavourful
  Leptoquarks}},
  \href{https://doi.org/10.1140/epjc/s10052-019-7047-2}{\emph{Eur. Phys. J. C}
  {\bfseries 79} (2019) 536}
  [\href{https://arxiv.org/abs/1901.10484}{{\ttfamily 1901.10484}}].

\bibitem{deMedeirosVarzielas:2015yxm}
I.~de~Medeiros~Varzielas and G.~Hiller, \emph{{Clues for flavor from rare
  lepton and quark decays}},
  \href{https://doi.org/10.1007/JHEP06(2015)072}{\emph{JHEP} {\bfseries 06}
  (2015) 072} [\href{https://arxiv.org/abs/1503.01084}{{\ttfamily
  1503.01084}}].

\bibitem{Crivellin:2019dwb}
A.~Crivellin, D.~M\"uller and F.~Saturnino, \emph{{Flavor Phenomenology of the
  Leptoquark Singlet-Triplet Model}},
  \href{https://doi.org/10.1007/JHEP06(2020)020}{\emph{JHEP} {\bfseries 06}
  (2020) 020} [\href{https://arxiv.org/abs/1912.04224}{{\ttfamily
  1912.04224}}].

\bibitem{Saad:2020ihm}
S.~Saad, \emph{{Combined explanations of $(g-2)_{\mu}$, $R_{D^{(*)}}$,
  $R_{K^{(*)}}$ anomalies in a two-loop radiative neutrino mass model}},
  \href{https://doi.org/10.1103/PhysRevD.102.015019}{\emph{Phys. Rev. D}
  {\bfseries 102} (2020) 015019}
  [\href{https://arxiv.org/abs/2005.04352}{{\ttfamily 2005.04352}}].

\bibitem{Saad:2020ucl}
S.~Saad and A.~Thapa, \emph{{Common origin of neutrino masses and
  $R_{D^{(\ast)}}$, $R_{K^{(\ast)}}$ anomalies}},
  \href{https://doi.org/10.1103/PhysRevD.102.015014}{\emph{Phys. Rev. D}
  {\bfseries 102} (2020) 015014}
  [\href{https://arxiv.org/abs/2004.07880}{{\ttfamily 2004.07880}}].

\bibitem{Gherardi:2020qhc}
V.~Gherardi, D.~Marzocca and E.~Venturini, \emph{{Low-energy phenomenology of
  scalar leptoquarks at one-loop accuracy}},
  \href{https://doi.org/10.1007/JHEP01(2021)138}{\emph{JHEP} {\bfseries 01}
  (2021) 138} [\href{https://arxiv.org/abs/2008.09548}{{\ttfamily
  2008.09548}}].

\bibitem{DaRold:2020bib}
L.~Da~Rold and F.~Lamagna, \emph{{Model for the singlet-triplet leptoquarks}},
  \href{https://doi.org/10.1103/PhysRevD.103.115007}{\emph{Phys. Rev. D}
  {\bfseries 103} (2021) 115007}
  [\href{https://arxiv.org/abs/2011.10061}{{\ttfamily 2011.10061}}].

\bibitem{Greljo:2021xmg}
A.~Greljo, P.~Stangl and A.E.~Thomsen, \emph{{A model of muon anomalies}},
  \href{https://doi.org/10.1016/j.physletb.2021.136554}{\emph{Phys. Lett. B}
  {\bfseries 820} (2021) 136554}
  [\href{https://arxiv.org/abs/2103.13991}{{\ttfamily 2103.13991}}].

\bibitem{Bordone:2017bld}
M.~Bordone, C.~Cornella, J.~Fuentes-Martin and G.~Isidori, \emph{{A three-site
  gauge model for flavor hierarchies and flavor anomalies}},
  \href{https://doi.org/10.1016/j.physletb.2018.02.011}{\emph{Phys. Lett. B}
  {\bfseries 779} (2018) 317}
  [\href{https://arxiv.org/abs/1712.01368}{{\ttfamily 1712.01368}}].

\bibitem{Biswas:2018snp}
A.~Biswas, D.~Kumar~Ghosh, N.~Ghosh, A.~Shaw and A.K.~Swain, \emph{{Collider
  signature of $U_1$ Leptoquark and constraints from $b \to c$ observables}},
  \href{https://doi.org/10.1088/1361-6471/ab6948}{\emph{J. Phys. G} {\bfseries
  47} (2020) 045005} [\href{https://arxiv.org/abs/1808.04169}{{\ttfamily
  1808.04169}}].

\bibitem{Heeck:2018ntp}
J.~Heeck and D.~Teresi, \emph{{Pati-Salam explanations of the B-meson
  anomalies}}, \href{https://doi.org/10.1007/JHEP12(2018)103}{\emph{JHEP}
  {\bfseries 12} (2018) 103}
  [\href{https://arxiv.org/abs/1808.07492}{{\ttfamily 1808.07492}}].

\bibitem{Sahoo:2015wya}
S.~Sahoo and R.~Mohanta, \emph{{Scalar leptoquarks and the rare $B$ meson
  decays}}, \href{https://doi.org/10.1103/PhysRevD.91.094019}{\emph{Phys. Rev.
  D} {\bfseries 91} (2015) 094019}
  [\href{https://arxiv.org/abs/1501.05193}{{\ttfamily 1501.05193}}].

\bibitem{Chen:2016dip}
C.-H.~Chen, T.~Nomura and H.~Okada, \emph{{Explanation of $B \to K^{(*)} \ell^+
  \ell^-$ and muon $g-2$, and implications at the LHC}},
  \href{https://doi.org/10.1103/PhysRevD.94.115005}{\emph{Phys. Rev. D}
  {\bfseries 94} (2016) 115005}
  [\href{https://arxiv.org/abs/1607.04857}{{\ttfamily 1607.04857}}].

\bibitem{Dey:2017ede}
U.K.~Dey, D.~Kar, M.~Mitra, M.~Spannowsky and A.C.~Vincent, \emph{{Searching
  for Leptoquarks at IceCube and the LHC}},
  \href{https://doi.org/10.1103/PhysRevD.98.035014}{\emph{Phys. Rev. D}
  {\bfseries 98} (2018) 035014}
  [\href{https://arxiv.org/abs/1709.02009}{{\ttfamily 1709.02009}}].

\bibitem{Becirevic:2017jtw}
D.~Be\v{c}irevi\'c and O.~Sumensari, \emph{{A leptoquark model to accommodate
  $R_K^\mathrm{exp} < R_K^\mathrm{SM}$ and $R_{K^\ast}^\mathrm{exp} <
  R_{K^\ast}^\mathrm{SM}$}},
  \href{https://doi.org/10.1007/JHEP08(2017)104}{\emph{JHEP} {\bfseries 08}
  (2017) 104} [\href{https://arxiv.org/abs/1704.05835}{{\ttfamily
  1704.05835}}].

\bibitem{Chauhan:2017ndd}
B.~Chauhan, B.~Kindra and A.~Narang, \emph{{Discrepancies in simultaneous
  explanation of flavor anomalies and IceCube PeV events using leptoquarks}},
  \href{https://doi.org/10.1103/PhysRevD.97.095007}{\emph{Phys. Rev. D}
  {\bfseries 97} (2018) 095007}
  [\href{https://arxiv.org/abs/1706.04598}{{\ttfamily 1706.04598}}].

\bibitem{Becirevic:2018afm}
D.~Be\v{c}irevi\'c, I.~Dor\v{s}ner, S.~Fajfer, N.~Ko\v{s}nik, D.A.~Faroughy and
  O.~Sumensari, \emph{{Scalar leptoquarks from grand unified theories to
  accommodate the $B$-physics anomalies}},
  \href{https://doi.org/10.1103/PhysRevD.98.055003}{\emph{Phys. Rev. D}
  {\bfseries 98} (2018) 055003}
  [\href{https://arxiv.org/abs/1806.05689}{{\ttfamily 1806.05689}}].

\bibitem{Fajfer:2012jt}
S.~Fajfer, J.F.~Kamenik, I.~Nisandzic and J.~Zupan, \emph{{Implications of
  Lepton Flavor Universality Violations in B Decays}},
  \href{https://doi.org/10.1103/PhysRevLett.109.161801}{\emph{Phys. Rev. Lett.}
  {\bfseries 109} (2012) 161801}
  [\href{https://arxiv.org/abs/1206.1872}{{\ttfamily 1206.1872}}].

\bibitem{Deshpande:2012rr}
N.G.~Deshpande and A.~Menon, \emph{{Hints of R-parity violation in B decays
  into $\tau \nu$}}, \href{https://doi.org/10.1007/JHEP01(2013)025}{\emph{JHEP}
  {\bfseries 01} (2013) 025} [\href{https://arxiv.org/abs/1208.4134}{{\ttfamily
  1208.4134}}].

\bibitem{Freytsis:2015qca}
M.~Freytsis, Z.~Ligeti and J.T.~Ruderman, \emph{{Flavor models for $\bar{B} \to
  D^{(*)} \tau \bar{\nu}$}},
  \href{https://doi.org/10.1103/PhysRevD.92.054018}{\emph{Phys. Rev. D}
  {\bfseries 92} (2015) 054018}
  [\href{https://arxiv.org/abs/1506.08896}{{\ttfamily 1506.08896}}].

\bibitem{Bauer:2015knc}
M.~Bauer and M.~Neubert, \emph{{Minimal Leptoquark Explanation for the
  $R_{D^{(*)}}$ , $R_K$ , and $(g-2)_\mu$ Anomalies}},
  \href{https://doi.org/10.1103/PhysRevLett.116.141802}{\emph{Phys. Rev. Lett.}
  {\bfseries 116} (2016) 141802}
  [\href{https://arxiv.org/abs/1511.01900}{{\ttfamily 1511.01900}}].

\bibitem{Li:2016vvp}
X.-Q.~Li, Y.-D.~Yang and X.~Zhang, \emph{{Revisiting the one leptoquark
  solution to the R(D$^{(\ast)}$) anomalies and its phenomenological
  implications}}, \href{https://doi.org/10.1007/JHEP08(2016)054}{\emph{JHEP}
  {\bfseries 08} (2016) 054}
  [\href{https://arxiv.org/abs/1605.09308}{{\ttfamily 1605.09308}}].

\bibitem{Zhu:2016xdg}
J.~Zhu, H.-M.~Gan, R.-M.~Wang, Y.-Y.~Fan, Q.~Chang and Y.-G.~Xu, \emph{{Probing
  the R-parity violating supersymmetric effects in the exclusive $b \to
  c\ell^-\bar{\nu}_\ell$ decays}},
  \href{https://doi.org/10.1103/PhysRevD.93.094023}{\emph{Phys. Rev. D}
  {\bfseries 93} (2016) 094023}
  [\href{https://arxiv.org/abs/1602.06491}{{\ttfamily 1602.06491}}].

\bibitem{Popov:2016fzr}
O.~Popov and G.A.~White, \emph{{One Leptoquark to unify them? Neutrino masses
  and unification in the light of $(g-2)_\mu$, $R_{D^{(\star)}}$ and $R_K$
  anomalies}},
  \href{https://doi.org/10.1016/j.nuclphysb.2017.08.007}{\emph{Nucl. Phys. B}
  {\bfseries 923} (2017) 324}
  [\href{https://arxiv.org/abs/1611.04566}{{\ttfamily 1611.04566}}].

\bibitem{Deshpand:2016cpw}
N.G.~Deshpande and X.-G.~He, \emph{{Consequences of R-parity violating
  interactions for anomalies in $\bar B\to D^{(*)} \tau \bar \nu$ and $b\to s
  \mu^+\mu^-$}},
  \href{https://doi.org/10.1140/epjc/s10052-017-4707-y}{\emph{Eur. Phys. J. C}
  {\bfseries 77} (2017) 134}
  [\href{https://arxiv.org/abs/1608.04817}{{\ttfamily 1608.04817}}].

\bibitem{Becirevic:2016oho}
D.~Be\v{c}irevi\'c, N.~Ko\v{s}nik, O.~Sumensari and R.~Zukanovich~Funchal,
  \emph{{Palatable Leptoquark Scenarios for Lepton Flavor Violation in
  Exclusive $b\to s\ell_1\ell_2$ modes}},
  \href{https://doi.org/10.1007/JHEP11(2016)035}{\emph{JHEP} {\bfseries 11}
  (2016) 035} [\href{https://arxiv.org/abs/1608.07583}{{\ttfamily
  1608.07583}}].

\bibitem{Cai:2017wry}
Y.~Cai, J.~Gargalionis, M.A.~Schmidt and R.R.~Volkas, \emph{{Reconsidering the
  One Leptoquark solution: flavor anomalies and neutrino mass}},
  \href{https://doi.org/10.1007/JHEP10(2017)047}{\emph{JHEP} {\bfseries 10}
  (2017) 047} [\href{https://arxiv.org/abs/1704.05849}{{\ttfamily
  1704.05849}}].

\bibitem{Altmannshofer:2017poe}
W.~Altmannshofer, P.S.~Bhupal~Dev and A.~Soni, \emph{{$R_{D^{(*)}}$ anomaly: A
  possible hint for natural supersymmetry with $R$-parity violation}},
  \href{https://doi.org/10.1103/PhysRevD.96.095010}{\emph{Phys. Rev. D}
  {\bfseries 96} (2017) 095010}
  [\href{https://arxiv.org/abs/1704.06659}{{\ttfamily 1704.06659}}].

\bibitem{Kamali:2018fhr}
S.~Kamali, A.~Rashed and A.~Datta, \emph{{New physics in inclusive $B \to
  X_c\ell \bar{\nu}$ decay in light of $R(D^{(*)})$ measurements}},
  \href{https://doi.org/10.1103/PhysRevD.97.095034}{\emph{Phys. Rev. D}
  {\bfseries 97} (2018) 095034}
  [\href{https://arxiv.org/abs/1801.08259}{{\ttfamily 1801.08259}}].

\bibitem{Mandal:2018kau}
T.~Mandal, S.~Mitra and S.~Raz, \emph{{$R_{D^{(*)}}$ motivated $\mathcal{S}_1$
  leptoquark scenarios: Impact of interference on the exclusion limits from LHC
  data}}, \href{https://doi.org/10.1103/PhysRevD.99.055028}{\emph{Phys. Rev. D}
  {\bfseries 99} (2019) 055028}
  [\href{https://arxiv.org/abs/1811.03561}{{\ttfamily 1811.03561}}].

\bibitem{Azatov:2018knx}
A.~Azatov, D.~Bardhan, D.~Ghosh, F.~Sgarlata and E.~Venturini, \emph{{Anatomy
  of $b \to c \tau \nu$ anomalies}},
  \href{https://doi.org/10.1007/JHEP11(2018)187}{\emph{JHEP} {\bfseries 11}
  (2018) 187} [\href{https://arxiv.org/abs/1805.03209}{{\ttfamily
  1805.03209}}].

\bibitem{Wei:2018vmk}
J.~Zhu, B.~Wei, J.-H.~Sheng, R.-M.~Wang, Y.~Gao and G.-R.~Lu, \emph{{Probing
  the R-parity violating supersymmetric effects in $B_c\to
  J/\psi\ell^-\bar{\nu}_{\ell},\eta_c\ell^-\bar{\nu}_{\ell}$ and
  $\Lambda_b\to\Lambda_c\ell^-\bar{\nu}_{\ell}$ decays}},
  \href{https://doi.org/10.1016/j.nuclphysb.2018.07.011}{\emph{Nucl. Phys. B}
  {\bfseries 934} (2018) 380}
  [\href{https://arxiv.org/abs/1801.00917}{{\ttfamily 1801.00917}}].

\bibitem{Angelescu:2018tyl}
A.~Angelescu, D.~Be\v{c}irevi\'c, D.A.~Faroughy and O.~Sumensari,
  \emph{{Closing the window on single leptoquark solutions to the $B$-physics
  anomalies}}, \href{https://doi.org/10.1007/JHEP10(2018)183}{\emph{JHEP}
  {\bfseries 10} (2018) 183}
  [\href{https://arxiv.org/abs/1808.08179}{{\ttfamily 1808.08179}}].

\bibitem{Kim:2018oih}
T.J.~Kim, P.~Ko, J.~Li, J.~Park and P.~Wu, \emph{{Correlation between $
  {R}_{D^{\left(\ast \right)}} $ and top quark FCNC decays in leptoquark
  models}}, \href{https://doi.org/10.1007/JHEP07(2019)025}{\emph{JHEP}
  {\bfseries 07} (2019) 025}
  [\href{https://arxiv.org/abs/1812.08484}{{\ttfamily 1812.08484}}].

\bibitem{Aydemir:2019ynb}
U.~Aydemir, T.~Mandal and S.~Mitra, \emph{{Addressing the ${\mathbf
  R_{D^{(*)}}}$ anomalies with an ${\mathbf S_1}$ leptoquark from
  $\mathbf{SO(10)}$ grand unification}},
  \href{https://doi.org/10.1103/PhysRevD.101.015011}{\emph{Phys. Rev. D}
  {\bfseries 101} (2020) 015011}
  [\href{https://arxiv.org/abs/1902.08108}{{\ttfamily 1902.08108}}].

\bibitem{Crivellin:2019qnh}
A.~Crivellin and F.~Saturnino, \emph{{Correlating tauonic $B$ decays with the
  neutron electric dipole moment via a scalar leptoquark}},
  \href{https://doi.org/10.1103/PhysRevD.100.115014}{\emph{Phys. Rev. D}
  {\bfseries 100} (2019) 115014}
  [\href{https://arxiv.org/abs/1905.08257}{{\ttfamily 1905.08257}}].

\bibitem{Yan:2019hpm}
H.~Yan, Y.-D.~Yang and X.-B.~Yuan, \emph{{Phenomenology of $b\to c\tau\bar\nu$
  decays in a scalar leptoquark model}},
  \href{https://doi.org/10.1088/1674-1137/43/8/083105}{\emph{Chin. Phys. C}
  {\bfseries 43} (2019) 083105}
  [\href{https://arxiv.org/abs/1905.01795}{{\ttfamily 1905.01795}}].

\bibitem{Crivellin:2017zlb}
A.~Crivellin, D.~M\"uller and T.~Ota, \emph{{Simultaneous explanation of
  $R(D^{(\ast)})$ and $b\to s \mu^+\mu^-$: the last scalar leptoquarks
  standing}}, \href{https://doi.org/10.1007/JHEP09(2017)040}{\emph{JHEP}
  {\bfseries 09} (2017) 040}
  [\href{https://arxiv.org/abs/1703.09226}{{\ttfamily 1703.09226}}].

\bibitem{Marzocca:2018wcf}
D.~Marzocca, \emph{{Addressing the B-physics anomalies in a fundamental
  Composite Higgs Model}},
  \href{https://doi.org/10.1007/JHEP07(2018)121}{\emph{JHEP} {\bfseries 07}
  (2018) 121} [\href{https://arxiv.org/abs/1803.10972}{{\ttfamily
  1803.10972}}].

\bibitem{Fuentes-Martin:2020pww}
J.~Fuentes-Martin, G.~Isidori, J.~Pag\`es and B.A.~Stefanek, \emph{{Flavor
  non-universal Pati-Salam unification and neutrino masses}},
  \href{https://doi.org/10.1016/j.physletb.2021.136484}{\emph{Phys. Lett. B}
  {\bfseries 820} (2021) 136484}
  [\href{https://arxiv.org/abs/2012.10492}{{\ttfamily 2012.10492}}].

\bibitem{Bigaran:2019bqv}
I.~Bigaran, J.~Gargalionis and R.R.~Volkas, \emph{{A near-minimal leptoquark
  model for reconciling flavour anomalies and generating radiative neutrino
  masses}}, \href{https://doi.org/10.1007/JHEP10(2019)106}{\emph{JHEP}
  {\bfseries 10} (2019) 106}
  [\href{https://arxiv.org/abs/1906.01870}{{\ttfamily 1906.01870}}].

\bibitem{Dev:2020qet}
P.S.~Bhupal~Dev, R.~Mohanta, S.~Patra and S.~Sahoo, \emph{{Unified explanation
  of flavor anomalies, radiative neutrino masses, and ANITA anomalous events in
  a vector leptoquark model}},
  \href{https://doi.org/10.1103/PhysRevD.102.095012}{\emph{Phys. Rev. D}
  {\bfseries 102} (2020) 095012}
  [\href{https://arxiv.org/abs/2004.09464}{{\ttfamily 2004.09464}}].

\bibitem{Altmannshofer:2020axr}
W.~Altmannshofer, P.S.B.~Dev, A.~Soni and Y.~Sui, \emph{{Addressing
  R$_{D^{(*)}}$, R$_{K^{(*)}}$, muon $g-2$ and ANITA anomalies in a minimal
  $R$-parity violating supersymmetric framework}},
  \href{https://doi.org/10.1103/PhysRevD.102.015031}{\emph{Phys. Rev. D}
  {\bfseries 102} (2020) 015031}
  [\href{https://arxiv.org/abs/2002.12910}{{\ttfamily 2002.12910}}].

\bibitem{Fuentes-Martin:2020bnh}
J.~Fuentes-Mart\'\i{}n and P.~Stangl, \emph{{Third-family quark-lepton
  unification with a fundamental composite Higgs}},
  \href{https://doi.org/10.1016/j.physletb.2020.135953}{\emph{Phys. Lett. B}
  {\bfseries 811} (2020) 135953}
  [\href{https://arxiv.org/abs/2004.11376}{{\ttfamily 2004.11376}}].

\bibitem{Endo:2021lhi}
M.~Endo, S.~Iguro, T.~Kitahara, M.~Takeuchi and R.~Watanabe,
  \emph{{Non-resonant new physics search at the LHC for the b \textrightarrow{}
  c\ensuremath{\tau}\ensuremath{\nu} anomalies}},
  \href{https://doi.org/10.1007/JHEP02(2022)106}{\emph{JHEP} {\bfseries 02}
  (2022) 106} [\href{https://arxiv.org/abs/2111.04748}{{\ttfamily
  2111.04748}}].

\bibitem{Belanger:2021smw}
G.~Belanger et~al., \emph{{Leptoquark manoeuvres in the dark: a simultaneous
  solution of the dark matter problem and the $ {R}_{D^{\left(\ast \right)}} $
  anomalies}}, \href{https://doi.org/10.1007/JHEP02(2022)042}{\emph{JHEP}
  {\bfseries 02} (2022) 042}
  [\href{https://arxiv.org/abs/2111.08027}{{\ttfamily 2111.08027}}].

\bibitem{Lee:2021jdr}
H.M.~Lee, \emph{{Leptoquark option for B-meson anomalies and leptonic
  signatures}}, \href{https://doi.org/10.1103/PhysRevD.104.015007}{\emph{Phys.
  Rev. D} {\bfseries 104} (2021) 015007}
  [\href{https://arxiv.org/abs/2104.02982}{{\ttfamily 2104.02982}}].

\bibitem{Djouadi:1989md}
A.~Djouadi, T.~Kohler, M.~Spira and J.~Tutas, \emph{{(e b), (e t) TYPE
  LEPTOQUARKS AT e p COLLIDERS}},
  \href{https://doi.org/10.1007/BF01560270}{\emph{Z. Phys. C} {\bfseries 46}
  (1990) 679}.

\bibitem{Chakraverty:2001yg}
D.~Chakraverty, D.~Choudhury and A.~Datta, \emph{{A Nonsupersymmetric
  resolution of the anomalous muon magnetic moment}},
  \href{https://doi.org/10.1016/S0370-2693(01)00419-1}{\emph{Phys. Lett. B}
  {\bfseries 506} (2001) 103}
  [\href{https://arxiv.org/abs/hep-ph/0102180}{{\ttfamily hep-ph/0102180}}].

\bibitem{Cheung:2001ip}
K.-m.~Cheung, \emph{{Muon anomalous magnetic moment and leptoquark solutions}},
  \href{https://doi.org/10.1103/PhysRevD.64.033001}{\emph{Phys. Rev. D}
  {\bfseries 64} (2001) 033001}
  [\href{https://arxiv.org/abs/hep-ph/0102238}{{\ttfamily hep-ph/0102238}}].

\bibitem{Biggio:2016wyy}
C.~Biggio, M.~Bordone, L.~Di~Luzio and G.~Ridolfi, \emph{{Massive vectors and
  loop observables: the $g-2$ case}},
  \href{https://doi.org/10.1007/JHEP10(2016)002}{\emph{JHEP} {\bfseries 10}
  (2016) 002} [\href{https://arxiv.org/abs/1607.07621}{{\ttfamily
  1607.07621}}].

\bibitem{Davidson:1993qk}
S.~Davidson, D.C.~Bailey and B.A.~Campbell, \emph{{Model independent
  constraints on leptoquarks from rare processes}},
  \href{https://doi.org/10.1007/BF01552629}{\emph{Z. Phys. C} {\bfseries 61}
  (1994) 613} [\href{https://arxiv.org/abs/hep-ph/9309310}{{\ttfamily
  hep-ph/9309310}}].

\bibitem{Couture:1995he}
G.~Couture and H.~Konig, \emph{{Bounds on second generation scalar leptoquarks
  from the anomalous magnetic moment of the muon}},
  \href{https://doi.org/10.1103/PhysRevD.53.555}{\emph{Phys. Rev. D} {\bfseries
  53} (1996) 555} [\href{https://arxiv.org/abs/hep-ph/9507263}{{\ttfamily
  hep-ph/9507263}}].

\bibitem{Mahanta:2001yc}
U.~Mahanta, \emph{{Implications of BNL measurement of delta a(mu) on a class of
  scalar leptoquark interactions}},
  \href{https://doi.org/10.1007/s100520100705}{\emph{Eur. Phys. J. C}
  {\bfseries 21} (2001) 171}
  [\href{https://arxiv.org/abs/hep-ph/0102176}{{\ttfamily hep-ph/0102176}}].

\bibitem{Queiroz:2014pra}
F.S.~Queiroz, K.~Sinha and A.~Strumia, \emph{{Leptoquarks, Dark Matter, and
  Anomalous LHC Events}},
  \href{https://doi.org/10.1103/PhysRevD.91.035006}{\emph{Phys. Rev. D}
  {\bfseries 91} (2015) 035006}
  [\href{https://arxiv.org/abs/1409.6301}{{\ttfamily 1409.6301}}].

\bibitem{ColuccioLeskow:2016dox}
E.~Coluccio~Leskow, G.~D'Ambrosio, A.~Crivellin and D.~M\"uller,
  \emph{{$(g-2)\mu$, lepton flavor violation, and $Z$ decays with leptoquarks:
  Correlations and future prospects}},
  \href{https://doi.org/10.1103/PhysRevD.95.055018}{\emph{Phys. Rev. D}
  {\bfseries 95} (2017) 055018}
  [\href{https://arxiv.org/abs/1612.06858}{{\ttfamily 1612.06858}}].

\bibitem{Chen:2017hir}
C.-H.~Chen, T.~Nomura and H.~Okada, \emph{{Excesses of muon $g-2$,
  $R_{D^{(\ast)}}$, and $R_K$ in a leptoquark model}},
  \href{https://doi.org/10.1016/j.physletb.2017.10.005}{\emph{Phys. Lett. B}
  {\bfseries 774} (2017) 456}
  [\href{https://arxiv.org/abs/1703.03251}{{\ttfamily 1703.03251}}].

\bibitem{Das:2016vkr}
D.~Das, C.~Hati, G.~Kumar and N.~Mahajan, \emph{{Towards a unified explanation
  of $R_{D^{(\ast)}}$, $R_{K}$ and $(g-2)_{\mu}$ anomalies in a left-right
  model with leptoquarks}},
  \href{https://doi.org/10.1103/PhysRevD.94.055034}{\emph{Phys. Rev. D}
  {\bfseries 94} (2016) 055034}
  [\href{https://arxiv.org/abs/1605.06313}{{\ttfamily 1605.06313}}].

\bibitem{Crivellin:2018qmi}
A.~Crivellin, M.~Hoferichter and P.~Schmidt-Wellenburg, \emph{{Combined
  explanations of $(g-2)_{\mu,e}$ and implications for a large muon EDM}},
  \href{https://doi.org/10.1103/PhysRevD.98.113002}{\emph{Phys. Rev. D}
  {\bfseries 98} (2018) 113002}
  [\href{https://arxiv.org/abs/1807.11484}{{\ttfamily 1807.11484}}].

\bibitem{Kowalska:2018ulj}
K.~Kowalska, E.M.~Sessolo and Y.~Yamamoto, \emph{{Constraints on charmphilic
  solutions to the muon g-2 with leptoquarks}},
  \href{https://doi.org/10.1103/PhysRevD.99.055007}{\emph{Phys. Rev. D}
  {\bfseries 99} (2019) 055007}
  [\href{https://arxiv.org/abs/1812.06851}{{\ttfamily 1812.06851}}].

\bibitem{Dorsner:2019itg}
I.~Dor\v{s}ner, S.~Fajfer and O.~Sumensari, \emph{{Muon $g-2$ and scalar
  leptoquark mixing}},
  \href{https://doi.org/10.1007/JHEP06(2020)089}{\emph{JHEP} {\bfseries 06}
  (2020) 089} [\href{https://arxiv.org/abs/1910.03877}{{\ttfamily
  1910.03877}}].

\bibitem{DelleRose:2020qak}
L.~Delle~Rose, C.~Marzo and L.~Marzola, \emph{{Simplified leptoquark models for
  precision $l_i \rightarrow l_f \gamma$ experiments: two-loop structure of
  $O(\alpha_S Y^2)$ corrections}},
  \href{https://doi.org/10.1103/PhysRevD.102.115020}{\emph{Phys. Rev. D}
  {\bfseries 102} (2020) 115020}
  [\href{https://arxiv.org/abs/2005.12389}{{\ttfamily 2005.12389}}].

\bibitem{Bigaran:2020jil}
I.~Bigaran and R.R.~Volkas, \emph{{Getting chirality right: Single scalar
  leptoquark solutions to the $(g-2)_{e,\mu}$ puzzle}},
  \href{https://doi.org/10.1103/PhysRevD.102.075037}{\emph{Phys. Rev. D}
  {\bfseries 102} (2020) 075037}
  [\href{https://arxiv.org/abs/2002.12544}{{\ttfamily 2002.12544}}].

\bibitem{Dorsner:2020aaz}
I.~Dor\v{s}ner, S.~Fajfer and S.~Saad, \emph{{$\mu \to e \gamma$ selecting
  scalar leptoquark solutions for the $(g-2)_{e,\mu}$ puzzles}},
  \href{https://doi.org/10.1103/PhysRevD.102.075007}{\emph{Phys. Rev. D}
  {\bfseries 102} (2020) 075007}
  [\href{https://arxiv.org/abs/2006.11624}{{\ttfamily 2006.11624}}].

\bibitem{Babu:2020hun}
K.S.~Babu, P.S.B.~Dev, S.~Jana and A.~Thapa, \emph{{Unified framework for
  $B$-anomalies, muon $g - 2$ and neutrino masses}},
  \href{https://doi.org/10.1007/JHEP03(2021)179}{\emph{JHEP} {\bfseries 03}
  (2021) 179} [\href{https://arxiv.org/abs/2009.01771}{{\ttfamily
  2009.01771}}].

\bibitem{Crivellin:2020tsz}
A.~Crivellin, D.~Mueller and F.~Saturnino, \emph{{Correlating $h\to \mu^+
  \mu^-$ to the Anomalous Magnetic Moment of the Muon via Leptoquarks}},
  \href{https://doi.org/10.1103/PhysRevLett.127.021801}{\emph{Phys. Rev. Lett.}
  {\bfseries 127} (2021) 021801}
  [\href{https://arxiv.org/abs/2008.02643}{{\ttfamily 2008.02643}}].

\bibitem{Marzocca:2021azj}
D.~Marzocca and S.~Trifinopoulos, \emph{{Minimal Explanation of Flavor
  Anomalies: B-Meson Decays, Muon Magnetic Moment, and the Cabibbo Angle}},
  \href{https://doi.org/10.1103/PhysRevLett.127.061803}{\emph{Phys. Rev. Lett.}
  {\bfseries 127} (2021) 061803}
  [\href{https://arxiv.org/abs/2104.05730}{{\ttfamily 2104.05730}}].

\bibitem{Wang:2021uqz}
X.~Wang, \emph{{Muon $(g-2)$ and Flavor Puzzles in the $U(1)^{}_{X}$-gauged
  Leptoquark Model}},  \href{https://arxiv.org/abs/2108.01279}{{\ttfamily
  2108.01279}}.

\bibitem{Perez:2021ddi}
P.F.~Perez, C.~Murgui and A.D.~Plascencia, \emph{{Leptoquarks and matter
  unification: Flavor anomalies and the muon g-2}},
  \href{https://doi.org/10.1103/PhysRevD.104.035041}{\emph{Phys. Rev. D}
  {\bfseries 104} (2021) 035041}
  [\href{https://arxiv.org/abs/2104.11229}{{\ttfamily 2104.11229}}].

\bibitem{Crivellin:2021egp}
A.~Crivellin, D.~M\"uller and L.~Schnell, \emph{{Combined constraints on first
  generation leptoquarks}},
  \href{https://doi.org/10.1103/PhysRevD.103.115023}{\emph{Phys. Rev. D}
  {\bfseries 103} (2021) 115023}
  [\href{https://arxiv.org/abs/2104.06417}{{\ttfamily 2104.06417}}].

\bibitem{Crivellin:2021bkd}
A.~Crivellin, M.~Hoferichter, M.~Kirk, C.A.~Manzari and L.~Schnell,
  \emph{{First-generation new physics in simplified models: from low-energy
  parity violation to the LHC}},
  \href{https://doi.org/10.1007/JHEP10(2021)221}{\emph{JHEP} {\bfseries 10}
  (2021) 221} [\href{https://arxiv.org/abs/2107.13569}{{\ttfamily
  2107.13569}}].

\bibitem{Crivellin:2021rbf}
A.~Crivellin, C.A.~Manzari and M.~Montull, \emph{{Correlating nonresonant
  di-electron searches at the LHC to the Cabibbo-angle anomaly and lepton
  flavor universality violation}},
  \href{https://doi.org/10.1103/PhysRevD.104.115016}{\emph{Phys. Rev. D}
  {\bfseries 104} (2021) 115016}
  [\href{https://arxiv.org/abs/2103.12003}{{\ttfamily 2103.12003}}].

\bibitem{CMS:2021ctt}
{\scshape CMS} collaboration, \emph{{Search for resonant and nonresonant new
  phenomena in high-mass dilepton final states at $ \sqrt{s} $ = 13 TeV}},
  \href{https://doi.org/10.1007/JHEP07(2021)208}{\emph{JHEP} {\bfseries 07}
  (2021) 208} [\href{https://arxiv.org/abs/2103.02708}{{\ttfamily
  2103.02708}}].

\bibitem{Belle:2018ezy}
{\scshape Belle} collaboration, \emph{{Measurement of the CKM matrix element
  $|V_{cb}|$ from $B^0\to D^{*-}\ell^ {+} \nu_\ell$ at Belle}},
  \href{https://doi.org/10.1103/PhysRevD.100.052007}{\emph{Phys. Rev. D}
  {\bfseries 100} (2019) 052007}
  [\href{https://arxiv.org/abs/1809.03290}{{\ttfamily 1809.03290}}].

\bibitem{Bobeth:2021lya}
C.~Bobeth, M.~Bordone, N.~Gubernari, M.~Jung and D.~van Dyk,
  \emph{{Lepton-flavour non-universality of ${\bar{B}}\rightarrow D^*\ell
  {{\bar{\nu }}}$ angular distributions in and beyond the Standard Model}},
  \href{https://doi.org/10.1140/epjc/s10052-021-09724-2}{\emph{Eur. Phys. J. C}
  {\bfseries 81} (2021) 984}
  [\href{https://arxiv.org/abs/2104.02094}{{\ttfamily 2104.02094}}].

\bibitem{Crivellin:2020mjs}
A.~Crivellin, C.~Greub, D.~M\"uller and F.~Saturnino, \emph{{Scalar Leptoquarks
  in Leptonic Processes}},
  \href{https://doi.org/10.1007/JHEP02(2021)182}{\emph{JHEP} {\bfseries 02}
  (2021) 182} [\href{https://arxiv.org/abs/2010.06593}{{\ttfamily
  2010.06593}}].

\bibitem{Crivellin:2020ukd}
A.~Crivellin, D.~M\"uller and F.~Saturnino, \emph{{Leptoquarks in oblique
  corrections and Higgs signal strength: status and prospects}},
  \href{https://doi.org/10.1007/JHEP11(2020)094}{\emph{JHEP} {\bfseries 11}
  (2020) 094} [\href{https://arxiv.org/abs/2006.10758}{{\ttfamily
  2006.10758}}].

\bibitem{deBlas:2021wap}
J.~de~Blas, M.~Ciuchini, E.~Franco, A.~Goncalves, S.~Mishima, M.~Pierini
  et~al., \emph{{Global analysis of electroweak data in the Standard Model}},
  \href{https://arxiv.org/abs/2112.07274}{{\ttfamily 2112.07274}}.

\bibitem{Crivellin:2021ejk}
A.~Crivellin and L.~Schnell, \emph{{Complete Lagrangian and set of Feynman
  rules for scalar leptoquarks}},
  \href{https://doi.org/10.1016/j.cpc.2021.108188}{\emph{Comput. Phys. Commun.}
  {\bfseries 271} (2022) 108188}
  [\href{https://arxiv.org/abs/2105.04844}{{\ttfamily 2105.04844}}].

\bibitem{Barger:1989rk}
V.D.~Barger, G.F.~Giudice and T.~Han, \emph{{Some New Aspects of Supersymmetry
  R-Parity Violating Interactions}},
  \href{https://doi.org/10.1103/PhysRevD.40.2987}{\emph{Phys. Rev. D}
  {\bfseries 40} (1989) 2987}.

\bibitem{Barbier:2004ez}
R.~Barbier et~al., \emph{{R-parity violating supersymmetry}},
  \href{https://doi.org/10.1016/j.physrep.2005.08.006}{\emph{Phys. Rept.}
  {\bfseries 420} (2005) 1}
  [\href{https://arxiv.org/abs/hep-ph/0406039}{{\ttfamily hep-ph/0406039}}].

\bibitem{Angelescu:2021lln}
A.~Angelescu, D.~Be\v{c}irevi\'c, D.A.~Faroughy, F.~Jaffredo and O.~Sumensari,
  \emph{{Single leptoquark solutions to the B-physics anomalies}},
  \href{https://doi.org/10.1103/PhysRevD.104.055017}{\emph{Phys. Rev. D}
  {\bfseries 104} (2021) 055017}
  [\href{https://arxiv.org/abs/2103.12504}{{\ttfamily 2103.12504}}].

\bibitem{He:2021yck}
S.-P.~He, \emph{{Leptoquark and vectorlike quark extended models as the
  explanation of the muon g-2 anomaly}},
  \href{https://doi.org/10.1103/PhysRevD.105.035017}{\emph{Phys. Rev. D}
  {\bfseries 105} (2022) 035017}
  [\href{https://arxiv.org/abs/2112.13490}{{\ttfamily 2112.13490}}].

\bibitem{Bigaran:2021kmn}
I.~Bigaran and R.R.~Volkas, \emph{{Reflecting on chirality: CP-violating
  extensions of the single scalar-leptoquark solutions for the
  (g-2)e,\ensuremath{\mu} puzzles and their implications for lepton EDMs}},
  \href{https://doi.org/10.1103/PhysRevD.105.015002}{\emph{Phys. Rev. D}
  {\bfseries 105} (2022) 015002}
  [\href{https://arxiv.org/abs/2110.03707}{{\ttfamily 2110.03707}}].

\bibitem{Iguro:2020keo}
S.~Iguro, M.~Takeuchi and R.~Watanabe, \emph{{Testing leptoquark/EFT in
  ${\bar{B}} \rightarrow {D^{(*)}}l{\bar{\nu }}$ at the LHC}},
  \href{https://doi.org/10.1140/epjc/s10052-021-09125-5}{\emph{Eur. Phys. J. C}
  {\bfseries 81} (2021) 406}
  [\href{https://arxiv.org/abs/2011.02486}{{\ttfamily 2011.02486}}].

\bibitem{Davighi:2020qqa}
J.~Davighi, M.~Kirk and M.~Nardecchia, \emph{{Anomalies and accidental
  symmetries: charging the scalar leptoquark under L$_{\mu}$ \ensuremath{-}
  L$_{\tau}$}}, \href{https://doi.org/10.1007/JHEP12(2020)111}{\emph{JHEP}
  {\bfseries 12} (2020) 111}
  [\href{https://arxiv.org/abs/2007.15016}{{\ttfamily 2007.15016}}].

\bibitem{Greljo:2021npi}
A.~Greljo, Y.~Soreq, P.~Stangl, A.E.~Thomsen and J.~Zupan, \emph{{Muonic Force
  Behind Flavor Anomalies}},
  \href{https://arxiv.org/abs/2107.07518}{{\ttfamily 2107.07518}}.

\bibitem{He:1990pn}
X.G.~He, G.C.~Joshi, H.~Lew and R.R.~Volkas, \emph{{NEW Z-prime
  PHENOMENOLOGY}}, \href{https://doi.org/10.1103/PhysRevD.43.R22}{\emph{Phys.
  Rev. D} {\bfseries 43} (1991) 22}.

\bibitem{Foot:1990mn}
R.~Foot, \emph{{New Physics From Electric Charge Quantization?}},
  \href{https://doi.org/10.1142/S0217732391000543}{\emph{Mod. Phys. Lett. A}
  {\bfseries 6} (1991) 527}.

\bibitem{He:1991qd}
X.-G.~He, G.C.~Joshi, H.~Lew and R.R.~Volkas, \emph{{Simplest Z-prime model}},
  \href{https://doi.org/10.1103/PhysRevD.44.2118}{\emph{Phys. Rev. D}
  {\bfseries 44} (1991) 2118}.

\bibitem{Straub:2018kue}
D.M.~Straub, \emph{{flavio: a Python package for flavour and precision
  phenomenology in the Standard Model and beyond}},
  \href{https://arxiv.org/abs/1810.08132}{{\ttfamily 1810.08132}}.

\bibitem{Huschle:2015rga}
{\scshape Belle} collaboration, \emph{{Measurement of the branching ratio of
  $\bar{B} \to D^{(\ast)} \tau^- \bar{\nu}_\tau$ relative to $\bar{B} \to
  D^{(\ast)} \ell^- \bar{\nu}_\ell$ decays with hadronic tagging at Belle}},
  \href{https://doi.org/10.1103/PhysRevD.92.072014}{\emph{Phys. Rev. D}
  {\bfseries 92} (2015) 072014}
  [\href{https://arxiv.org/abs/1507.03233}{{\ttfamily 1507.03233}}].

\bibitem{Hirose:2016wfn}
{\scshape Belle} collaboration, \emph{{Measurement of the $\tau$ lepton
  polarization and $R(D^*)$ in the decay $\bar{B} \to D^* \tau^-
  \bar{\nu}_\tau$}},
  \href{https://doi.org/10.1103/PhysRevLett.118.211801}{\emph{Phys. Rev. Lett.}
  {\bfseries 118} (2017) 211801}
  [\href{https://arxiv.org/abs/1612.00529}{{\ttfamily 1612.00529}}].

\bibitem{MILC:2015uhg}
{\scshape MILC} collaboration,
  \emph{{B\textrightarrow{}D\ensuremath{\ell}\ensuremath{\nu} form factors at
  nonzero recoil and |V$_{cb}$| from 2+1-flavor lattice QCD}},
  \href{https://doi.org/10.1103/PhysRevD.92.034506}{\emph{Phys. Rev. D}
  {\bfseries 92} (2015) 034506}
  [\href{https://arxiv.org/abs/1503.07237}{{\ttfamily 1503.07237}}].

\bibitem{Na:2015kha}
{\scshape HPQCD} collaboration, \emph{{$B \rightarrow D l \nu$ form factors at
  nonzero recoil and extraction of $|V_{cb}|$}},
  \href{https://doi.org/10.1103/PhysRevD.93.119906}{\emph{Phys. Rev. D}
  {\bfseries 92} (2015) 054510}
  [\href{https://arxiv.org/abs/1505.03925}{{\ttfamily 1505.03925}}].

\bibitem{Fajfer:2012vx}
S.~Fajfer, J.F.~Kamenik and I.~Nisandzic, \emph{{On the $B \to D^* \tau \bar
  \nu_{\tau}$ Sensitivity to New Physics}},
  \href{https://doi.org/10.1103/PhysRevD.85.094025}{\emph{Phys. Rev. D}
  {\bfseries 85} (2012) 094025}
  [\href{https://arxiv.org/abs/1203.2654}{{\ttfamily 1203.2654}}].

\bibitem{FlavourLatticeAveragingGroup:2019iem}
{\scshape Flavour Lattice Averaging Group} collaboration, \emph{{FLAG Review
  2019: Flavour Lattice Averaging Group (FLAG)}},
  \href{https://doi.org/10.1140/epjc/s10052-019-7354-7}{\emph{Eur. Phys. J. C}
  {\bfseries 80} (2020) 113}
  [\href{https://arxiv.org/abs/1902.08191}{{\ttfamily 1902.08191}}].

\bibitem{Bigi:2016mdz}
D.~Bigi and P.~Gambino, \emph{{Revisiting $B\to D \ell \nu$}},
  \href{https://doi.org/10.1103/PhysRevD.94.094008}{\emph{Phys. Rev. D}
  {\bfseries 94} (2016) 094008}
  [\href{https://arxiv.org/abs/1606.08030}{{\ttfamily 1606.08030}}].

\bibitem{Gambino:2019sif}
P.~Gambino, M.~Jung and S.~Schacht, \emph{{The $V_{cb}$ puzzle: An update}},
  \href{https://doi.org/10.1016/j.physletb.2019.06.039}{\emph{Phys. Lett. B}
  {\bfseries 795} (2019) 386}
  [\href{https://arxiv.org/abs/1905.08209}{{\ttfamily 1905.08209}}].

\bibitem{Bordone:2019vic}
M.~Bordone, M.~Jung and D.~van Dyk, \emph{{Theory determination of $\bar{B}\to
  D^{(*)}\ell^-\bar\nu$ form factors at $\mathcal{O}(1/m_c^2)$}},
  \href{https://doi.org/10.1140/epjc/s10052-020-7616-4}{\emph{Eur. Phys. J. C}
  {\bfseries 80} (2020) 74} [\href{https://arxiv.org/abs/1908.09398}{{\ttfamily
  1908.09398}}].

\bibitem{Aebischer:2018bkb}
J.~Aebischer, J.~Kumar and D.M.~Straub, \emph{{Wilson: a Python package for the
  running and matching of Wilson coefficients above and below the electroweak
  scale}}, \href{https://doi.org/10.1140/epjc/s10052-018-6492-7}{\emph{Eur.
  Phys. J. C} {\bfseries 78} (2018) 1026}
  [\href{https://arxiv.org/abs/1804.05033}{{\ttfamily 1804.05033}}].

\bibitem{Bordone:2016gaq}
M.~Bordone, G.~Isidori and A.~Pattori, \emph{{On the Standard Model predictions
  for $R_K$ and $R_{K^*}$}},
  \href{https://doi.org/10.1140/epjc/s10052-016-4274-7}{\emph{Eur. Phys. J. C}
  {\bfseries 76} (2016) 440}
  [\href{https://arxiv.org/abs/1605.07633}{{\ttfamily 1605.07633}}].

\bibitem{Isidori:2020acz}
G.~Isidori, S.~Nabeebaccus and R.~Zwicky, \emph{{QED corrections in $
  \overline{B}\to \overline{K}{\mathrm{\ell}}^{+}{\mathrm{\ell}}^{-} $ at the
  double-differential level}},
  \href{https://doi.org/10.1007/JHEP12(2020)104}{\emph{JHEP} {\bfseries 12}
  (2020) 104} [\href{https://arxiv.org/abs/2009.00929}{{\ttfamily
  2009.00929}}].

\bibitem{Serra:2016ivr}
N.~Serra, R.~Silva~Coutinho and D.~van Dyk, \emph{{Measuring the breaking of
  lepton flavor universality in $B\to K^*\ell^+\ell^-$}},
  \href{https://doi.org/10.1103/PhysRevD.95.035029}{\emph{Phys. Rev. D}
  {\bfseries 95} (2017) 035029}
  [\href{https://arxiv.org/abs/1610.08761}{{\ttfamily 1610.08761}}].

\bibitem{Capdevila:2017ert}
B.~Capdevila, S.~Descotes-Genon, L.~Hofer and J.~Matias, \emph{{Hadronic
  uncertainties in $B \to K^* \mu^+ \mu^-$: a state-of-the-art analysis}},
  \href{https://doi.org/10.1007/JHEP04(2017)016}{\emph{JHEP} {\bfseries 04}
  (2017) 016} [\href{https://arxiv.org/abs/1701.08672}{{\ttfamily
  1701.08672}}].

\bibitem{Bharucha:2015bzk}
A.~Bharucha, D.M.~Straub and R.~Zwicky, \emph{{$B\to V\ell^+\ell^-$ in the
  Standard Model from light-cone sum rules}},
  \href{https://doi.org/10.1007/JHEP08(2016)098}{\emph{JHEP} {\bfseries 08}
  (2016) 098} [\href{https://arxiv.org/abs/1503.05534}{{\ttfamily
  1503.05534}}].

\bibitem{Jager:2014rwa}
S.~J\"ager and J.~Martin~Camalich, \emph{{Reassessing the discovery potential
  of the $B \to K^{*} \ell^+\ell^-$ decays in the large-recoil region: SM
  challenges and BSM opportunities}},
  \href{https://doi.org/10.1103/PhysRevD.93.014028}{\emph{Phys. Rev. D}
  {\bfseries 93} (2016) 014028}
  [\href{https://arxiv.org/abs/1412.3183}{{\ttfamily 1412.3183}}].

\bibitem{Aebischer:2018iyb}
J.~Aebischer, J.~Kumar, P.~Stangl and D.M.~Straub, \emph{{A Global Likelihood
  for Precision Constraints and Flavour Anomalies}},
  \href{https://doi.org/10.1140/epjc/s10052-019-6977-z}{\emph{Eur. Phys. J. C}
  {\bfseries 79} (2019) 509}
  [\href{https://arxiv.org/abs/1810.07698}{{\ttfamily 1810.07698}}].

\bibitem{LHCb:2021lvy}
{\scshape LHCb} collaboration, \emph{{Tests of lepton universality using
  $B^0\to K^0_S \ell^+ \ell^-$ and $B^+\to K^{*+} \ell^+ \ell^-$ decays}},
  \href{https://arxiv.org/abs/2110.09501}{{\ttfamily 2110.09501}}.

\bibitem{Capdevila:2017iqn}
B.~Capdevila, A.~Crivellin, S.~Descotes-Genon, L.~Hofer and J.~Matias,
  \emph{{Searching for New Physics with $b\to s\tau^+\tau^-$ processes}},
  \href{https://doi.org/10.1103/PhysRevLett.120.181802}{\emph{Phys. Rev. Lett.}
  {\bfseries 120} (2018) 181802}
  [\href{https://arxiv.org/abs/1712.01919}{{\ttfamily 1712.01919}}].

\bibitem{Carvunis:2021dss}
A.~Carvunis, A.~Crivellin, D.~Guadagnoli and S.~Gangal, \emph{{Forward-backward
  asymmetry in B\textrightarrow{}D*\ensuremath{\ell}\ensuremath{\nu}: One more
  hint for scalar leptoquarks?}},
  \href{https://doi.org/10.1103/PhysRevD.105.L031701}{\emph{Phys. Rev. D}
  {\bfseries 105} (2022) L031701}
  [\href{https://arxiv.org/abs/2106.09610}{{\ttfamily 2106.09610}}].

\bibitem{Liptak:2021opc}
Z.~Liptak, M.~Kuriki and J.M.~Roney, \emph{{Possibilities for Upgrading to
  Polarized SuperKEKB}},
  \href{https://doi.org/10.18429/JACoW-IPAC2021-THPAB022}{\emph{JACoW}
  {\bfseries IPAC2021} (2021) THPAB022}.

\bibitem{Bernabeu:2007rr}
J.~Bernabeu, G.A.~Gonzalez-Sprinberg, J.~Papavassiliou and J.~Vidal, \emph{{Tau
  anomalous magnetic moment form-factor at super B/flavor factories}},
  \href{https://doi.org/10.1016/j.nuclphysb.2007.09.001}{\emph{Nucl. Phys. B}
  {\bfseries 790} (2008) 160}
  [\href{https://arxiv.org/abs/0707.2496}{{\ttfamily 0707.2496}}].

\bibitem{Bernabeu:2008ii}
J.~Bernabeu, G.A.~Gonzalez-Sprinberg and J.~Vidal, \emph{{Tau spin correlations
  and the anomalous magnetic moment}},
  \href{https://doi.org/10.1088/1126-6708/2009/01/062}{\emph{JHEP} {\bfseries
  01} (2009) 062} [\href{https://arxiv.org/abs/0807.2366}{{\ttfamily
  0807.2366}}].

\bibitem{Crivellin:2021spu}
A.~Crivellin, M.~Hoferichter and J.M.~Roney, \emph{{Towards testing the
  magnetic moment of the tau at one part per million}},
  \href{https://arxiv.org/abs/2111.10378}{{\ttfamily 2111.10378}}.

\bibitem{ParticleDataGroup:2020ssz}
{\scshape Particle Data Group} collaboration, \emph{{Review of Particle
  Physics}}, \href{https://doi.org/10.1093/ptep/ptaa104}{\emph{PTEP} {\bfseries
  2020} (2020) 083C01}.

\bibitem{Parker:2018vye}
R.H.~Parker, C.~Yu, W.~Zhong, B.~Estey and H.~M\"uller, \emph{{Measurement of
  the fine-structure constant as a test of the Standard Model}},
  \href{https://doi.org/10.1126/science.aap7706}{\emph{Science} {\bfseries 360}
  (2018) 191} [\href{https://arxiv.org/abs/1812.04130}{{\ttfamily
  1812.04130}}].

\bibitem{Morel:2020dww}
L.~Morel, Z.~Yao, P.~Clad\'e and S.~Guellati-Kh\'elifa, \emph{{Determination of
  the fine-structure constant with an accuracy of 81 parts per trillion}},
  \href{https://doi.org/10.1038/s41586-020-2964-7}{\emph{Nature} {\bfseries
  588} (2020) 61}.

\bibitem{Adelmann:2021udj}
A.~Adelmann et~al., \emph{{Search for a muon EDM using the frozen-spin
  technique}},  \href{https://arxiv.org/abs/2102.08838}{{\ttfamily
  2102.08838}}.

\bibitem{Aiba:2021bxe}
M.~Aiba et~al., \emph{{Science Case for the new High-Intensity Muon Beams HIMB
  at PSI}},  \href{https://arxiv.org/abs/2111.05788}{{\ttfamily 2111.05788}}.

\bibitem{Aebischer:2021uvt}
J.~Aebischer, W.~Dekens, E.E.~Jenkins, A.V.~Manohar, D.~Sengupta and
  P.~Stoffer, \emph{{Effective field theory interpretation of lepton magnetic
  and electric dipole moments}},
  \href{https://doi.org/10.1007/JHEP07(2021)107}{\emph{JHEP} {\bfseries 07}
  (2021) 107} [\href{https://arxiv.org/abs/2102.08954}{{\ttfamily
  2102.08954}}].

\bibitem{Dorsner:2016wpm}
I.~Dor\v{s}ner, S.~Fajfer, A.~Greljo, J.F.~Kamenik and N.~Ko\v{s}nik,
  \emph{{Physics of leptoquarks in precision experiments and at particle
  colliders}}, \href{https://doi.org/10.1016/j.physrep.2016.06.001}{\emph{Phys.
  Rept.} {\bfseries 641} (2016) 1}
  [\href{https://arxiv.org/abs/1603.04993}{{\ttfamily 1603.04993}}].

\bibitem{Qweak:2014xey}
{\scshape Qweak} collaboration, \emph{{The Q$_{weak}$ experimental apparatus}},
  \href{https://doi.org/10.1016/j.nima.2015.01.023}{\emph{Nucl. Instrum. Meth.
  A} {\bfseries 781} (2015) 105}
  [\href{https://arxiv.org/abs/1409.7100}{{\ttfamily 1409.7100}}].

\bibitem{Carlini:2019ksi}
R.D.~Carlini, W.T.H.~van Oers, M.L.~Pitt and G.R.~Smith, \emph{{Determination
  of the Proton's Weak Charge and Its Constraints on the Standard Model}},
  \href{https://doi.org/10.1146/annurev-nucl-101918-023633}{\emph{Ann. Rev.
  Nucl. Part. Sci.} {\bfseries 69} (2019) 191}.

\bibitem{Qweak:2018tjf}
{\scshape Qweak} collaboration, \emph{{Precision measurement of the weak charge
  of the proton}},
  \href{https://doi.org/10.1038/s41586-018-0096-0}{\emph{Nature} {\bfseries
  557} (2018) 207} [\href{https://arxiv.org/abs/1905.08283}{{\ttfamily
  1905.08283}}].

\bibitem{Wood:1997zq}
C.S.~Wood, S.C.~Bennett, D.~Cho, B.P.~Masterson, J.L.~Roberts, C.E.~Tanner
  et~al., \emph{{Measurement of parity nonconservation and an anapole moment in
  cesium}}, \href{https://doi.org/10.1126/science.275.5307.1759}{\emph{Science}
  {\bfseries 275} (1997) 1759}.

\bibitem{Guena:2004sq}
J.~Guena, M.~Lintz and M.A.~Bouchiat, \emph{{Measurement of the parity
  violating 6S-7S transition amplitude in cesium achieved within 2 x 10(-13)
  atomic-unit accuracy by stimulated-emission detection}},
  \href{https://doi.org/10.1103/PhysRevA.71.042108}{\emph{Phys. Rev. A}
  {\bfseries 71} (2005) 042108}
  [\href{https://arxiv.org/abs/physics/0412017}{{\ttfamily physics/0412017}}].

\bibitem{Cadeddu:2021dqx}
M.~Cadeddu, N.~Cargioli, F.~Dordei, C.~Giunti and E.~Picciau, \emph{{Muon and
  electron g-2 and proton and cesium weak charges implications on dark Zd
  models}}, \href{https://doi.org/10.1103/PhysRevD.104.L011701}{\emph{Phys.
  Rev. D} {\bfseries 104} (2021) 011701}
  [\href{https://arxiv.org/abs/2104.03280}{{\ttfamily 2104.03280}}].

\bibitem{Erler:2013xha}
J.~Erler and S.~Su, \emph{{The Weak Neutral Current}},
  \href{https://doi.org/10.1016/j.ppnp.2013.03.004}{\emph{Prog. Part. Nucl.
  Phys.} {\bfseries 71} (2013) 119}
  [\href{https://arxiv.org/abs/1303.5522}{{\ttfamily 1303.5522}}].

\bibitem{ALEPH:2005ab}
{\scshape ALEPH, DELPHI, L3, OPAL, SLD, LEP Electroweak Working Group, SLD
  Electroweak Group, SLD Heavy Flavour Group} collaboration, \emph{{Precision
  electroweak measurements on the $Z$ resonance}},
  \href{https://doi.org/10.1016/j.physrep.2005.12.006}{\emph{Phys. Rept.}
  {\bfseries 427} (2006) 257}
  [\href{https://arxiv.org/abs/hep-ex/0509008}{{\ttfamily hep-ex/0509008}}].

\bibitem{Arnan:2019olv}
P.~Arnan, D.~Becirevic, F.~Mescia and O.~Sumensari, \emph{{Probing low energy
  scalar leptoquarks by the leptonic $W$ and $Z$ couplings}},
  \href{https://doi.org/10.1007/JHEP02(2019)109}{\emph{JHEP} {\bfseries 02}
  (2019) 109} [\href{https://arxiv.org/abs/1901.06315}{{\ttfamily
  1901.06315}}].

\bibitem{Aoki:2021kgd}
Y.~Aoki et~al., \emph{{FLAG Review 2021}},
  \href{https://arxiv.org/abs/2111.09849}{{\ttfamily 2111.09849}}.

\bibitem{Bona:2017gut}
{\scshape Utfit} collaboration, \emph{{Unitarity Triangle Analysis and D meson
  mixing in the Standard Model and Beyond}},
  \href{https://doi.org/10.22323/1.314.0205}{\emph{PoS} {\bfseries EPS-HEP2017}
  (2017) 205}.

\bibitem{UTfit:2006onp}
{\scshape UTfit} collaboration, \emph{{Constraints on new physics from the
  quark mixing unitarity triangle}},
  \href{https://doi.org/10.1103/PhysRevLett.97.151803}{\emph{Phys. Rev. Lett.}
  {\bfseries 97} (2006) 151803}
  [\href{https://arxiv.org/abs/hep-ph/0605213}{{\ttfamily hep-ph/0605213}}].

\bibitem{UTfit:2007eik}
{\scshape UTfit} collaboration, \emph{{Model-independent constraints on $\Delta
  F=2$ operators and the scale of new physics}},
  \href{https://doi.org/10.1088/1126-6708/2008/03/049}{\emph{JHEP} {\bfseries
  03} (2008) 049} [\href{https://arxiv.org/abs/0707.0636}{{\ttfamily
  0707.0636}}].

\bibitem{Crivellin:2021lix}
A.~Crivellin, J.F.~Eguren and J.~Virto, \emph{{Next-to-Leading-Order QCD
  Matching for $\Delta F=2$ Processes in Scalar Leptoquark Models}},
  \href{https://arxiv.org/abs/2109.13600}{{\ttfamily 2109.13600}}.

\bibitem{Ellis:2016jkw}
J.~Ellis, \emph{{TikZ-Feynman: Feynman diagrams with TikZ}},
  \href{https://doi.org/10.1016/j.cpc.2016.08.019}{\emph{Comput. Phys. Commun.}
  {\bfseries 210} (2017) 103}
  [\href{https://arxiv.org/abs/1601.05437}{{\ttfamily 1601.05437}}].

\bibitem{Greljo:2017vvb}
A.~Greljo and D.~Marzocca, \emph{{High-$p_T$ dilepton tails and flavor
  physics}}, \href{https://doi.org/10.1140/epjc/s10052-017-5119-8}{\emph{Eur.
  Phys. J. C} {\bfseries 77} (2017) 548}
  [\href{https://arxiv.org/abs/1704.09015}{{\ttfamily 1704.09015}}].

\bibitem{Alwall:2014hca}
J.~Alwall, R.~Frederix, S.~Frixione, V.~Hirschi, F.~Maltoni, O.~Mattelaer
  et~al., \emph{{The automated computation of tree-level and next-to-leading
  order differential cross sections, and their matching to parton shower
  simulations}}, \href{https://doi.org/10.1007/JHEP07(2014)079}{\emph{JHEP}
  {\bfseries 07} (2014) 079} [\href{https://arxiv.org/abs/1405.0301}{{\ttfamily
  1405.0301}}].

\bibitem{Degrande:2011ua}
C.~Degrande, C.~Duhr, B.~Fuks, D.~Grellscheid, O.~Mattelaer and T.~Reiter,
  \emph{{UFO - The Universal FeynRules Output}},
  \href{https://doi.org/10.1016/j.cpc.2012.01.022}{\emph{Comput. Phys. Commun.}
  {\bfseries 183} (2012) 1201}
  [\href{https://arxiv.org/abs/1108.2040}{{\ttfamily 1108.2040}}].

\bibitem{Borschensky:2020hot}
C.~Borschensky, B.~Fuks, A.~Kulesza and D.~Schwartl\"ander, \emph{{Scalar
  leptoquark pair production at hadron colliders}},
  \href{https://doi.org/10.1103/PhysRevD.101.115017}{\emph{Phys. Rev. D}
  {\bfseries 101} (2020) 115017}
  [\href{https://arxiv.org/abs/2002.08971}{{\ttfamily 2002.08971}}].

\bibitem{Borschensky:2021jyk}
C.~Borschensky, B.~Fuks, A.~Kulesza and D.~Schwartl\"ander, \emph{{Precision
  predictions for scalar leptoquark pair production at the LHC}},  in
  \emph{{European Physical Society Conference on High Energy Physics 2021}},
  10, 2021 [\href{https://arxiv.org/abs/2110.15324}{{\ttfamily 2110.15324}}].

\bibitem{Ball:2021leu}
R.D.~Ball et~al., \emph{{The Path to Proton Structure at One-Percent
  Accuracy}},  \href{https://arxiv.org/abs/2109.02653}{{\ttfamily 2109.02653}}.

\bibitem{ATLAS:2020yat}
{\scshape ATLAS} collaboration, \emph{{Search for new non-resonant phenomena in
  high-mass dilepton final states with the ATLAS detector}},
  \href{https://doi.org/10.1007/JHEP11(2020)005}{\emph{JHEP} {\bfseries 11}
  (2020) 005} [\href{https://arxiv.org/abs/2006.12946}{{\ttfamily
  2006.12946}}].

\bibitem{ATLAS:2020zms}
{\scshape ATLAS} collaboration, \emph{{Search for heavy Higgs bosons decaying
  into two tau leptons with the ATLAS detector using $pp$ collisions at
  $\sqrt{s}=13$ TeV}},
  \href{https://doi.org/10.1103/PhysRevLett.125.051801}{\emph{Phys. Rev. Lett.}
  {\bfseries 125} (2020) 051801}
  [\href{https://arxiv.org/abs/2002.12223}{{\ttfamily 2002.12223}}].

\bibitem{Jaffredo:2021ymt}
F.~Jaffredo, \emph{{Revisiting mono-$\tau$ tails at the LHC}},
  \href{https://arxiv.org/abs/2112.14604}{{\ttfamily 2112.14604}}.

\bibitem{ATLAS:2021bjk}
{\scshape ATLAS} collaboration, \emph{{Search for high-mass resonances in final
  states with a tau lepton and missing transverse momentum with the ATLAS
  detector}}, .

\bibitem{Buonocore:2020erb}
L.~Buonocore, U.~Haisch, P.~Nason, F.~Tramontano and G.~Zanderighi,
  \emph{{Lepton-Quark Collisions at the Large Hadron Collider}},
  \href{https://doi.org/10.1103/PhysRevLett.125.231804}{\emph{Phys. Rev. Lett.}
  {\bfseries 125} (2020) 231804}
  [\href{https://arxiv.org/abs/2005.06475}{{\ttfamily 2005.06475}}].

\bibitem{ATLAS:2020dsk}
{\scshape ATLAS} collaboration, \emph{{Search for pairs of scalar leptoquarks
  decaying into quarks and electrons or muons in $ \sqrt{s} $ = 13 TeV $pp$
  collisions with the ATLAS detector}},
  \href{https://doi.org/10.1007/JHEP10(2020)112}{\emph{JHEP} {\bfseries 10}
  (2020) 112} [\href{https://arxiv.org/abs/2006.05872}{{\ttfamily
  2006.05872}}].

\bibitem{ATLAS:2020xov}
{\scshape ATLAS} collaboration, \emph{{Search for pair production of scalar
  leptoquarks decaying into first- or second-generation leptons and top quarks
  in proton\textendash{}proton collisions at $\sqrt{s}$ = 13 TeV with the ATLAS
  detector}}, \href{https://doi.org/10.1140/epjc/s10052-021-09009-8}{\emph{Eur.
  Phys. J. C} {\bfseries 81} (2021) 313}
  [\href{https://arxiv.org/abs/2010.02098}{{\ttfamily 2010.02098}}].

\bibitem{ATLAS:2021oiz}
{\scshape ATLAS} collaboration, \emph{{Search for pair production of
  third-generation scalar leptoquarks decaying into a top quark and a
  $\tau$-lepton in $pp$ collisions at $ \sqrt{s} $ = 13 TeV with the ATLAS
  detector}}, \href{https://doi.org/10.1007/JHEP06(2021)179}{\emph{JHEP}
  {\bfseries 06} (2021) 179}
  [\href{https://arxiv.org/abs/2101.11582}{{\ttfamily 2101.11582}}].

\bibitem{ATLAS:2019qpq}
{\scshape ATLAS} collaboration, \emph{{Searches for third-generation scalar
  leptoquarks in $\sqrt{s}$ = 13 TeV pp collisions with the ATLAS detector}},
  \href{https://doi.org/10.1007/JHEP06(2019)144}{\emph{JHEP} {\bfseries 06}
  (2019) 144} [\href{https://arxiv.org/abs/1902.08103}{{\ttfamily
  1902.08103}}].

\bibitem{Peskin:1990zt}
M.E.~Peskin and T.~Takeuchi, \emph{{A New constraint on a strongly interacting
  Higgs sector}}, \href{https://doi.org/10.1103/PhysRevLett.65.964}{\emph{Phys.
  Rev. Lett.} {\bfseries 65} (1990) 964}.

\bibitem{Ellis:2018gqa}
J.~Ellis, C.W.~Murphy, V.~Sanz and T.~You, \emph{{Updated Global SMEFT Fit to
  Higgs, Diboson and Electroweak Data}},
  \href{https://doi.org/10.1007/JHEP06(2018)146}{\emph{JHEP} {\bfseries 06}
  (2018) 146} [\href{https://arxiv.org/abs/1803.03252}{{\ttfamily
  1803.03252}}].

\bibitem{CDF:2013dpa}
{\scshape CDF, D0} collaboration, \emph{{Combination of CDF and D0 $W$-Boson
  Mass Measurements}},
  \href{https://doi.org/10.1103/PhysRevD.88.052018}{\emph{Phys. Rev. D}
  {\bfseries 88} (2013) 052018}
  [\href{https://arxiv.org/abs/1307.7627}{{\ttfamily 1307.7627}}].

\bibitem{ATLAS:2019fyu}
{\scshape ATLAS} collaboration, \emph{{Measurement of $W^{\pm }$-boson and
  Z-boson production cross-sections in pp collisions at $\sqrt{s}=2.76$ TeV
  with the ATLAS detector}},
  \href{https://doi.org/10.1140/epjc/s10052-019-7399-7}{\emph{Eur. Phys. J. C}
  {\bfseries 79} (2019) 901}
  [\href{https://arxiv.org/abs/1907.03567}{{\ttfamily 1907.03567}}].

\bibitem{AbdusSalam:2020rdj}
S.S.~AbdusSalam et~al., \emph{{Simple and statistically sound strategies for
  analysing physical theories}},
  \href{https://arxiv.org/abs/2012.09874}{{\ttfamily 2012.09874}}.

\bibitem{Virtanen:2019joe}
P.~Virtanen et~al., \emph{{SciPy 1.0--Fundamental Algorithms for Scientific
  Computing in Python}},
  \href{https://doi.org/10.1038/s41592-019-0686-2}{\emph{Nature Meth.}
  {\bfseries 17} (2020) 261}
  [\href{https://arxiv.org/abs/1907.10121}{{\ttfamily 1907.10121}}].

\bibitem{Alguero:2018nvb}
M.~Alguer\'o, B.~Capdevila, S.~Descotes-Genon, P.~Masjuan and J.~Matias,
  \emph{{Are we overlooking lepton flavour universal new physics in $b\to
  s\ell\ell$ ?}}, \href{https://doi.org/10.1103/PhysRevD.99.075017}{\emph{Phys.
  Rev. D} {\bfseries 99} (2019) 075017}
  [\href{https://arxiv.org/abs/1809.08447}{{\ttfamily 1809.08447}}].

\bibitem{Alguero:2019ptt}
M.~Alguer\'o, B.~Capdevila, A.~Crivellin, S.~Descotes-Genon, P.~Masjuan,
  J.~Matias et~al., \emph{{Emerging patterns of New Physics with and without
  Lepton Flavour Universal contributions}},
  \href{https://doi.org/10.1140/epjc/s10052-019-7216-3}{\emph{Eur. Phys. J. C}
  {\bfseries 79} (2019) 714}
  [\href{https://arxiv.org/abs/1903.09578}{{\ttfamily 1903.09578}}].

\bibitem{Crivellin:2021rbq}
A.~Crivellin and M.~Hoferichter, \emph{{Consequences of chirally enhanced
  explanations of $(g-2)_{\mu}$ for $h\to\mu\mu$ and $Z\to\mu\mu$}},
  \href{https://doi.org/10.1007/JHEP07(2021)135}{\emph{JHEP} {\bfseries 07}
  (2021) 135} [\href{https://arxiv.org/abs/2104.03202}{{\ttfamily
  2104.03202}}].

\bibitem{FCC:2018vvp}
{\scshape FCC} collaboration, \emph{{FCC-hh: The Hadron Collider}: {Future
  Circular Collider Conceptual Design Report Volume 3}},
  \href{https://doi.org/10.1140/epjst/e2019-900087-0}{\emph{Eur. Phys. J. ST}
  {\bfseries 228} (2019) 755}.

\bibitem{Borzumati:1999sp}
F.~Borzumati, G.R.~Farrar, N.~Polonsky and S.D.~Thomas, \emph{{Soft Yukawa
  couplings in supersymmetric theories}},
  \href{https://doi.org/10.1016/S0550-3213(99)00328-4}{\emph{Nucl. Phys. B}
  {\bfseries 555} (1999) 53}
  [\href{https://arxiv.org/abs/hep-ph/9902443}{{\ttfamily hep-ph/9902443}}].

\bibitem{Crivellin:2010gw}
A.~Crivellin and J.~Girrbach, \emph{{Constraining the MSSM sfermion mass
  matrices with light fermion masses}},
  \href{https://doi.org/10.1103/PhysRevD.81.076001}{\emph{Phys. Rev. D}
  {\bfseries 81} (2010) 076001}
  [\href{https://arxiv.org/abs/1002.0227}{{\ttfamily 1002.0227}}].

\bibitem{Djouadi:1989me}
A.~Djouadi and M.~Spira, \emph{{MEASURING STATIC QUARK PROPERTIES AT LEP}},
  \href{https://doi.org/10.1016/0370-2693(89)91575-X}{\emph{Phys. Lett. B}
  {\bfseries 228} (1989) 443}.

\bibitem{Sirunyan:2021khd}
{\scshape CMS} collaboration, \emph{{Search for resonant and nonresonant new
  phenomena in high-mass dilepton final states at $ \sqrt{s} $ = 13 TeV}},
  \href{https://doi.org/10.1007/JHEP07(2021)208}{\emph{JHEP} {\bfseries 07}
  (2021) 208} [\href{https://arxiv.org/abs/2103.02708}{{\ttfamily
  2103.02708}}].

\bibitem{Junk:1999kv}
T.~Junk, \emph{{Confidence level computation for combining searches with small
  statistics}},
  \href{https://doi.org/10.1016/S0168-9002(99)00498-2}{\emph{Nucl. Instrum.
  Meth. A} {\bfseries 434} (1999) 435}
  [\href{https://arxiv.org/abs/hep-ex/9902006}{{\ttfamily hep-ex/9902006}}].

\bibitem{Sjostrand:2014zea}
T.~Sj\"ostrand, S.~Ask, J.R.~Christiansen, R.~Corke, N.~Desai, P.~Ilten et~al.,
  \emph{{An introduction to PYTHIA 8.2}},
  \href{https://doi.org/10.1016/j.cpc.2015.01.024}{\emph{Comput. Phys. Commun.}
  {\bfseries 191} (2015) 159}
  [\href{https://arxiv.org/abs/1410.3012}{{\ttfamily 1410.3012}}].

\bibitem{Conte:2012fm}
E.~Conte, B.~Fuks and G.~Serret, \emph{{MadAnalysis 5, A User-Friendly
  Framework for Collider Phenomenology}},
  \href{https://doi.org/10.1016/j.cpc.2012.09.009}{\emph{Comput. Phys. Commun.}
  {\bfseries 184} (2013) 222}
  [\href{https://arxiv.org/abs/1206.1599}{{\ttfamily 1206.1599}}].

\bibitem{Conte:2014zja}
E.~Conte, B.~Dumont, B.~Fuks and C.~Wymant, \emph{{Designing and recasting LHC
  analyses with MadAnalysis 5}},
  \href{https://doi.org/10.1140/epjc/s10052-014-3103-0}{\emph{Eur. Phys. J. C}
  {\bfseries 74} (2014) 3103}
  [\href{https://arxiv.org/abs/1405.3982}{{\ttfamily 1405.3982}}].

\bibitem{Conte:2018vmg}
E.~Conte and B.~Fuks, \emph{{Confronting new physics theories to LHC data with
  MADANALYSIS 5}}, \href{https://doi.org/10.1142/S0217751X18300272}{\emph{Int.
  J. Mod. Phys. A} {\bfseries 33} (2018) 1830027}
  [\href{https://arxiv.org/abs/1808.00480}{{\ttfamily 1808.00480}}].

\bibitem{Cacciari:2011ma}
M.~Cacciari, G.P.~Salam and G.~Soyez, \emph{{FastJet User Manual}},
  \href{https://doi.org/10.1140/epjc/s10052-012-1896-2}{\emph{Eur. Phys. J. C}
  {\bfseries 72} (2012) 1896}
  [\href{https://arxiv.org/abs/1111.6097}{{\ttfamily 1111.6097}}].

\bibitem{Araz:2020lnp}
J.Y.~Araz, B.~Fuks and G.~Polykratis, \emph{{Simplified fast detector
  simulation in MADANALYSIS 5}},
  \href{https://doi.org/10.1140/epjc/s10052-021-09052-5}{\emph{Eur. Phys. J. C}
  {\bfseries 81} (2021) 329}
  [\href{https://arxiv.org/abs/2006.09387}{{\ttfamily 2006.09387}}].

\bibitem{Pendlebury:2015lrz}
J.M.~Pendlebury et~al., \emph{{Revised experimental upper limit on the electric
  dipole moment of the neutron}},
  \href{https://doi.org/10.1103/PhysRevD.92.092003}{\emph{Phys. Rev. D}
  {\bfseries 92} (2015) 092003}
  [\href{https://arxiv.org/abs/1509.04411}{{\ttfamily 1509.04411}}].

\bibitem{Baker:2006ts}
C.A.~Baker et~al., \emph{{An Improved experimental limit on the electric dipole
  moment of the neutron}},
  \href{https://doi.org/10.1103/PhysRevLett.97.131801}{\emph{Phys. Rev. Lett.}
  {\bfseries 97} (2006) 131801}
  [\href{https://arxiv.org/abs/hep-ex/0602020}{{\ttfamily hep-ex/0602020}}].

\bibitem{nEDM:2020crw}
C.~Abel et~al., \emph{{Measurement of the Permanent Electric Dipole Moment of
  the Neutron}},
  \href{https://doi.org/10.1103/PhysRevLett.124.081803}{\emph{Phys. Rev. Lett.}
  {\bfseries 124} (2020) 081803}
  [\href{https://arxiv.org/abs/2001.11966}{{\ttfamily 2001.11966}}].

\bibitem{Griffith:2009zz}
W.C.~Griffith, M.D.~Swallows, T.H.~Loftus, M.V.~Romalis, B.R.~Heckel and
  E.N.~Fortson, \emph{{Improved Limit on the Permanent Electric Dipole Moment
  of Hg-199}},
  \href{https://doi.org/10.1103/PhysRevLett.102.101601}{\emph{Phys. Rev. Lett.}
  {\bfseries 102} (2009) 101601}
  [\href{https://arxiv.org/abs/0901.2328}{{\ttfamily 0901.2328}}].

\bibitem{Graner:2016ses}
B.~Graner, Y.~Chen, E.G.~Lindahl and B.R.~Heckel, \emph{{Reduced Limit on the
  Permanent Electric Dipole Moment of Hg199}},
  \href{https://doi.org/10.1103/PhysRevLett.116.161601}{\emph{Phys. Rev. Lett.}
  {\bfseries 116} (2016) 161601}
  [\href{https://arxiv.org/abs/1601.04339}{{\ttfamily 1601.04339}}].

\bibitem{Dekens:2018bci}
W.~Dekens, J.~de~Vries, M.~Jung and K.K.~Vos, \emph{{The phenomenology of
  electric dipole moments in models of scalar leptoquarks}},
  \href{https://doi.org/10.1007/JHEP01(2019)069}{\emph{JHEP} {\bfseries 01}
  (2019) 069} [\href{https://arxiv.org/abs/1809.09114}{{\ttfamily
  1809.09114}}].

\end{thebibliography}\endgroup

\end{document}